\documentclass[12pt]{article}
\usepackage{amsmath}
\usepackage{amssymb}
\begin{document}
\begin{center}
{\bf {\Large   {Modified  2D Proca Theory: Revisited Under BRST and (Anti-)Chiral Superfield  Formalisms}}}

\vskip 2.0cm

{\sf B. Chauhan$^{(a)}$, S. Kumar$^{(a)}$, A. Tripathi$^{(a)}$, R. P. Malik$^{(a,b)}$}\\
$^{(a)}$ {\it Physics Department, Center of Advance Studies, Institute of Science,}\\
{\it Banaras Hindu University, Varanasi - 221 005, (U.P.), India}\\

\vskip 0.1cm

\vskip 0.1cm

$^{(b)}$ {\it DST Center for Interdisciplinary Mathematical Sciences,}\\
{\it Institute of Science, Banaras Hindu University, Varanasi - 221 005, India}\\
{\small {\sf {e-mails: bchauhan501@gmail.com; sunil.bhu93@gmail.com;\\ ankur1793@gmail.com; rpmalik1995@gmail.com}}}

\end{center}

\vskip 1.5cm

\noindent
{\bf Abstract:} Within the framework of Becchi-Rouet-Stora-Tyutin (BRST) approach, we discuss {\it mainly} 
the fermionic (i.e. off-shell nilpotent)  (anti-)BRST, (anti-)co-BRST 
and some discrete dual-symmetries of the appropriate Lagrangian densities for a  two (1+1)-dimensional (2D) modified 
Proca (i.e. a {\it massive} Abelian 1-form)  theory {\it without} any interaction with matter fields. One of the {\it novel} observations 
of our present investigation is the existence of some kinds of restrictions in the case of 
our present St\"{u}ckelberg-modified version of the 2D Proca theory which are {\it not} like  the 
standard Curci-Ferrari (CF)-condition of a non-Abelian 1-form gauge theory. Some kinds of similarities and a few differences between {\it them} have been pointed 
out in our present investigation. To establish the sanctity of the above off-shell nilpotent (anti-)BRST and (anti-)co-BRST symmetries, 
we derive {\it them} by using our {\it newly} proposed (anti-)chiral superfield formalism where a few {\it specific} and appropriate sets of  invariant quantities play a decisive role. We express 
the (anti-)BRST and (anti-)co-BRST conserved charges in terms of the superfields that are 
obtained after the applications of (anti-)BRST and (anti-)co-BRST invariant restrictions and prove their off-shell
nilpotency and absolute anticommutativity properties, too. Finally, we make some comments on (i) the novelty of our  restrictions/obstructions, and
  (ii) the physics behind the  negative kinetic term associated with the  pseudo-scalar  field of our present theory.  \\

\vskip 1.4 cm


\vskip 0.25cm


\newpage
\section{Introduction}

One of the simplest gauge theories is the well-known Maxwell $U(1)$ gauge theory which can be generalized to the Proca theory by incorporating a  mass term
in the Lagrangian density for the bosonic field (thereby rendering the {\it latter} field to acquire {\it three} degrees of freedom in the 
{\it physical} four (3 + 1)-dimensional (4D) flat Minkowskian spacetime). The beautiful gauge symmetry of the Maxwell theory (generated by the first-class constraints) is {\it not} respected by the Proca theory because the {\it latter} is endowed with the second-class constraints in the terminology of Dirac's prescription 
for the classification scheme of constraints (see, e.g. [1-3] for details). By exploiting the theoretical potential and power of the celebrated St\"{u}ckelberg 
formalism (see, e.g. [4]), the beautiful gauge symmetry can be restored by invoking a new  pure real scalar field in the theory. This happens 
 because the second-class constraints of the Proca theory get converted into the first-class constraints which generate the gauge 
symmetry transformations (see, e.g. [5, 6]) for the St\"{u}ckelberg-modified version of the Proca theory in any arbitrary dimension of spacetime.
 As a consequence, the modified version of the Proca theory is an example of the {\it massive} gauge theory.

The purpose of our present investigation is to concentrate on the two (1 + 1)-dimensional (2D) St\"{u}ckelberg-modified version of the Proca theory 
within the framework of Becchi-Rouet-Stora-Tyutin (BRST) formalism and show the existence of {\it fermionic} (anti-)BRST and (anti-)co-BRST
symmetry transformations as well as other kinds of {\it discrete} and {\it continuous} symmetries which provide the physical realizations
 of the de Rham cohomological operators\footnote{On a compact manifold without a boundary, the set of three operators $(d,\, \delta, \,\Delta)$ constitute the de 
Rham cohomological operators of differential geometry [7-11] where $(\delta)d$  are the (co-)exterior derivatives (with $d^2 = \delta ^2 = 0$) and 
$\Delta =  (d + \delta)^2 = \{d, \delta\}$ is the Laplacian  operator with an underlying algebra: $d^2 = \delta ^2 = 0, \;\Delta  = \{d, \delta\},\;
[\Delta, d] = [\Delta, \delta] = 0$ which is popularly known as the Hodge algebra.} of differential geometry [7-11]. In other words, we prove that the {\it{massive}} 2D Abelian 1-form gauge 
theory (i.e. the  St\"{u}ckelberg-modified version of the 2D Proca theory) is a field-theoretic example of Hodge theory. In this context, it is pertinent to
 point out that we have {\it already} shown, in our earlier work [12], that the above modified 2D Proca theory is a tractable field-theoretic model for the Hodge theory. 
However, the fermionic (anti-)BRST and (anti-)co-BRST symmetries of the theory have been shown to be nilpotent and absolutely anticommuting in
 nature ({\it only} on the on-shell). The question of the existence of the {\it off-shell} nilpotent {\it fermionic} symmetries has {\it not} been discussed, in detail, in our previous works. We accomplish  {\it this} goal cogently  in our present endeavor.

Against the backdrop of the discussions on the models for the Hodge theory, we would like to state that we have established that any 
arbitrary Abelian $p$-form ($p$ = 1, 2, 3...) gauge theory is a model for the Hodge theory in $D = 2\,p$ dimensions of spacetime 
(see, e.g. [13-15] for details). However, these models are for the {\it massless} fields because these are field-theoretic
examples of {\it gauge} theories. In addition, we have shown that the $ \mathcal {N}  = 2$ supersymmetric quantum 
mechanical models [16-20] are {\it also} examples for  the Hodge theory. These latter  models are, however, {\it massive} but they are {\it not} gauge theories
because these are {\it not} endowed with the first-class constraints in the terminology of Dirac's classification scheme for
constraints  (see, e.g., [1-3]). Thus, the St\"{u}ckelberg-modified 2D Proca theory is very  {\it special}  because, for this field-theoretic  
model, {\it mass} and {\it gauge} invariance 
co-exist {\it together} at the {\it classical} level and, at the {\it quantum} level, many discrete and continuous {\it internal} symmetries exist for this theory
 within the framework of BRST formalism. We discuss these symmetries {\it  extensively} in our present endeavor.

In our present investigation, we have demonstrated the existence of two {\it equivalent} Lagrangian densities for the 2D Proca theory (within the framework of BRST formalism) which respect the {\it off-shell} nilpotent and absolutely anticommuting (anti-)BRST and (anti-)co-BRST symmetry transformations (separately and independently). We have also shown, for the first time, the existence of some restrictions in the case of our present 2D {\it massive}
{\it  Abelian} 1-form gauge theory which are distinctly different from the {\it usual} CF-condition that exists for the {\it non-Abelian} 1-form gauge theory [21]. We have obtained the correct expressions for the  conserved (anti-)BRST and (anti-)co-BRST charges which are found to be off-shell nilpotent and absolutely anticommuting (separately and independently). To verify the sanctity of the (anti-)BRST and (anti-)co-BRST symmetries (and corresponding conserved charges), we have applied our newly proposed  (anti-)chiral superfield approach to BRST formalism [22-25] and proven their nilpotency and absolute anticommutativity  properties. We have captured the existence of the {\it new} type of restrictions/obstructions  within the framework of (anti-)chiral superfield approach to BRST formalism {\it while} proving the invariance of the Lagrangian densities under the (anti-)BRST and (anti-) co-BRST symmetry transformations (cf. Sec. 6). We have {\it also} shown that the discrete and continuous symmetries of the {\it equivalent} Lagrangian densities are such that {\it both} of them represent the field-theoretic examples of Hodge theory
({\it independently} and {\it separately}).

We would like to state a few words about the geometrical superfield approach [26-33] to BRST formalism (SFABF)  which leads to the 
derivation of the (anti-)BRST symmetries and the CF-type condition [21] in the context of (non-)Abelian 1-form gauge theories. Within the framework of 
SFABF, a given $D$-dimensional gauge theory is generalized  onto a $(D, 2)$-dimensional supermanifold which is parameterized by the superspace coordinates
$Z^\mu = (x^\mu, \theta, \bar\theta)$ where $x^\mu$ (with $\mu =  0, 1,...D-1$) are the bosonic coordinates associated with the $D$-dimensional Minkowski 
space and a pair of Grassmannian variables $(\theta, \bar\theta)$ satisfy: $\theta^2 = \bar\theta^2 = 0, \theta\bar\theta + \bar\theta\theta = 0$.
We invoke the theoretical strength of celebrated horizontality condition to obtain the (anti-)BRST symmetries and  the CF-condition [21]. In the process,
 we {\it also} provide the geometrical basis for the {\it abstract} mathematical properties (i.e. nilpotency and absolute anticommutativity) that are associated with the (anti-)BRST symmetries and corresponding conserved charges. In our recent works [22-25], we have simplified the above SFABF by considering {\it only} the (anti-)chiral superfields on the $(D, 1)$-dimensional super-submanifolds of the general $(D, 2)$-dimensional supermanifold and obtained the  (anti-)BRST symmetries by demanding the Grassmannian independence of the (anti-)BRST invariant quantities at the {\it{quantum}} level. The {\it novel} observation, in this context,  has been the result that the conserved (anti-)BRST charges turn out to be absolutely anticommuting\footnote {The absolute anticommutativity 
property of the conserved charges  is obvious when we take the {\it full} expansions of the superfields that are defined on the $(D, 2)$-dimensional supermanifold.} {\it{even}} within the framework of the (anti-)chiral superfield approach to BRST formalism [22-25] where only {\it one} Grassmannian variable is taken into account. 
This observation should be contrasted  with the applications of the (anti-)chiral supervariable approach to the $\mathcal N = 2$ SUSY quantum mechanical models 
where the {\it absolute anticommutativity} is {\it not} respected.

Our present investigation is essential and interesting on the following counts. First and foremost, we wish to 
discuss the {\it off-shell} nilpotent (anti-)BRST and (anti-)co-BRST symmetries (in detail) for our present modified version of
2D Proca theory in contrast to our earlier work [12] where we have discussed {\it only} the {\it on-shell} nilpotent version of the 
above {\it fermionic} symmetries. Second, there are some very interesting discrete symmetries in the theory which have {\it not} been discussed 
in [12]. These discrete symmetries are essential for the proof of equivalence of the coupled Lagrangian densities of our 2D 
{\it massive} Abelian  1-form gauge theory. Third, for the first time, we find a set of non-trivial  restrictions/obstructions 
in our Stueckelberg-modified version of the  Proca (i.e.  {\it massive} Abelian 1-form) theory  which are {\it not} 
like the {\it usual} CF-condition [21] of the non-Abelian 1-form gauge theory.
We dwell briefly on the key differences and  some kinds of  similarities of these {\it different} restrictions. Fourth, we apply the (anti-)chiral
superfield approach to derive the nilpotent (anti-)BRST and (anti-)co-BRST symmetries to {\it prove} the 
sanctity of these nilpotent transformations. Fifth, the proof of {\it absolute anticommutativity} of the nilpotent (anti-)BRST and
(anti-)co-BRST symmetries is a {\it novel} observation within the framework of (anti-)chiral superfield approach.
Sixth, the existence of a pseudo-scalar  field with a negative kinetic term and its physical relevances are pointed out at the fag ends of Secs. 7 and 8. 
 Finally, there are some {\it novel} observations in our present investigation that
we point out at the fag end of our present paper. At the moment, we do {\it not} know the reasons behind the existence of these {\it novel} features 
in the context of our St\"{u}ckelberg-modified version of the 2D Proca gauge theory (cf. Sec. 7 below).

Our present paper is organized as follows. First of all, to set the notions, we recapitulate the bare essentials of our earlier work [12]
and discuss the on-shell nilpotent symmetries of the theory in the Lagrangian formulation. We also show the existence of the
equivalent {\it two} Lagrangian densities for our modified version of 2D Proca (i.e. a {\it massive} Abelian 1-form) gauge theory in Sec. 2.
Our Sec. 3 is  devoted to the discussion of the off-shell nilpotent version of  (anti-)BRST, (anti-)co-BRST symmetries
and the existence of  restrictions/obstructions  on the theory.
In Sec. 4, we derive the conserved currents and corresponding charges. We also prove the off-shell nilpotency and absolute anticommutativity properties associated with them. Our Sec. 5 deals with the derivations of {\it all} the conserved and nilpotent charges and their proof of the
off-shell nilpotency and absolute anticommutativity within the framework of our {\it newly} proposed (anti-)chiral superfield approach [22-25]. Sec. 6 contains the proof of the invariance(s) of the Lagrangian densities within the framework of (anti-)chiral superfield approach. In this section, we prove the sanctity of the underlying CF-type restrictions
our theory, too. We devote time, in Sec. 7, on the discussion of our {\it new}  restrictions for the coupled Lagrangian densities and discuss their some kinds of 
similarities   and distinct differences with the standard CF-condition that exists in the case of non-Abelian 1-form gauge theory [21]. We also briefly comment on the negative kinetic term (associated with the pseudo-scalar field of our  present modified version of 2D Proca theory). Finally,
we summarize the key results of our present investigation  and point out a few future directions for further investigation(s) in Sec. 8.

In our Appendices A, B and C, we discuss a few explicit computations. The essence of {\it these}  has been incorporated in the main body 
of our text. The contents   of these   Appendices are essential for the full appreciation of the key results of our present paper.
Our Appendix D is devoted to a concise discussion of bosonic and ghost symmetries of the two {\it equivalent} Lagrangian
densities of our theory to prove that {\it both} of them represent models for the Hodge theory provided we consider {\it all}
the discrete and continuous symmetries  {\it together}.\\

\noindent
{\it Convention and Notations}: 
We choose the background 2D Minkowskian flat spacetime metric $(\eta_{\mu\nu}$)  with the signatures 
$(+1, -1)$ so that $P\cdot Q = P_\mu\,Q^\mu = \eta_{\mu\nu} P^\mu\,Q^\nu\equiv (P_0\,Q_0 -P_i\,Q_i)$ 
for the non-null 2D vectors $P_\mu$ and $Q_\mu$ where the Greek indices $\mu,\,\nu,\,\lambda,... = 0,1$ and Latin indices
$i,j,k,... = 1$ (because there is only {\it one} space direction in our theory). We also take the  
Levi-Civita tensor $\varepsilon_{\mu\nu}$ such that $\varepsilon_{01}=  \varepsilon^{10} = +1$ and
$\varepsilon_{\mu\nu}\varepsilon^{\nu\lambda} = \delta^\lambda_\mu$, $\varepsilon_{\mu\nu}\varepsilon^{\mu\nu} = -\,2!$, etc.
We denote, in the whole body of our text, the (anti-)BRST 
and (anti-)co-BRST symmetries of {\it all} varieties  (and in {\it all} contexts) by the symbols $s_{(a)b}$ and $s_{(a)d}$, respectively. 
We also adopt the convention of the left-derivative w.r.t. all the {\it fermionic} fields and use the
notations $\Box = \partial_0^2- \partial_1^2 $ and $\dot\Psi = \frac{\partial \Psi}{\partial t}$, etc, for a generic field $\Psi$. We focus 
only on the {\it internal} symmetries of our 2D theory and spacetime symmetries of the 2D
Minkowskian spacetime manifold do {\it not} play any crucial role in our whole discussion.

\section{Preliminaries: Lagrangian Formulation and Various Kinds of  Symmetries}

We begin with the celebrated Proca (i.e. a {\it massive}  Abelian 1-form) theory in any arbitrary $D$-dimension  of spacetime. 
This theory, with the rest mass $m$  for the vector boson, is described by the following Lagrangian density (see, e.g. [4]) 
\begin{eqnarray}
{\cal L}_{(p)} = -\frac {1}{4} F_{\mu\nu} F^{\mu\nu} +  \frac {m^2}{2}  A_\mu \, A^\mu,
\end{eqnarray}
 where the antisymmetric field strength  tensor  $F_{\mu\nu}=\partial_\mu A_\nu - \partial_\nu A_\mu$
 is derived from the 2-form [$F^{(2)} = \frac{1} {2!}\, (d x^\mu \wedge  d x^\nu) = d A^{(1)}$]
 where the nilpotent ($d^2  = 0$) exterior derivative $d = d x^\mu\partial_\mu$ (with $\mu = 0, 1...D-1$) acts on a 1-form 
 ($A^{(1)} =  d x^\mu A_\mu$) to produce the 2-form $F^{(2)}$ w.r.t. to the vector potential $A_\mu$.
This theory is endowed with the second-class constraints  and, therefore, it does {\it not} respect any kind of {\it gauge} symmetry.
However, one can exploit the theoretical strength of the Stueckelberg formalism [4] and replace $A_\mu$ by 
\begin{eqnarray}
A_\mu \longrightarrow A_\mu \mp \frac {1}{m}\, \partial_\mu \phi,
\end{eqnarray}
where $\phi$ is a pure scalar field. The resulting Stueckelberg's modified Lagrangian density
\begin{eqnarray}
{\cal L}_{(s)} = -\frac {1}{4} F_{\mu\nu} F^{\mu\nu} +  \frac {m^2}{2}  A_\mu \, A^\mu 
+ \frac {1}{2}\partial_\mu\phi \, \partial^\mu\phi \mp m \, A_\mu \partial^\mu \phi,
\end{eqnarray}
is endowed with the first-class constraints (with $\Pi ^0 = 0, \Pi ^i = E_i, \partial_i\,E_i = \vec\nabla\cdot \vec E)$ 
\begin{eqnarray}
\Pi ^0 \approx 0, \qquad \vec\nabla\cdot\vec E \mp  m \,\Pi _{\phi} \approx 0,  
\end{eqnarray}
where $\Pi ^\mu = - F^{0\mu}$ and $\Pi _{\phi} = \dot \phi \mp \,m\, A_0$ are the momenta w.r.t. $A_\mu$ and $\phi$ and
$\vec E$ is the electric field (present as a component in $F_{\mu\nu}$). The generator of the infinitesimal gauge transformations
($\delta_g$) can be written, in terms of the above first-class constraints, as [5, 6]
\begin{eqnarray}
G = \int d^{(D-1)}x\, \Big[ \dot\Sigma \,( \Pi ^0 ) - \Sigma \,\,(\vec\nabla\cdot\vec E \mp  m \,\Pi _{\phi})\Big], 
\end{eqnarray}
where $\Sigma (x)$ is the gauge transformation parameter (with   $\dot\Sigma = \frac{\partial \Sigma }{\partial t})$. The above generator
leads to the following gauge transformation for a generic field $\Psi$, namely;
\begin{eqnarray}
\delta_g\,\Psi = - \,i\,[\Psi, G],~~~~~~~~~~~~~\Psi = A_\mu,\, \phi,
\end{eqnarray}
where we have to use the following equal-time canonical commutators  (with $\hbar  = c = 1)$
\begin{eqnarray}
&&[A_0(\vec x, t), \,\,\Pi _0 (\vec y, t)] \,= \;i\, \delta ^{(D-1))}(\vec x- \vec y),\nonumber\\
&&[A_i(\vec x, t), \, \,E_j (\vec y, t)] \,\,= \;i\, \delta_{ij}\,\delta ^{(D-1))}(\vec x- \vec y),\nonumber\\
&&[\phi(\vec x, t), \, \,\Pi _{\phi} (\vec y, t)] \,\;\;= \;i\, \delta ^{(D-1))}(\vec x- \vec y),
\end{eqnarray}
and the rest of the equal-time commutators are taken to be zero. Ultimately, we obtain the following infinitesimal gauge transformations $(\delta_g)$, namely;
\begin{eqnarray}
\delta_g\, A_\mu = \partial_\mu\,\Sigma,~~~~~~~~~~~~~~~~~~~~ \delta_g\,\phi = \pm\, m\, \Sigma,
\end{eqnarray}   
which are valid in any arbitrary $D$-dimension of spacetime.

For the definition of the propagator for the {\it massive} vector field $A_\mu$ {\it and}  for the purpose of quantization of the 
Stueckelberg-modified Lagrangian density ${\cal L}_{(s)}$, we have to incorporate the gauge-fixing term. The ensuing
Lagrangian density   ${\cal L}_{(s)}^{(g)}$ is
\begin{eqnarray}
{\cal L}_{(s)}^{(g)} = -\frac {1}{4} F_{\mu\nu} F^{\mu\nu} +  \frac {m^2}{2}  A_\mu \, A^\mu 
+ \frac {1}{2}\partial_\mu\phi \, \partial^\mu\phi \mp m \, A_\mu \partial^\mu \phi- \frac {1}{2}(\partial\cdot A \pm m\,\phi)^2,
\end{eqnarray} 
which {\it does not} respect the gauge symmetry transformations (8) unless we put a restriction from {\it outside} equal to $(\Box + m^2)\,\Sigma = 0$. 
In the special case of two (1 + 1)-dimensional (2D) theory, the Lagrangian density (9) takes the following form:
\begin{eqnarray}
{\cal L}_{(s)}^{(2D)} = \frac {1}{2} \,E^2 +  \frac {m^2}{2}  A_\mu \, A^\mu 
+ \frac {1}{2}\partial_\mu\phi \, \partial^\mu\phi \mp m \, A_\mu \partial^\mu \phi- \frac {1}{2}(\partial\cdot A \pm m\,\phi)^2,
\end{eqnarray}
because, in 2D spacetime, we have only $F_{01} = - F_{10} =  E = -\,\varepsilon^{\mu\nu}\partial_\mu A_\nu$ as the existing 
(non-zero) component of $F_{\mu\nu}$ (because  there is no magnetic field in this  theory). The above gauge-fixed Lagrangian density has the following {\it generalized}
form (see, e.g. [12]):
\begin{eqnarray} 
{\cal L}^{(2D)} &=& \frac {1}{2}\, {(E \mp m\,\tilde\phi)}^2 \pm m\, E\,\tilde\phi - \frac {1}{2}\,\partial_\mu \,\tilde\phi\,\,\partial^\mu\,\tilde\phi 
+ \frac {m^2}{2} A_\mu\, A^ \mu + \frac {1}{2}\, \partial_\mu\, \phi\, \partial^\mu\, \phi 
\nonumber\\ &\mp& m \,A_\mu \,\partial^\mu\, \phi
 -\frac {1}{2}\,(\partial\cdot A \pm m\,\phi)^2.
\end{eqnarray}  
In the above, we have generalized  ($\frac {E ^2 }{2})$ in the {\it same} manner as the Stueckelberg formalism generalizes $(\frac {m^2}{2}  A_\mu \, A^\mu )$ 
term in the Lagrangian density (3). To be precise, we have incorporated a pseudo-scalar field $(\tilde\phi)$ in our theory because the electric field $E$
is a pseudo-scalar in 2D spacetime. It will be worthwhile to point out that all the {\it basic}  fields of our 2D theory (i.e. $A_\mu, \phi, \tilde\phi)$   
have mass dimension {\it zero} in the natural units (where $\hbar = c = 1$).

\subsection{Discrete Symmetries  and (Dual-)Gauge Symmetries}

\noindent
We shall now concentrate on the most generalized version of the 2D Lagrangian density (11) for our further discussions. In this connection, it can be checked that
under the following discrete symmetry transformations
\begin{eqnarray}
A_\mu \to \pm\, i\, \varepsilon_{\mu\nu} A^\nu, \qquad \phi \to \pm\, i\, \tilde \phi, 
\qquad \tilde \phi \to \pm \,i \, \phi,
\end{eqnarray} 
the 2D Lagrangian density ${\cal L}^{(2D)}$ remains invariant (because $E \to \mp\, i\, (\partial \cdot A), \,
(\partial \cdot A) \to \mp \,i\, E$ due to   $A_\mu \to \pm\, i\, \varepsilon_{\mu\nu} A^\nu $) modulo some total spacetime derivaties.
Furthermore, it is very interesting to point out that under the following (dual-)gauge transformations ($\delta _{(d)g}$) 
\begin{eqnarray}
&& \delta_{dg} A_\mu = - \varepsilon_{\mu\nu} \partial^\nu \Omega , \qquad \delta_{dg} \tilde \phi = \mp\,m\,\Omega, \qquad \delta_{dg} \phi = 0,
\qquad \delta_{dg} (\partial \cdot A \pm m \phi) = 0\nonumber\\
&& \delta_{dg} E = \Box \,\Omega, \qquad \delta_{dg} (E \mp m \,\tilde \phi) = (\Box + m^2)\,\Omega,  \nonumber\\
&& \delta_{g} A_\mu = \partial_\mu \Sigma , \qquad \delta_{g} \phi = \pm \,m \, \Sigma , \qquad \delta_{g} \tilde \phi = 0,
\qquad  \delta_{g} E = 0,\nonumber\\
&&\delta_{g} (\partial \cdot A) = \Box \,\Sigma, \qquad \delta_{g} (\partial \cdot A \pm m \phi) = (\Box + m^2)\, \Sigma,  
\end{eqnarray}
the 2D Lagrangian density transforms as
\begin{eqnarray}
\delta_{dg} {\cal L}^{(2D)} &=& \partial_\mu \Bigl [ m \,\varepsilon^{\mu\nu}\; 
\bigl (m \,A_\nu \,\Omega 
\pm \,\phi\, \partial_\nu \,\Omega  \bigr ) \pm m \,\tilde \phi \,\partial^\mu \,\Omega  \Bigr ]
+ \,(E \mp m \,\tilde \phi) \, (\Box + m^2) \, \Omega , \nonumber\\
\delta_{g} {\cal L}^{(2D)} &=& - (\partial \cdot A\,  \pm \,m \, \phi) \, (\Box + m^2) \, \Sigma,
\end{eqnarray}
where $\Sigma (x)$ and $\Omega (x)$ are the infinitesimal gauge and dual-gauge transformation 
parameters. In other words, $\Sigma (x)$ and $\Omega (x)$ are the pure scalar and pseudo-scalar, respectively.

At this stage a few comments are in order. First of all, there are two {\it equivalent} gauge-fixed Lagrangian densities that
are hidden in (11), namely;   
\begin{eqnarray} 
{\cal L}_{(1)} &=& \frac {1}{2}\, {(E - m\,\tilde\phi)}^2 + m\, E\,\tilde\phi - \frac {1}{2}\,\partial_\mu \,\tilde\phi\,\,\partial^\mu\,\tilde\phi 
+ \frac {m^2}{2} A_\mu\, A^ \mu + \frac {1}{2}\, \partial_\mu\, \phi\, \partial^\mu\, \phi 
\nonumber\\ &-& m \,A_\mu \,\partial^\mu\, \phi
 -\frac {1}{2}\,(\partial\cdot A + m\,\phi)^2,
\end{eqnarray} 
\begin{eqnarray} 
{\cal L}_{(2)} &=& \frac {1}{2}\, {(E  + m\,\tilde\phi)}^2 - m\, E\,\tilde\phi - \frac {1}{2}\,\partial_\mu \,\tilde\phi\,\,\partial^\mu\,\tilde\phi 
+ \frac {m^2}{2} A_\mu\, A^ \mu + \frac {1}{2}\, \partial_\mu\, \phi\, \partial^\mu\, \phi 
\nonumber\\ & + & m \,A_\mu \,\partial^\mu\, \phi
 -\frac {1}{2}\,(\partial\cdot A - m\,\phi)^2,
\end{eqnarray}
which are connected to each-other by a discrete symmetry transformations: $\phi\rightarrow  -\,\phi, \tilde\phi\rightarrow  -\,\tilde\phi,
A_\mu\rightarrow  A_\mu.$ Second, it is obvious that the (dual-)gauge transformation parametere $\Omega $ and $\Sigma$ are constrained  by 
the {\it same} type of restrictions (i.e. ($\Box + m^2)\;\Omega   = 0, (\Box + m^2)\;\Sigma  = 0$) from {\it outside} if we wish
to have  perfect (dual-)gauge symmetries  
in the theory. Third, we note that {\it only} one pair of ghost and anti-ghost fields would be good enough to take care of {\it these}
 restrictions for the {\it perfect} {\it``quantum"} (dual-)gauge  (i.e. BRST-type) symmetries within the framework of BRST  formalism. Fourth, one of the decisive features of the (dual-)gauge symmetries is the observation that the {\it gauge-fixing} and 
{\it kinetic terms}  of our 2D theory remain invariant under {\it these} symmetries, respectively.\\

\vskip 1 cm

\subsection{On-Shell Nilpotent Symmetries and Discrete Symmetries}

In our-earlier work [12], we have taken up one of the above Lagrangian densities (i.e. (15)) for the generalizations of 
the (dual-)gauge symmetries at the ``quantum" level within the framework of BRST formalism. For instance, the following (anti-)BRST symmetries
 (which are the generalizations of the gauge symmetries (13)), namely;
\begin{eqnarray}
&& s_{ab} \,A_\mu = \partial_\mu \bar C,\quad s_{ab} C  = i\, (\partial\cdot A  + m\, \phi),\; \qquad s_{ab} \phi = m\,\bar C,\nonumber\\
&&s_{ab} \bar C = 0, \quad s_{ab} E = s_{ab} \tilde \phi = 0,\quad s_{ab} \,(\partial \cdot A + m \phi) = (\Box + m^2)\,\bar C, \nonumber\\
&& s_b \,A_\mu = \partial_\mu  C,\quad s_b \bar C  = -\,i\, (\partial\cdot A  + m\, \phi), \qquad s_b \phi = m\,C,\nonumber\\
&&s_b  C = 0, \quad s_b E = s_b \tilde \phi = 0,\quad s_b \,(\partial \cdot A + m \phi) = (\Box + m^2)\, C, 
\end{eqnarray} 
leave the following Lagrangian density invariant (modulo a total spacetime derivative)   
\begin{eqnarray} 
{\cal L}_{(B_{1})} &=& \frac {1}{2}\, {(E - m\,\tilde\phi)}^2 + m\, E\,\tilde\phi - \frac {1}{2}\,\partial_\mu \,\tilde\phi\,\,\partial^\mu\,\tilde\phi 
+ \frac {m^2}{2} A_\mu\, A^ \mu + \frac {1}{2}\, \partial_\mu\, \phi\, \partial^\mu\, \phi 
\nonumber\\ &-& m \,A_\mu \,\partial^\mu\, \phi
 -\frac {1}{2}\,(\partial\cdot A + m\,\phi)^2 - i\, \partial_\mu \bar C\,\partial^\mu C + i\, m^2 \bar C \,C,
\end{eqnarray} 
which is a generalization of the gauge-fixed 2D Lagrangian density ${\cal L}_{(1)}$ to the ``quantum" level (within the framework 
of BRST formalism where the last two terms, in the Lagrangian density (18), are the Faddeev-Popov ghost terms). It should be noted that 
the fermionic (i.e. $C^2 = \bar C^2 = 0, \, C \,\bar C +  \bar C \, C = 0$) (anti-)ghost fields $(\bar C)C$ are introduced in the theory to maintain the unitarity 
at any arbitrary order of perturbative computations.

A few comments are in order at this juncture. First, we note that the {\it total} kinetic term (associated with the gauge field)  remains invariant 
(i.e. $s_{(a)b} \, E = s_{(a)b} \tilde \phi = 0$) under the nilpotent (anti-)BRST symmetry
transformations $s_{(a)b}$. Second, the (anti-)BRST symmetries are on-shell nilpotent $(s^2_{(a)b} = 0)$ as we have to use the relevant EOMs: 
$(\Box + m^2)\, C  = 0, (\Box + m^2)\, \bar C  = 0$ for the proof of the nilpotency property. Third, the Lagrangian density ${\cal L}_{(B_2)}$,
as the generalized version of (16),  can {\it also} be obtained from ${\cal L}_{(B_1)}$ by the replacements: $\phi \to - \phi, \, \tilde \phi
\to - \tilde \phi, \, A_\mu \to A_\mu, \, C \to C, \, \bar C \to \bar C$. Fourth, the (anti-)BRST symmetries for the Lagrangina density ${\cal L}_{(B_2)}$
can {\it also} be obtained from (17) by the above replacements (i.e. $\phi \to - \phi, \, \tilde \phi
\to - \tilde \phi, \, A_\mu \to A_\mu, \, C \to C, \, \bar C \to \bar C$). Finally, we conclude that {\it both} the Lagrangian densities ${\cal L}_{(B_2)}$
and ${\cal L}_{(B_1)}$ are {\it equivalent} and they describe the {\it same} 2D Stueckelberg-modified massive Abelian 1-form gauge theory.

In addition to the on-shell nilpotent (anti-)BRST symmetries (17), there is another set of on-shell nilpotent ($s_{(a)d}^2 = 0$) (anti-)co-BRST
(or (anti-)dual BRST)  symmetries $s_{(a)d}$ in our theory because under these (i.e. $s_{(a)d}$) transformations 
\begin{eqnarray}
&& s_{ad} \,A_\mu = - \,\varepsilon_{\mu\nu} \, \partial^\nu C,\quad s_{ad} \bar C  = +\, i\,(E  - m\, \tilde \phi),\; \qquad s_{ad} \phi = 0,\nonumber\\
&&s_{ad}  C = 0, \quad s_{ad} E = \Box C,\quad s_{ad} \,(\partial \cdot A + m \phi) = 0, \quad s_{ad} \tilde \phi = - m \,C,\nonumber\\
&& s_d \,A_\mu = - \varepsilon_{\mu\nu}  \partial^\nu \bar C,\quad s_d C  = -\,i\, (E  - m\, \tilde \phi), \qquad s_d \phi = 0,\nonumber\\
&&s_d  \bar C = 0, \quad s_d E = \Box \bar C,\quad s_d \,(\partial \cdot A + m \phi) = 0, \quad s_d \tilde \phi = - m \, \bar C, 
\end{eqnarray}
the Lagrangian density ${\cal L}_{(B_1)}$ (cf. Eq. (18)) remains invariant, modulo some total spacetime  derivatives, as listed below:
\begin{eqnarray}
s_{ad} \,  {\cal L}_{(B_1)} &=&  \partial_\mu \Bigl [ m \,\varepsilon^{\mu\nu}\; 
\bigl (m \,A_\nu \, C
+ \,\phi\, \partial_\nu \, C  \bigr ) + E \,\partial^\mu \,C  \Bigr ],\nonumber\\
s_d \,  {\cal L}_{(B_1)} &=& \partial_\mu \Bigl [ m \,\varepsilon^{\mu\nu}\; 
\bigl (m \,A_\nu \, \bar C
+ \,\phi\, \partial_\nu \, \bar C  \bigr ) + E\,\partial^\mu \,\bar C  \Bigr ].
\end{eqnarray}
As a consequence, the action integral $S = \int d^2 x\,  {\cal L}_{(B_1)}$ remains perfectly invariant under the on-shell nilpotent (anti-)co-BRST
symmetry transformations.

We comment on some of the salient features of the (anti-)co-BRST symmetries at this specific point of our discussion. First, we note that
the {\it total gauge-fixing} term of the Lagrangian densities remains invariant under the (anti-)co-BRST symmetry transformations (i.e. 
$s_{(a)d}\,(\partial\cdot A) = 0, s_{(a)d}\,\phi = 0$). Second, the
mathematical origin of the gauge-fixing term (corresponding to the gauge field) is hidden in the co-exterior derivative of  differential geometry
because we note that $\delta \, A^{(1)} = - \, * \, d\, * \, A^{(1)} = (\partial \cdot A)$  where $\delta = - *\, d\,*$ is the co-exterior derivative
and $*$ is the Hodge duality operation on the 2D Minkowskian spacetime manifold. The other part of the gauge-fixing term (i.e. $ m \,\phi$)
has been added/subtracted on the dimensional ground (in the natural units). Third, the (anti-)co-BRST symmetries are absolutely anticommuting 
and nilpotent of order two provided we take the advantage of EOMs. Finally, we note that the other Lagrangian density ${\cal L}_{(B_2)}$
and its corresponding (anti-)co-BRST symmetries can be obtained from ${\cal L}_{(B_1)}$ and Eq. (19) by the replacements: $\phi \to - \phi, \, 
\tilde \phi \to - \tilde \phi, \, A_\mu \to A_\mu, \, C \to C, \, \bar C \to \bar C$. These latter (anti-)co-BRST symmetries are also found to be 
on-shell nilpotent and absolutely anticommuting in nature (provided we take into account the validity of EOMs derived from the Lagrangian density ${\cal L}_{(B_2)}$).

It is very interesting to note that the following discrete symmetries, namely;
\begin{eqnarray}
A_\mu \to \pm\, i\, \varepsilon_{\mu\nu} A^\nu, \quad \phi \to \pm\, i\, \tilde \phi, 
\quad \tilde \phi \to \pm \,i \, \phi,
\quad C \to \mp\, i\, \bar C, \quad \bar C \to \mp\, i\, C, 
\end{eqnarray}
leave the Lagrangian densities ${\cal L}_{(B_2)}$ and ${\cal L}_{(B_1)}$ invariant (modulo some total spacetime derivatives). The  existence of
these discrete symmetries is very important for us as these symmetries provide the physical realizations of the Hodge duality $*$ operation
of the differential geometry because we note that the following interesting relationships 
\begin{eqnarray}
s_{(a)d} = \pm\, *\, s_{(a)b}\, *, \qquad s_{(a)b} = \mp\, *\, s_{(a)d}\,*,
\end{eqnarray}
are true provided we take the above mathematical connections in their operator form. In the above relationships, the $*$ is nothing but the discrete
symmetry transformations (21). Thus, we note that it is the interplay between the discrete and continuous symmetries of our 2D BRST invariant
theory that provides the physical realizations of the celebrated relationship of differential geometry where the (co-)exterior derivatives
are connected to each-other by the relationships: $\delta = \pm\, *\; d \; *$ [7-11]. There is another very important relationship that is governed
and dictated by the discrete symmetry transformations in (21). For instance, it can be checked that the direct application of the
discrete symmetry transformations (21) on (17) and (19) leads to the following mappings:
\begin{eqnarray}
s_b \Longleftrightarrow s_d, \qquad s_{ad} \Longleftrightarrow s_{ab}.
\end{eqnarray}
In other words, the (anti-)co-BRST and (anti-)BRST symmetries (that have been listed in (19) and (17)) are {\it  also} connected with each-other
by the direct application of the discrete symmetry transformations (21). Let us take an example to illustrate {\it this} point clearly. We
note that $s_b A_\mu = \partial_\mu\, C$. Now we apply {\it directly} the discrete symmetry transformations (21) on it. Taking into account
the mapping listed in (23),  we have to take  $s_b \to s_d$ and, after that, we obtain the
 following (from $s_b A_\mu = \partial_\mu\, C$), namely;
\begin{eqnarray}
s_d\, (*\, A_\mu) = \partial_\mu \, (*\, C) \qquad \Longrightarrow \qquad s_d \,(\pm\, i\, \varepsilon_{\mu\nu}\, A^\nu) = \partial_\mu \,( \mp \,i\, \bar C),
\end{eqnarray}
where $*$ is nothing but the discrete symmetry transformations (21). From the above relationship, it is obvious that we have obtained
the dual-BRST symmetry transformation $s_d$ (from the given BRST symmetry transformation $s_b$) on the gauge field of our theory which amounts
 to $s_d A_\mu = -\, \varepsilon_{\mu\nu} \, \partial^\nu\, \bar C$. Thus, the discrete symmetry transformations (21) provide a {\it direct} relationships
between $s_{(a)d}$ and $s_{(a)b}$. It can be checked that the mappings, given in Eq. (23), are correct and very useful.

We end this section with the remarks that there are
various kinds of discrete symmetries in the theory which connect equivalent Lagrangian densities ${\cal L}_{(B_2)}$ and ${\cal L}_{(B_1)}$ as well as
the on-shell nilpotent and absolutely anticommuting (anti-)BRST and (anti-)co-BRST symmetry transformations. In the next section, we shall discuss
about the coupled (but equivalent) Lagrangian densities, {\it off-shell} nilpotent {\it fermionic} symmetries and the corresponding CF-type restrictions.

\section{Off-Shell Nilpotent Symmetries, Discrete Symmetries and Some Kinds of Restrictions}

We have seen that the Lagrangian densities ${\cal L}_{(B_1)}$ and ${\cal L}_{(B_2)}$ respect the {\it on-shell}
nilpotent (anti-)BRST and (anti-)co-BRST symmetry transformations. These Lagrangian densities 
can be generalized in the following fashion (i.e. ${\cal L}_{(B_1)}\rightarrow {\cal L}_{(b_1)}, {\cal L}_{(B_2)}\rightarrow {\cal L}_{(b_2)}$):
\begin{eqnarray} 
{\cal L}_{(b_1)} &=& {\cal B} \, {(E - m\,\tilde\phi)}  - \frac{1}{2} \, {\cal B}^2
+ m\, E\,\tilde\phi - \frac {1}{2}\,\partial_\mu \,\tilde\phi\,\,\partial^\mu\,\tilde\phi 
+ \frac {m^2}{2} A_\mu\, A^ \mu + \frac {1}{2}\, \partial_\mu\, \phi\, \partial^\mu\, \phi 
\nonumber\\ &-& m \, A_\mu \,\partial^\mu\, \phi
 + B\,(\partial\cdot A +  m\,\phi) + \frac{1}{2}\, B^2 - i\,\partial_\mu\, \bar C \,
\partial^\mu \,C + i \,m^2 \,\bar C\, C,\nonumber\\
{\cal L}_{(b_2)} &=& \bar {\cal B} \, {(E + m\,\tilde\phi)}  - \frac{1}{2} \, \bar {\cal B}^2
- m\, E\,\tilde\phi - \frac {1}{2}\,\partial_\mu \,\tilde\phi\,\,\partial^\mu\,\tilde\phi 
+ \frac {m^2}{2} A_\mu\, A^ \mu + \frac {1}{2}\, \partial_\mu\, \phi\, \partial^\mu\, \phi 
\nonumber\\ &+& m \, A_\mu \,\partial^\mu\, \phi
 + \bar B\,(\partial\cdot A -  m\,\phi) + \frac{1}{2}\, \bar B^2 
- i\,\partial_\mu\, \bar C \,\partial^\mu \,C + i\, m^2 \,\bar C\, C.
\end{eqnarray} 
In the above, we have linearized the kinetic term as well as the gauge-fixing term by invoking 
the Nakanishi-Lautrup type auxiliary fields (${\cal B}, \bar {\cal B}, B, \bar B$). It is elementary 
to check that the following (anti-)BRST symmetries $(s^{(1)}_{(a)b})$, namely;
\begin{eqnarray}
&& s^{(1)}_{ab} \,A_\mu = \partial_\mu \bar C, \qquad 
s^{(1)}_{ab} E = s^{(1)}_{ab} \tilde \phi = s^{(1)}_{ab} \bar C = 0, 
\qquad  
s^{(1)}_{ab} B = s^{(1)}_{ab} {\cal B} = 0,\nonumber\\
&&  s^{(1)}_{ab} \phi = + \,m \, \bar C, \qquad  s^{(1)}_{ab}  C = -\, i\, B, \qquad 
s^{(1)}_{ab} \,(\partial \cdot A + m \phi) = (\Box + m^2) \, \bar C, \nonumber\\
&& s^{(1)}_{b} \,A_\mu = \partial_\mu  C, \qquad 
s^{(1)}_{b} E = s^{(1)}_{b} \tilde \phi = s^{(1)}_{b} C = 0, 
\qquad  
s^{(1)}_{b} B = s^{(1)}_{b} {\cal B} = 0,\nonumber\\
&&  s^{(1)}_{b} \phi = + \,m \,  C, \qquad  s^{(1)}_{b} \bar  C = +\, i\, B, \qquad 
s^{(1)}_{b} \,(\partial \cdot A + m \phi) = (\Box + m^2) \, C, 
\end{eqnarray}
leave the action integral $S = \int d^2 x\, {\cal L}_{(b_1)} $ invariant because 
the Lagrangian density ${\cal L}_{(b_1)}$ transforms to a total spacetime derivative 
(i.e. $s_b^{(1)}\, {\cal L}_{(b_1)} = \partial_\mu\,[B\, \partial^\mu C], \;\;s_{ab}^{(1)}\; {\cal L}_{(b_1)} = \partial_\mu\,[B\, \partial^\mu \bar C]$).
 We note that the above (anti-)BRST
symmetry transformations $s_{(a)b}^{(1)}$ are {\it off-shell} nilpotent [$(s_{(a)b}^{(1)})^{2} = 0$] and absolutely 
anticommuting ($s_{b}^{(1)}\,s_{ab}^{(1)} + s_{ab}^{(1)}\,s_{b}^{(1)} = 0$) in nature. They leave the total kinetic terms
$[{\cal B} \, {(E - m\,\tilde\phi)}  - \frac{1}{2} \, {\cal B}^2
+ m\, E\,\tilde\phi - \frac {1}{2}\,\partial_\mu \,\tilde\phi\,\,\partial^\mu\,\tilde\phi]$ for the 1-form gauge field and
a pseudo-scalar field  invariant.  We recall here that the kinetic term of the gauge field has its origin in 
the exterior derivative $d$.

There is another set of (anti-)BRST symmetry transformations $(s_{(a)b}^{(2)})$ that leave the action integral
 $S = \int d^2 x\, {\cal L}_{(b_2)}$ invariant because the Lagrangian density ${\cal L}_{(b_2)}$ respects the following 
 off-shell nilpotent [$(s_{(a)b}^{(2)})^{2} = 0$] and absolutely anticommuting  ($s_{b}^{(2)}\,s_{ab}^{(2)} + s_{ab}^{(2)}\,s_{b}^{(2)} = 0$)
(anti-)BRST symmetry transformations $(s_{(a)b}^{(2)})$, namely;
 \begin{eqnarray}
&& s^{(2)}_{ab} \,A_\mu = \partial_\mu \bar C, \qquad 
s^{(2)}_{ab} E = s^{(2)}_{ab} \tilde \phi = s^{(2)}_{ab} \bar C = 0, \qquad  
s^{(2)}_{ab} \bar B = s^{(2)}_{ab} \bar {\cal B} = 0,\nonumber\\
&&  s^{(2)}_{ab} \phi = - \,m \, \bar C, \qquad  s^{(2)}_{ab}  C = -\, i\, \bar B, \qquad 
s^{(2)}_{ab} \,(\partial \cdot A - m \phi) = (\Box + m^2) \, \bar C, \nonumber\\
&& s^{(2)}_{b} \,A_\mu = \partial_\mu  C, \qquad 
s^{(2)}_{b} E = s^{(2)}_{b} \tilde \phi = s^{(2)}_{b} C = 0, 
\qquad  
s^{(2)}_{b} \bar B = s^{(2)}_{b} \bar {\cal B} = 0,\nonumber\\
&&  s^{(2)}_{b} \phi = - \,m \,  C, \qquad  s^{(2)}_{b} \bar  C = +\, i\, \bar B, \qquad 
s^{(2)}_{b} \,(\partial \cdot A - m \phi) = (\Box + m^2) \, C, 
\end{eqnarray}
because the Lagrangian density ${\cal L}_{(b_2)}$ transforms to a total spacetime derivative (under the (anti-)BRST symmetry 
transformations $(s_{(a)b}^{(2)})$). It can be, once again, checked that the total kinetic terms $[\bar {\cal B}
 \, {(E + m\,\tilde\phi)}  - \frac{1}{2} \, \bar {\cal B}^2 - m\, E\,\tilde\phi - \frac {1}{2}\,\partial_\mu
 \,\tilde\phi\,\,\partial^\mu\,\tilde\phi]$ for the Abelian
1-form gauge field and pseudo-scalar field  remain invariant under the (anti-)BRST  transformations $s_{(a)b}^{(2)}$.

There is an interesting  discrete symmetry in the theory which relates ${\cal L}_{(b_1)}$ with ${\cal L}_{(b_2)}$ and 
$s_{(a)b}^{(1)}$ with $s_{(a)b}^{(2)}$. These symmetry transformations are:
\begin{eqnarray} 
B \leftrightarrow \bar B,\quad\quad {\cal B} \leftrightarrow \bar {\cal B},\quad \phi \leftrightarrow -\,
 \phi,\quad \tilde\phi\leftrightarrow -\,\tilde\phi,\quad
A_\mu \leftrightarrow A_\mu,\quad C\leftrightarrow C,\quad \bar C\leftrightarrow \bar C.  
\end{eqnarray} 
In other words, only the auxiliary fields and analogues of Stueckelberg's fields transform {\it but} the original basic fields 
$(A_\mu, C, \bar C)$ do {\it not} transform {\it at all} under the discrete transformations (28). Thus, we note that the 
Lagrangian densities ${\cal L}_{(b_1)}$ and  ${\cal L}_{(b_2)}$ are {\it equivalent}
due to the existence of the discrete symmetry transformations in (28). It would be very 
interesting  to  apply the (anti-)BRST symmetry transformations $s_{(a)b}^{(1)}$ on ${\cal L}_{(b_2)}$
and $s_{(a)b}^{(2)}$ on ${\cal L}_{(b_1)}$. In this context, we note that the following are true, namely;  
\begin{eqnarray*}
s^{(1)}_b {\cal L}_{(b_2)} &=& \partial_\mu \, \Bigl [ B \,\partial^\mu \,C 
+ 2\, m^2\, A^\mu \,C + 2 \,m \,\phi \,\partial^\mu \,C \, \Bigr ] \nonumber\\
&-& \,\Bigl [ B + \bar B + 2 \,\,(\partial \cdot A) \Bigr ] \,( \Box + m^2)\, C, \nonumber\\
s^{(1)}_{ab} {\cal L}_{(b_2)} &=& \partial_\mu \, \Bigl [ B \,\partial^\mu \,\bar C 
+ 2\, m^2\, A^\mu \,\bar C + 2 \,m \,\phi \,\partial^\mu \,\bar C \, \Bigr ] \nonumber\\
&-& \,\Bigl [ B + \bar B + 2 \,\,(\partial \cdot A) \Bigr ] \,( \Box + m^2)\, \bar C,\nonumber\\
s^{(2)}_{b} {\cal L}_{(b_1)} &=& \partial_\mu \, \Bigl [ \bar B \,\partial^\mu \, C 
+ 2 \,m^2\, A^\mu\,  C - 2\, m \,\phi \,\partial^\mu \, C \, \Bigr ] \nonumber\\
&-& \,\Bigl [ B + \bar B + 2 \,\,(\partial \cdot A) \Bigr ]\, ( \Box + m^2)\, C, \nonumber\\
\end{eqnarray*}
\begin{eqnarray}
s^{(2)}_{ab} {\cal L}_{(b_1)} &=& \partial_\mu \, \Bigl [ \bar B \,\partial^\mu \, \bar C 
+ 2 \,m^2\, A^\mu\, \bar  C - 2\, m \,\phi \,\partial^\mu \, \bar C \, \Bigr ] \nonumber\\
&-& \,\Bigl [ B + \bar B + 2 \,\,(\partial \cdot A) \Bigr ]\, ( \Box + m^2)\, \bar C,
\end{eqnarray}
where we have used the following nilpotent transformations 
\begin{eqnarray}
&& s^{(1)}_b \bar B = - 2\,\Box\,C, \qquad s^{(1)}_{ab} \bar B = - 2\,\Box\, \bar C, \qquad
s^{(1)}_b \bar {\cal B} = 0, \qquad s^{(1)}_{ab} \bar {\cal B} = 0, \nonumber\\
&& s^{(2)}_b  B = - 2\,\Box\,C, \qquad s^{(2)}_{ab}  B = - 2\,\Box\, \bar C, \qquad
s^{(2)}_b  {\cal B} = 0, \qquad s^{(2)}_{ab}  {\cal B} = 0,
\end{eqnarray}
in addition to the (anti-)BRST symmetry transformations (26) and (27).
We note that if we impose the following restriction
\begin{eqnarray}
B + \bar B + 2 \;\;(\partial \cdot A) = 0 ,
\end{eqnarray}
we find that the Lagrangian densities ${\cal L}_{(b_1)}$ and  ${\cal L}_{(b_2)}$ {\it both}
respect {\it both} types of (anti-)BRST symmetry transformations $s_{(a)b}^{(1)}$ as well as
$s_{(a)b}^{(2)}$ in a beautiful fashion. It should be pointed out that, at this stage, we can {\it not}
use the EOM $((\Box + m^2)\, C = (\Box + m^2)\,\bar C =  0)$.

We provide here the origin of the restriction (31) as well as the 
(anti-)BRST symmetry transformations (30) (in addition to the (anti-)BRST symmetry transformations
listed in (26) and (27)). First of all, we note that Lagrangian densities ${\cal L}_{(b_1)}$ and  ${\cal L}_{(b_2)}$
lead to the following EL-EOMs, namely;
\begin{eqnarray}
{\cal B} = E - m\, \tilde \phi, \quad  
\bar {\cal B} = E + m\, \tilde \phi,   \quad
B = \,- \,[(\partial \cdot A) - m\, \phi],  \quad 
\bar B = \,- \,[(\partial \cdot A) + m\, \phi],
\end{eqnarray}  
which result in the following combinations of restrictions:
\begin{eqnarray}
&& B + \bar B + 2 \;(\partial \cdot A) = 0, \qquad B - \bar B + 2\, m\, \phi = 0, \nonumber\\
&& {\cal B} + \bar {\cal B} - 2 \, E = 0, \;\,\qquad \qquad {\cal B} 
- \bar {\cal B} + 2\, m\, \tilde \phi = 0.
\end{eqnarray} 
If these restriction are to be imposed from {\it outside}, these have to be 
(anti-)BRST invariant. This requirement leads to the derivation of the 
(anti-)BRST symmetry transformations listed in (30).
We would like to comment that, on the constrained hypersurface in the 2D Minkowskian spacetime
manifold where the restriction   $B + \bar B + 2\, (\partial \cdot A) = 0$
is valid, we obtain the following (anti-)BRST symmetry transformations:
\begin{eqnarray}
s^{(1)}_b {\cal L}_{(b_2)} &=& \partial_\mu \, \Bigl [ B \,\partial^\mu \,C 
+ 2\, m^2\, A^\mu \,C + 2 \,m \,\phi \,\partial^\mu \,C \, \Bigr ], \nonumber\\
s^{(1)}_{ab} {\cal L}_{(b_2)} &=& \partial_\mu \, \Bigl [ B \,\partial^\mu \,\bar C 
+ 2\, m^2\, A^\mu \,\bar C + 2 \,m \,\phi \,\partial^\mu \,\bar C \, \Bigr ], 
 \nonumber\\
s^{(2)}_{b} {\cal L}_{(b_1)} &=& \partial_\mu \, \Bigl [ \bar B \,\partial^\mu \, C 
+ 2 \,m^2\, A^\mu\,  C - 2\, m \,\phi \,\partial^\mu \, C \, \Bigr ], \nonumber\\
s^{(2)}_{ab} {\cal L}_{(b_1)} &= &\partial_\mu \, \Bigl [ \bar B \,\partial^\mu \, \bar C 
+ 2 \,m^2\, A^\mu\, \bar  C - 2\, m \,\phi \,\partial^\mu \, \bar C \, \Bigr ],\nonumber\\
s^{(1)}_b \,{\cal L}_{(b_1)} &=& \partial_\mu \,\Bigl [ \,B \,\partial^\mu \,C \, \Bigr ], 
\quad\quad s^{(1)}_{ab} \,{\cal L}_{(b_1)} = \partial_\mu \,\Bigl [ \,B \,\partial^\mu \, \bar C \, \Bigr ],\nonumber\\
s^{(2)}_b \,{\cal L}_{( b_2)} & = &\partial_\mu \,\Bigl [ \,\bar B \,\partial^\mu \,C \,\Bigr ],  \quad \quad
s^{(2)}_{ab} \,{\cal L}_{(b_2)} = \partial_\mu \,\Bigl [ \,\bar B \,\partial^\mu \,
\bar C \,\Bigr ].
\end{eqnarray}
Hence, we note that {\it both} the Lagrangian densities ${\cal L}_{(b_1)}$ and  ${\cal L}_{(b_2)}$
respect {\it both} the nilpotent (anti-)BRST symmetries $s_{(a)b}^{(1)}$ as well as
$s_{(a)b}^{(2)}$ provided we use the  restriction: 
$B + \bar B + 2 \,(\partial \cdot A) = 0.$ In other words, the action 
integrals $S_1 = \int d^2 x \;{\cal L}_{(b_1)}$ and  $S_2 = \int d^2 x \;{\cal L}_{(b_2)}$ 
are invariant under $s_{(a)b}^{(1)}$ as well as
$s_{(a)b}^{(2)}$ on a hypersurface in the 2D Minkowskian space which is defined by the 
field equation $B + \bar B + 2\,(\partial \cdot A) = 0$. We shall discuss more about 
{\it this}  restriction in our Sec. 7 (see below).

In addition to the above (anti-)BRST symmetry transformations $s_{(a)b}^{(1,2)}$, 
the Lagrangian density  ${\cal L}_{(b_1)}$ respects the following off-shell nilpotent 
$[(s_{(a)d}^{(1)})^2 = 0]$  and absolutely anticommuting $(s_{ad}^{(1)} s_d^{(1)}+ s_{d}^{(1)}  s_{ad}^{(1)} = 0)$
(anti-)co-BRST symmetry transformations $(s_{(a)d}^{(1)})$:
\begin{eqnarray}
&& s^{(1)}_{ad} \,A_\mu = - \varepsilon_{\mu\nu} \partial^\nu  C, \qquad 
s^{(1)}_{ad} (\partial \cdot A) = s^{(1)}_{ad} \phi = s^{(1)}_{ad}  C = 0, 
\qquad  
s^{(1)}_{ad} B = s^{(1)}_{ad} {\cal B} = 0,\nonumber\\
&&  s^{(1)}_{ad} \tilde \phi = - \,m \,  C, \quad  s^{(1)}_{ad} \bar C = +\, i\, {\cal B}, \quad 
s^{(1)}_{ad} \,(E - m \tilde \phi) = (\Box + m^2) \,  C, \quad s^{(1)}_{ad} E = \Box C, \nonumber\\
&& s^{(1)}_{d} \,A_\mu = - \varepsilon_{\mu\nu} \partial^\nu  \bar C, \qquad 
s^{(1)}_{d} (\partial \cdot A) = s^{(1)}_{d} \phi = s^{(1)}_{d}  \bar C = 0, 
\qquad  
s^{(1)}_{d} B = s^{(1)}_{d} {\cal B} = 0,\nonumber\\
&&  s^{(1)}_{d} \tilde \phi = - \,m \,  \bar C, \quad  s^{(1)}_{d}  C = -\, i\, {\cal B}, \quad 
s^{(1)}_{d} \,(E - m \tilde \phi) = (\Box + m^2) \, \bar C, \quad s^{(1)}_{d} E = \Box \bar C.
\end{eqnarray}
A few noteworthy points, at this stage, are as follows. First, we note that the total gauge-fixing term remains invariant [$s_{(a)d}^{(1)} \;(\partial\cdot A + m\, \phi) = 0]$
which owes its origin to the co-exterior  derivative $\delta  = - \,\ast \, d\,\ast $ (because $\delta \, A^{(1)}  = (\partial\cdot A)$
and the extra term $m\phi$ has been added to it on the dimensional ground). Second, we note that the Lagrangian density ${\cal L}_{(b_1)}$ transforms, under 
the (anti-)co-BRST symmetry transformations, as:   
\begin{eqnarray}
&&s^{(1)}_{ad} \,{\cal L}_{(b_1)} = \partial_\mu \,\Bigl [ \,{\cal B} \,\partial^\mu \,C 
+ m \, \varepsilon^{\mu\nu}\, \bigl (m \,A_\nu \,C + \phi\, \partial_\nu \,C \bigr )
+ m \,\tilde \phi \,\partial^\mu \,C \Bigr ], \nonumber\\ 
&& s^{(1)}_d \,{\cal L}_{(b_1)} = \partial_\mu \,\Bigl [ \,{\cal B}\, \partial^\mu \,\bar C 
+ m \, \varepsilon^{\mu\nu}\, \bigl (m\, A_\nu \,\bar C 
+ \phi \,\partial_\nu \,\bar C \bigr )
+ m \, \tilde \phi\, \partial^\mu\, \bar C \Bigr ].
\end{eqnarray}
As a consequence, we observe  that the action integral $S  =  \int d^2 x\; {\cal L}_{(b_1)}$
respects the (anti-) co-BRST symmetry transformations $s_{(a)d}^{(1)}$.

It can be checked that the following (anti-)co-BRST symmetry transformation $s_{(a)d}^{(2)}$ which
are off-shell nilpotent $[(s_{(a)d}^{(2)})^2 = 0]$ and absolutely anticommuting $(s_{ad}^{(2)} s_d^{(2)}+ s_{d}^{(2)}  s_{ad}^{(2)} = 0)$ in nature,
namely;
\begin{eqnarray}
&& s^{(2)}_{ad} \,A_\mu = - \varepsilon_{\mu\nu} \partial^\nu  C, \qquad 
s^{(2)}_{ad} (\partial \cdot A) = s^{(2)}_{ad} \phi = s^{(2)}_{ad}  C = 0, 
\qquad  
s^{(2)}_{ad} \bar B = s^{(2)}_{ad} \bar {\cal B} = 0,\nonumber\\
&&  s^{(2)}_{ad} \tilde \phi = + \,m \,  C, \quad  s^{(2)}_{ad} \bar C = +\, i\, \bar {\cal B}, \quad 
s^{(2)}_{ad} \,(E + m \tilde \phi) = (\Box + m^2) \,  C, \quad s^{(2)}_{ad} E = \Box C, \nonumber\\
&& s^{(2)}_{d} \,A_\mu = - \varepsilon_{\mu\nu} \partial^\nu  \bar C, \qquad 
s^{(2)}_{d} (\partial \cdot A) = s^{(2)}_{d} \phi = s^{(2)}_{d}  \bar C = 0, 
\qquad  
s^{(2)}_{d} \bar B = s^{(1)}_{d} \bar {\cal B} = 0,\nonumber\\
&&  s^{(2)}_{d} \tilde \phi = + \,m \,  \bar C, \quad  s^{(2)}_{d}  C = -\, i\, \bar {\cal B}, \quad 
s^{(2)}_{d} \,(E + m \tilde \phi) = (\Box + m^2) \, \bar C, \quad s^{(2)}_{d} E = \Box \bar C,
\end{eqnarray}
leave the action integral $S  =  \int d^2 x \;{\cal L}_{(b_2)}$ invariant because the Lagrangian density
${\cal L}_{(b_2)}$ transforms, under the above fermionic symmetry transformations $s^{(2)}_{(a)d},$ as 
 \begin{eqnarray}
&&s^{(2)}_{ad} \,{\cal L}_{(b_2)} = \partial_\mu \,\Bigl [ \,\bar {\cal B}\, \partial^\mu \,C 
+ m \, \varepsilon^{\mu\nu}\, \bigl (m \,A_\nu \,C - \phi \,\partial_\nu\, C \bigr )
- m\, \tilde \phi\, \partial^\mu\, C \Bigr ], \nonumber\\ 
&& s^{(2)}_d \,{\cal L}_{(b_2)} = \partial_\mu \,\Bigl [ \,\bar {\cal B}\, \partial^\mu \,\bar C 
+ m \, \varepsilon^{\mu\nu}\, \bigl ( m \,A_\nu \,\bar C - \phi\, \partial_\nu\, \bar C \bigr )
- m \,\tilde \phi \,\partial^\mu \,\bar C \Bigr ],
\end{eqnarray}
because {\it all} the well-defined physical fields vanish at $x = \pm\,\infty$ due to the Gauss
divergence theorem. We note that, once again, the gauge-fixing term for the Abelian 1-form gauge field, owing its origin to the 
 co-exterior derivative $\delta  = -\, \ast d\, \ast$ (with $\delta ^2 = 0$), remains invariant [$s_{(a)d}^{(2)} (\partial\cdot A - m\, \phi) = 0$ ]  
under the (anti-)co-BRST symmetry transformation $s_{(a)d}^{(2)}$.

As we have done for the (anti-)BRST symmetry transformations $s_{(a)b}^{(1,2)}$, it would be very 
interesting to find out the applications of  $s_{(a)d}^{(1)}$ on the Lagrangian density ${\cal L}_{(b_2)}$
and $s_{(a)d}^{(2)}$ on the Lagrangian density ${\cal L}_{(b_1)}$. With the following inputs, namely; 
\begin{eqnarray}
&& s^{(1)}_d \bar B = 0, \qquad s^{(1)}_{ad}  \bar B = 0, \qquad
s^{(1)}_d \bar {\cal B} = 2 \, \Box\,\bar C, \qquad s^{(1)}_{ad} \bar {\cal B} = 2 \,\Box \,C,
\nonumber\\ 
&& s^{(2)}_d  B = 0, \qquad s^{(2)}_{ad}  B = 0, \qquad
s^{(2)}_d  {\cal B} = 2 \, \Box\,\bar C, \qquad s^{(2)}_{ad} {\cal B} = 2 \,\Box \,C,
\end{eqnarray}
we obtain the following results:
\begin{eqnarray}
s^{(1)}_d \, {\cal L}_{(b_2)} &=& \partial_\mu \,\Bigl [ {\cal B}\, \partial^\mu \,\bar C
+ m \,\varepsilon^{\mu\nu} \, \bigl (m \,A_\nu \,\bar C -  \phi\, \partial_\nu \,\bar C \bigr ) 
+ m \,\tilde \phi\, \partial^\mu \bar C \Bigr ] 
\nonumber\\ &-& \Bigl [ {\cal B} + \bar {\cal B} - 2 E \Bigr ] \,
(\Box + m^2 )\; \bar C, \nonumber\\
s^{(1)}_{ad} \, {\cal L}_{(b_2)} &=& \partial_\mu \,\Bigl [  {\cal B}\, \partial^\mu \, C
+ m \,\varepsilon^{\mu\nu} \, \bigl (m \,A_\nu \, C -  \phi\, \partial_\nu \, C \bigr ) 
+ m \,\tilde \phi\, \partial^\mu \, C \Bigr ] 
\nonumber\\ &-& \Bigl [ {\cal B} + \bar {\cal B} - 2 E \Bigr ] \,
(\Box + m^2 ) \; C, \nonumber\\
s^{(2)}_d \, {\cal L}_{(b_1)} &=& \partial_\mu \Bigl [  \bar {\cal B}\, \partial^\mu\, \bar C
+ m \,\varepsilon^{\mu\nu}\,  \bigl (m\, A_\nu \,\bar C +  \phi\, \partial_\nu \,\bar C \bigr ) 
- m \,\tilde \phi\, \partial^\mu \, \bar C \Bigr ]
\nonumber\\ &-& \Bigl [ {\cal B} + \bar {\cal B} - 2 E \Bigr ] \,
(\Box + m^2 ) \;\bar C, \nonumber\\
s^{(2)}_{ad} \, {\cal L}_{(b_1)} &=& \partial_\mu \Bigl [ \bar {\cal B}\, \partial^\mu\,  C
+ m \,\varepsilon^{\mu\nu}\,  \bigl (m\, A_\nu \, C +  \phi\, \partial_\nu \, C \bigr ) 
- m \,\tilde \phi\, \partial^\mu \,  C \Bigr ]
\nonumber\\ &-& \Bigl [ {\cal B} + \bar {\cal B} - 2 E \Bigr ] \,
(\Box + m^2 )\; C. 
\end{eqnarray}
Thus, if we impose the restriction (${\cal B} + \bar {\cal B} - 2 \;E = 0$) from Eq. (33),
we shall be able to note that the {\it both} the Lagrangian densities ${\cal L}_{(b_1)}$ and ${\cal L}_{(b_2)}$
respect {\it both} the (anti-)co-BRST symmetry transformations $s_{(a)d}^{(1, 2)}$. In other words, on the 2D 
hypersurface (defined by the restrictions (33)), the Lagrangian densities ${\cal L}_{(b_1)}$ and ${\cal L}_{(b_2)}$
respect {\it both} the sets of (anti-)BRST  and (anti-)co-BRST symmetries. 
These symmetries are off-shell nilpotent $\big[ [s_{(a)b}^{(1, 2)}]^2 = [s_{(a)d}^{(1, 2)}]^2 = 0\big]$ and absolutely anticommuting 
in a couple of pairs (separately and independently). For a instance, we have the validity of the following:
\begin{eqnarray}
&&s_{b}^{(1)} s_{ab}^{(1)}+ s_{ab}^{(1)}  s_b^{(1)} = 0,  \qquad s_{b}^{(2)} s_{ab}^{(2)}+ s_{ab}^{(2)}  s_b^{(2)} = 0,\nonumber\\ 
&&s_{d}^{(1)} s_{ad}^{(1)}+ s_{ad}^{(1)}  s_d^{(1)} = 0,\qquad s_{d}^{(2)} s_{ad}^{(2)}+ s_{ad}^{(2)}  s_d^{(2)} = 0. 
\end{eqnarray}
We discuss, in our Appendix A, all the {\it other} combinations of the anticommutators which are {\it not} 
found to be zero. Thus, we note that ${\cal L}_{(b_1)}$ supports well-defined (anti-)BRST symmetries   ($s_{(a)b}^{(1)}$)
and (anti-)co-BRST symmetries ($s_{(a)d}^{(1)}$). On the other hand, the well-defined (i.e. off-shell nilpotent and absolutely
anticommuting) symmetry transformations  $s_{(a)b}^{(2)}$ and $s_{(a)d}^{(2)}$ are respected by the Lagrangian 
density ${\cal L}_{(b_2)}$ in a {\it perfect} manner. However, it is observed that 
$s_{(a)b}^{(1)}$ and $s_{(a)d}^{(1)}$ are the symmetry transformations for ${\cal L}_{(b_2)}$
(as well as  $s_{(a)b}^{(2)}$ and $s_{(a)d}^{(2)}$ are respected by ${\cal L}_{(b_1)}$) provided  we invoke the 
(anti-)BRST and (anti-)co-BRST invariant restrictions ($B + \bar B + 2 \,(\partial \cdot A) = 0,\quad {\cal B} + \bar {\cal B} - 2\; E = 0 $),
namely;
\begin{eqnarray}
&&s_{(a)b}^{(1,2)} \;[B + \bar B + 2 \,(\partial \cdot A)] = 0, \qquad s_{(a)d}^{(1,2)}\;[B + \bar B + 2 \,(\partial \cdot A)] = 0,\nonumber\\
&&s_{(a)b}^{(1,2)}\;[{\cal B} + \bar {\cal B} - 2\; E] = 0,\,\quad\qquad\quad s_{(a)d}^{(1,2)}\;[{\cal B} + \bar {\cal B} - 2 \;E] = 0,
\end{eqnarray}
which are the {\it physical} restrictions/conditions because of their invariance properties under the {\it basic} fermionic 
symmetries: $s_{(a)b}^{(1,2)}$ and $s_{(a)d}^{(1,2)}$. Furthermore, we lay emphasis on the fact that 
the  restrictions in (33) {\it also} remain invariant under the discrete symmetry transformations (28).
Hence, these restrictions are {\it physical} for our theory.

There are some discrete symmetries in our theory which provide the physical realizations 
of the Hodge duality operation of differential geometry. These are nothing but the 
generalization of discrete symmetries (21) that we have discussed in our previous section.  
We note the following discrete transformations, in this context, namely;
\begin{eqnarray}
&&A_\mu \to \pm\, i\, \varepsilon_{\mu\nu} A^\nu, \qquad \phi \to \pm\, i\, \tilde \phi, 
\qquad \tilde \phi \to \pm \,i \, \phi,\qquad C \to \mp i\,\bar C,\qquad \bar C \to \mp\, i\, C,\nonumber\\
&& B \to \pm\, i\, {\cal B}, \qquad \bar B \to \pm\, i\, \bar {\cal B},
\qquad {\cal  B} \to \pm\, i\, B, \qquad \bar {\cal  B} \to \pm\, i\, \bar B,
\end{eqnarray} 
leave the Lagrangian densities ${\cal L}_{(b_1)}$ and ${\cal L}_{(b_2)}$ invariant 
(separately and independently). It is evident that,
the transformations
 $(\partial \cdot A) \to \mp\; i\, E,\; E \to \mp \;i \,(\partial \cdot A)$  are {\it true} due to 
the discrete symmetry transformation $A_\mu \to \pm\, i\, \varepsilon_{\mu\nu} A^\nu$
on the basic gauge field $A_\mu$. As argued in the previous section, the discrete symmetry transformations (43)  {\it also} lead to 
\begin{eqnarray}
s_{(a)b}^{(1)}\longleftrightarrow s_{(a)d}^{(1)}, \qquad\qquad s_{(a)b}^{(2)}\longleftrightarrow s_{(a)d}^{(2)},
\end{eqnarray}
as can be explicitly checked  by taking into account Eqs. (26), (27), (35) and  (37). Furthermore, 
we {\it also} note that we have the validity of the following 
\begin{eqnarray}
s_{(a)d}^{(1)} = \pm\, *\, s_{(a)b}^{(1)}\, *, \qquad s_{(a)b}^{(2)} = \pm\, *\, s_{(a)d}^{(2)}\,*,
\end{eqnarray}
where $\ast$ is the discrete symmetry transformations in (43).\vskip 1cm

\vskip 1 cm

\section{Conserved Currents and Charges: Nilpotency and Absolute  Anticommutativity Properties}

In this section, first of all, we derive the conserved currents by exploiting the basic ideas behind the celebrated 
Noether theorem and deduce the simple forms of conserved charges corresponding to them.
In this context, we first concentrate on the Lagrangian density ${\cal L}_{(b_1)}$ and using the 
continuous, nilpotent  [$({s_{(a)b}^{(1)})^2} = 0, ({s_{(a)d}^{(1)})^2} = 0$] and absolutely anticommuting  
($s_{b}^{(1)} s_{ab}^{(1)}+ s_{ab}^{(1)}  s_{b}^{(1)} = 0, \;\; s_{d}^{(1)} s_{ad}^{(1)}+ s_{ad}^{(1)}  s_{d}^{(1)} = 0)$
(anti-)BRST and (anti-)co-BRST symmetry transformation [cf. Eqs. (26), (35)], we derive the following Noether currents:
\begin{eqnarray} 
&& J ^{\mu}_{(ab)} =  - \varepsilon^{\mu\nu} ({\cal B}  +  m\,\tilde \phi)\,\partial_{\nu} \bar C
+ B\,\partial^{\mu} \bar C + m\, \bar C\,\partial^{\mu} \phi - m^2 \, A^{\mu}\,\bar C, \nonumber\\ 
&& J^{\mu}_{(b)} = - \varepsilon^{\mu\nu} ({\cal B}  +  m \,\tilde \phi)\,\partial_{\nu} C
+ B\,\partial^{\mu} C + m\,C \,\partial^{\mu} \phi  - m^2 \, A^{\mu}C, \nonumber\\
&& J ^{\mu}_{(ad)} = -\varepsilon^{\mu\nu} (B\,\partial_{\nu}C +m^2\,A_{\nu}C + m\,\,\phi\,\partial_{\nu}C)
 +{\cal B}\,\partial^{\mu}\,C + m\,C\,\partial^{\mu}\tilde{\phi},\nonumber\\
&& J ^{\mu}_{(d)} = -\varepsilon^{\mu\nu}\,(B\,\partial_{\nu}\bar{C} + m^2\,A_{\nu}\bar{C} +
 m\,\phi\,\partial_{\nu}\bar{C}) + {\cal B}\,\partial^{\mu}\bar{C} + m\,\bar{C}\,\partial^{\mu}\tilde{\phi}.
\end{eqnarray}
The conservation law ($\partial_\mu\, J^{\mu}_{(r)} = 0,\;\; r = b, ab, d, ad $) of these Noether currents can be proven by the following 
Euler-Lagrange  (EL) equations  of motion (EOMs)
\begin{eqnarray}  
&&(\Box+ m^2)\,C = 0,\qquad (\Box+ m^2)\,\bar C = 0\qquad (\Box + m^2)\,B = 0,\nonumber\\ 
&& \Box\,\phi = m\,(\partial \cdot A)+ m\,B,\qquad\Box\,\tilde{\phi} = m \,{\cal B}- m\,E, \nonumber\\
&& \varepsilon^{\mu\nu}(\partial_{\nu} {\cal B} + m\,\partial_\nu \tilde \phi)
-\partial^{\mu}\,B + m^2\,A^{\mu}- m\,\partial^{\mu}\phi = 0,
\end{eqnarray}
that are derived from the variation of the action integral w.r.t. Lagrangian density ${\cal L}_{(b_1)}$.

The conserved currents of (46) lead to the following explicit expressions for the conserved charges $(Q_{(r)}^{(1)} = \int\;d\,x \;J _{(r)}^{(0)},
\;\; r = b, ab, d, ad $), namely;
\begin{eqnarray}  
&& Q_b ^{(1)} = \int\;d\,x\, J _{(b)}^{(0)} =  \int\, dx\;\big[{\cal B}\,\partial_1\, C + m\, \tilde{\phi}\,(\partial_1\,C)+ B\,\dot{C} + 
m\,C\,\dot{\phi}-m^2\,C\,A_{0}\big],\nonumber\\
&& Q_{ab} ^{(1)} = \int\;d\,x \,J _{(ab)}^{(0)}  =  \int dx\;\big[{\cal B}\,\partial_1\,\bar{C} +m\,\tilde{\phi}\,\,(\partial_1\,\bar{C})
+B\,\dot{\bar{C}}+ m\,\bar{C}\,\dot{\phi}-m^2\,\bar C\,A_{0}\big],                                \nonumber\\
&& Q_d ^{(1)} = \int\;d\,x\, J _{(d)}^{(0)} =   \int\, dx\;\big[{\cal B}\,\dot{\bar{C}}+B\,\partial_1\,\bar{C} + m\,\bar{C}\dot{\tilde{\phi}} 
+ m^2\,A_1\bar{C}+ m\;\phi\;\partial_1\,\bar{C}\big],\nonumber\\
&& Q_{ad} ^{(1)} = \int\;d\,x\, J _{(ad)}^{(0)} = \int dx\;\big[{\cal B}\,\dot{C}+B\,\partial_1\,C +m\,C\,\dot{\tilde{\phi}}
+m^2\,A_1\,C+m\,\phi\;\partial_1\;C\big],    
\end{eqnarray}
which reduce to the following {\it simple } forms by using the strength of El-EOMs (47):
\begin{eqnarray}  
&& Q_b ^{(1)} = \int d x \;\big[B\,\dot C - \dot B \,C\big],  \qquad Q_{ab} ^{(1)} = \int d x\; \big[B\,\dot{\bar C} - \dot B \,\bar C\big],  \nonumber\\
&& Q_d ^{(1)} = \int d x \;\big[{\cal B}\,\dot {\bar C} - \dot {\cal B} \,\bar C\big],\qquad 
Q_{ad} ^{(1)} = \int d x\;\big[{\cal B}\,\dot {C} - \dot {\cal B} \,C\big]. 
\end{eqnarray}
We would like to point out that, in the derivation of (49), we have used the Gauss divergence theorem to 
drop all the total {\it space} derivatives of terms and we have used the expressions for $\dot B$ and $\dot {\cal B}$
that are deduced from the last entry of EL-EOM in (47).     
The above conserved charges (48) and (49) are the generators of the continuous symmetry 
transformations (26) and (35) which can be verified using the generic definition (6) where we have to take into account 
$ G =  (Q_{(a)b}^{(1)},\;\; Q_{(a)d}^{(1)})$ and the generic field $\Psi = A_\mu, C, \bar C, B, {\cal B}, \phi, \tilde\phi$.

The absolute anticommutativity and off-shell nilpotency of the above conserved charges $Q_r^{(1)} (r = b, ab, d, ad)$ can be proven.
In this context, first of all, we prove the off-shell nilpotency property by using the following standard formula, namely;
\begin{eqnarray}
&& s_{b}^{(1)}\,Q_{b}^{(1)}\, = -\,i\,\lbrace{Q_{b}^{(1)},\,Q_{b}^{(1)}}\rbrace\, = \,0, \qquad \qquad 
s_{d}^{(1)}\,Q_{d}^{(1)}\, = -\,i\,\lbrace{Q_{d}^{(1)},\,Q_{d}^{(1)}}\rbrace\, = \,0, \nonumber\\
&& s_{ab}^{(1)}\,Q_{ab}^{(1)}\, = -\,i\,\lbrace{Q_{ab}^{(1)},\,Q_{ab}^{(1)}}\rbrace\, = \,0, \quad \quad\,\,\quad \;\;\,
s_{ad}^{(1)}\,Q_{ad}^{(1)}\, = -\,i\,\lbrace{Q_{ad}^{(1)},\,Q_{ad}^{(1)}}\rbrace\, = \,0, 
\end{eqnarray}
where we have exploited the basic definition of the generator for the continuous symmetry transformations (26) and (35).
In the above proof, it is straightforward to use the continuous symmetry transformations (26) and (35) and apply them {\it directly} 
on the concise forms of the conserved charges (49). In other words, we have to use the l.h.s. of the equations given in (50).
In exactly similar manner, to prove the absolute anticommutativity of the conserved charges $Q_{(a)b}^{(1)}$, we take
 into account the following expressions:
\begin{eqnarray}
&& s_{b}^{(1)}\,Q_{ab}^{(1)}\, = -\,i\,\lbrace{Q_{ab}^{(1)},\,Q_{b}^{(1)}}\rbrace\, = \,0\quad \equiv\quad s_{ab}^{(1)}\,Q_{b}^{(1)},\nonumber\\
&& s_{d}^{(1)}\,Q_{ad}^{(1)}\, = -\,i\,\lbrace{Q_{ad}^{(1)},\,Q_{d}^{(1)}}\rbrace\, = \,0 \quad\equiv\quad s_{ad}^{(1)}\,Q_{d}^{(1)}.
\end{eqnarray}
It is obvious that one can compute the expressions $s_{b}^{(1)}\,Q_{ab}^{(1)}, s_{ab}^{(1)}\,Q_{b}^{(1)}, s_{d}^{(1)}\,Q_{ad}^{(1)}$
and $s_{ad}^{(1)}\,Q_{d}^{(1)}$ from the {\it direct} applications of the transformations (26) and (35) to verify
 that the following anticommutativity properties of the conserved charges 
\begin{eqnarray}
Q_{b}^{(1)}\; Q_{ab}^{(1)} + Q_{ab}^{(1)}\; Q_{b}^{(1)} = 0, \qquad Q_{d}^{(1)}\; Q_{ad}^{(1)} + Q_{ad}^{(1)}\; Q_{d}^{(1)} = 0,
\end{eqnarray} 
 are satisfied. Thus, we have already demonstrated that, for the Lagrangian density ${\cal L}_{(b_1)}$,
 the (anti-)BRST and (anti-)co-BRST charges obey the off-shell nilpotency and absolute anticommutativity properties
in a {\it perfect} manner.

Now we focus on the (anti-)BRST and (anti-)co-BRST symmetries    (cf. Eq. (27), (37)) that are
associated with the Lagrangian density ${\cal L}_{(b_2)}$. It can be checked  that the Noether
theorem leads to the following expressions for the currents $({\cal J}^{\mu}_{(r)}, r  = b, ab, d, ad$), namely;
\begin{eqnarray} 
&& {\cal J} ^{\mu}_{(ab)} =  - \varepsilon^{\mu\nu} (\bar{\cal B}  -  m\,\tilde \phi)\,\partial_{\nu} \bar C
+ \bar B\,\partial^{\mu} \bar C - m\, \bar C\,\partial^{\mu} \phi - m^2 \, A^{\mu}\,\bar C, \nonumber\\ 
&& {\cal J}^{\mu}_{(b)} = - \varepsilon^{\mu\nu} (\bar {\cal B}  -  m \,\tilde \phi)\,\partial_{\nu} C
+ \bar B\,\partial^{\mu} C - m\,C \,\partial^{\mu} \phi  - m^2 \, A^{\mu}C, \nonumber\\
&& {\cal J} ^{\mu}_{(ad)} = -\varepsilon^{\mu\nu} (m^2\,A_{\nu}C - m\,\,\phi\,\partial_{\nu}C + \bar B\,\partial_{\nu}C )
 +\bar {\cal B}\,\partial^{\mu}\,C - m\,C\,\partial^{\mu}\tilde{\phi},\nonumber\\
&& {\cal J} ^{\mu}_{(d)} = -\varepsilon^{\mu\nu}\,(- \,m\,\phi\,\partial_{\nu}\bar{C}+ m^2\,A_{\nu}\bar{C}+\bar B\,\partial_{\nu}\bar{C} 
 ) + \bar {\cal B}\,\partial^{\mu}\bar{C} - m\,\bar{C}\,\partial^{\mu}\tilde{\phi},
\end{eqnarray}
where we have used the continuous symmetry transformations (27)and (37). 
The conservation law (i.e. $\partial_\mu\, {\cal J}^{\mu}_{(r)} = 0,\;\; r = b, ab, d, ad $) can be proven by using the following 
EL-EOMs 
\begin{eqnarray}  
&&(\Box+ m^2)\,C = 0,\qquad (\Box+ m^2)\,\bar C = 0,\qquad (\Box + m^2)\,\bar B = 0,\nonumber\\ 
&& \Box\,\phi = -\,m\,(\partial \cdot A)- m\,\bar B,\qquad \Box\,\tilde{\phi} = -\,m \,\bar {\cal B}+ m\,E, \nonumber\\
&& \varepsilon^{\mu\nu}(m\,\partial_\nu \tilde \phi - \partial_{\nu} \bar{\cal B} )
-\partial^{\mu}\,\bar B + m^2\,A^{\mu}+ m\,\partial^{\mu}\phi = 0,
\end{eqnarray}
which are derived from the Lagrangian density ${\cal L}_{(b_2)}$.
The above conserved currents ${\cal J}^{\mu}_{(r)}$ (with $r = b, ab, d, ad$) lead to the following expressions for charges 
\begin{eqnarray}  
&& Q_b ^{(2)} = \int\;dx\; {\cal J} _{(b)}^{(0)} =  \int\, dx\;\big[\bar{\cal B}\,\partial_1\, C - m\, \tilde{\phi}\,(\partial_1\,C)+ \bar B\,\dot{C} - 
m\,C\,\dot{\phi}-m^2\,C\,A_{0}\big],\nonumber\\
&& Q_{ab} ^{(2)} = \int\;dx \;{\cal J} _{(ab)}^{(0)}  =  \int dx\;\big[\bar{\cal B}\,\partial_1\,\bar{C} - m\,\tilde{\phi}\,\,(\partial_1\,\bar{C})
+ \bar B\,\dot{\bar{C}}-  m\,\bar{C}\,\dot{\phi}-m^2\,\bar C\,A_{0}\big],                                \nonumber\\
&& Q_d ^{(2)} = \int\;dx \;{\cal J} _{(d)}^{(0)} =   \int\, dx\;\big[\bar {\cal B}\,\dot{\bar{C}} + \bar B\,\partial_1\,\bar{C} - m\,\bar{C}\,\dot{\tilde{\phi}} 
+ m^2\,A_1\bar{C} - m\,\phi\,\partial_1\,\bar{C}\big],\nonumber\\
&& Q_{ad} ^{(2)} = \int\;dx \;{\cal J} _{(ad)}^{(0)} = \int dx\;\big[\bar {\cal B}\,\dot{C} + \bar B\,\partial_1\,C - m\,C\,\dot{\tilde{\phi}}
+ m^2\,A_1\,C - m\,\phi\,\partial_1\,C\big],    
\end{eqnarray}
which are the generators for the continuous symmetry transformations (27) and (37).
This statement can be  verified by replacing $G$ by the charges  ($Q_{(a)b}^{(2)}, Q_{(a)d}^{(2)}$) and the generic field 
$\Psi$ by the fields $A_\mu, C, \bar C,  \bar B, \bar {\cal B}, \phi, \tilde\phi$  in the {\it basic} definition (6).

The explicit expressions for the conserved charges (55) can be expressed in a concise form
by using the following EOMs that are derived form (54), namely;
\begin{eqnarray}  
&&\dot {\bar B} = \partial_1 \bar{\cal B} - m\, \partial_1 {\tilde\phi}  + m^2\, A_0 + m\, \dot \phi, \nonumber\\
&& {\dot{\bar {\cal B}}}= \partial_1  \bar B - m\, \partial_1 \phi  - m^2\, A_1 + m\, \dot {\tilde\phi}.   
\end{eqnarray}
At this stage, first of all, we use Gauss's divergence theorem and drop all the total {\it space} derivative terms. After this,
we use the  equations (56).    
The substitutions of the above equations, in the explicit forms of the conserved charges (55), lead to the following:
\begin{eqnarray}  
&& Q_b ^{(2)} = \int d x \;\big[\bar B\,\dot C - \dot {\bar B} \,C\big],  \qquad Q_{ab} ^{(2)} = \int d x\; \big[\bar B\,\dot{\bar C} - \dot{\bar B}\,\bar C\big],  \nonumber\\
&& Q_d ^{(2)} = \int d x \;\big[\bar {\cal B}\,\dot {\bar C} - {\dot{\bar {\cal B}}} \,\bar C\big],\qquad \;
Q_{ad} ^{(2)} = \int d x\;\big[\bar {\cal B}\,\dot {C} - {\dot{\bar {\cal B}}}\,C\big]. 
\end{eqnarray}
It is now straightforward to prove the off-shell nilpotency and absolute anticommutativity properties of the above charges
by exploiting  the {\it basic} ideas behind the relationship between the continuous symmetry transformations and their 
generators. For instance, it can be explicitly checked that the following are true, namely;
\begin{eqnarray}
&& s_{b}^{(2)}\,Q_{b}^{(2)}\, = -\,i\,\lbrace{Q_{b}^{(2)},\,Q_{b}^{(2)}}\rbrace\, = \,0,  \qquad 
s_{d}^{(2)}\,Q_{d}^{(2)}\, = -\,i\,\lbrace{Q_{d}^{(2)},\,Q_{d}^{(2)}}\rbrace\, = \,0, \nonumber\\
&& s_{ab}^{(2)}\,Q_{ab}^{(2)}\, = -\,i\,\lbrace{Q_{ab}^{(2)},\,Q_{ab}^{(2)}}\rbrace\, = \,0, \quad\quad 
s_{ad}^{(2)}\,Q_{ad}^{(2)}\, = -\,i\,\lbrace{Q_{ad}^{(2)},\,Q_{ad}^{(2)}}\rbrace\, = \,0,\nonumber\\ 
&& s_{b}^{(2)}\,Q_{ab}^{(2)}\, = -\,i\,\lbrace{Q_{ab}^{(2)},\,Q_{b}^{(2)}}\rbrace\, = \,0 \quad\equiv\quad s_{ab}^{(2)}\,Q_{b}^{(2)},\nonumber\\
&& s_{d}^{(2)}\,Q_{ad}^{(2)}\, = -\,i\,\lbrace{Q_{ad}^{(2)},\,Q_{d}^{(2)}}\rbrace\, = \,0 \quad\equiv\quad s_{ad}^{(2)}\,Q_{d}^{(2)}.
\end{eqnarray}
In the above, we note that it is elementary exercise to compute the l.h.s. of the expressions {\it directly}
 by taking into account the continuous symmetry transformations ((27), (37)) and expressions for the conserved charges (57).
 At the level of the conserved charges, the relations in (58) imply the following relationships
\begin{eqnarray}
[{Q_{(a)b}^{(2)}}]^2 = 0,\quad [{Q_{(a)d}^{(2)}}]^2 = 0,\quad
Q_{b}^{(2)}\; Q_{ab}^{(2)} + Q_{ab}^{(2)}\; Q_{b}^{(2)} = 0, \quad Q_{d}^{(2)}\; Q_{ad}^{(2)} + Q_{ad}^{(2)}\; Q_{d}^{(2)} = 0,
\end{eqnarray}  
 which prove the off-shell nilpotency of the conserved charges along with the absolute anticommutativity between the pairs $(Q_{b}^{(2)}, Q_{ab}^{(2)} )$
 and $(Q_{d}^{(2)}, Q_{ad}^{(2)})$.

 We end this section with the remarks that the pairs $(s_b ^{(1)},  s_{ab}^{(1)}), (s_d ^{(1)},  s_{ad}^{(1)}),(s_b ^{(2)},  s_{ab}^{(2)})$
 and $(s_d ^{(2)},  s_{ad}^{(2)})$ anticommute among themselves (separately and independently). However, it has been found that {\it even}
 $s_b ^{(1)}$ and $s_b ^{(2)}$ do {\it not} absolutely anticommute with each-other. We discuss all these, in detail, in our Appendix A 
where we  compute {\it all} the possible anticommutators among {\it all} this {\it fermionic} transformation operators $s_{(a)b}^{(1)}, s_{(a)d}^{(1)},s_{(a)b}^{(2)}$
and $s_{(a)d}^{(2)}$. As  a result of these observations, we find that the pairs of the conserved charges
$(Q_b^{(1)},Q_{ab}^{(1)}), (Q_d^{(1)},Q_{ad}^{(1)}), (Q_b^{(2)},Q_{ab}^{(2)})$ and $(Q_d^{(2)},Q_{ad}^{(2)})$ absolutely  anticommute but {\it other }
possible pairs of the conserved charges {\it do not} absolutely anticommute {\it even}  if we impose the restrictions  (33).
These  computations have  been incorporated in our  Appendix B.

\section{(Anti-)Chiral Superfield Approach: Nilpotent  Symmetries and Conserved Charges }

To {\it verify} the sanctity of all the off-shell nilpotent and absolutely anticommuting 
(anti-) BRST and (anti-)co-BRST symmetry transformations, we exploit the potential and power
 of our {\it newly} proposed (anti-)chiral superfield approach to BRST formalism [22-25].

\subsection {Off-Shell Nilpotent (Anti-)BRST Symmetries and Conserved Charges: (Anti-)Chiral Superfield Formalism}

First of all, we concentrate on the derivation of the off-shell nilpotent symmetries 
 $s_b^{(1)}$ for the Lagrangian density ${\cal L}_{(b_1)}$. Towards this goal in mind,
 we generalize the 2D basic and auxiliary fields $A_\mu , C, \bar C, \phi, \tilde\phi, B, {\cal B}$ (onto a (2, 1)-dimensional 
 {\it anti-chiral} supermanifold) as\footnote{To be precise, the (2, 1)-dimensional anti-chiral 
supermanifold is a {\it super-submanifold} of the general (2, 2)-dimensional
supermanifold   (parameterized by $Z^M = (x^\mu, \theta, \bar\theta)$) on which our 2D theory is generalized.}
\begin{eqnarray}
 &&A_{\mu} (x) \longrightarrow  B_\mu (x, \bar\theta) = A_{\mu}(x) + \bar\theta\, R_\mu (x),\quad
 C (x) \longrightarrow F(x, \bar\theta) = C(x) + i\,\bar\theta\, B_{1}(x),\nonumber\\
&& \bar C (x) \longrightarrow \bar F(x, \bar\theta)~~~ = \bar C_(x) + i\,\bar\theta\; B_{2} (x),\quad
 \phi (x)  \longrightarrow \Phi  (x, \bar\theta) = \phi (x) + \bar\theta\, f_1 (x),\nonumber\\
&& \tilde\phi (x)  \longrightarrow \tilde\Phi  (x, \bar\theta)~~~ = \tilde\phi (x) + \bar\theta\, f_2 (x),\qquad
 B (x)  \longrightarrow \tilde B (x, \bar\theta) = B(x) + \bar\theta\, f_3(x),\nonumber\\
&& {\cal B} (x)  \longrightarrow \tilde{\cal B}  (x, \bar\theta)~~~ = {\cal B} (x) + \bar\theta\, f_4 (x),
 \end{eqnarray}  
where the (2, 1)-dimensional {\it anti-chiral} super-submanifold is characterized by the superspace coordinates $Z^M = (x^\mu, \bar\theta)$.
The bosonic coordinates $x^\mu$ (with $\mu = 0, 1)$ describe the 2D Minkowskian spacetime manifold and $\bar\theta$
is a fermionic $(\bar\theta^2 = 0)$ Grassmannian variable. In the above 
expansions (60), the fields $(R_\mu, B_2, f_1, B_1, f_2, f_3, f_4)$ are called as the {\it secondary} fields
which are to be determined in terms of the basic and auxiliary
 fields of our 2D theory (described by the Lagrangian densities ${\cal L}_{(b_1)}$ and ${\cal L}_{(b_2)}$) by invoking one of the key ideas of the (anti-)chiral 
superfield formalism where we demand that {\it all} the BRST-invariant quantities (i.e. physical quantities at the {\it quantum} level) must be independent
of the Grassmannian variable $\bar\theta$ (which happens to be {\it merely} a mathematical artifact). The fermionic  nature of $\bar\theta$ ensures  that
$(R_\mu,  f_1,  f_2, f_3, f_4)$ are fermionic and $(B_1, B_2 )$ are the bosonic {\it secondary} fields in the 
expansion (60) for { \it all} the basic and auxiliary {\it anti-chiral} superfields (defined on the (2, 1)-dimensional {\it anti-chiral} super-submanifold
 of the {\it general} (2, 2)-dimensional supermanifold  as the generalizations of the  2D {\it ordinary} fields).

Towards our goal of determining the secondary fields in terms of the basic and auxiliary fields 
of the Lagrangian density ${\cal L}_{(b_1)}$, we note that the following very useful and 
interesting quantities (which are obtained from the symmetry transformations (26)), namely;
\begin{eqnarray}
&& s_b^{(1)} C = \; s_b ^{(1)}B = \; s_b^{(1)}{\cal B} = 0,\;\;\; s_b ^{(1)}(m\,A_\mu - \partial_\mu\,\phi) = 0,\;\;\quad s_b^{(1)} (C\;\phi) = 0,  \nonumber\\
&& s_b^{(1)}[A^{\mu}\;\partial_\mu B + i\;\partial_\mu \bar C\; \partial^\mu C] = 0,\quad
\quad s_b^{(1)}[m\, \bar C \, C - i\, B\, \phi] = 0,\nonumber\\
&& s_b^{(1)} (B\, \dot{\bar C} - \dot B\, \bar C) = 0,\;\;\;\;\; s_b^{(1)} (A^{\mu}\;\partial_\mu C) = 0,\;\;\;\;
 s_b^{(1)} (\tilde\phi\; ) = 0,
\end{eqnarray} 
are BRST invariant. As a consequence, these useful and interesting  quantities are {\it physical}    
at the {\it quantum} level (and, hence, at the classical level, they ought to be gauge invariant).
Such quantities, according to the basic tenets of (anti-)chiral superfield approach to BRST formalism [22-25],
must be independent of the Grassmannian $\bar\theta$ variable. For instance, we note that the following 
equalities are true, namely;
\begin{eqnarray}
s_b^{(1)} C = 0\quad\Longrightarrow \quad F^{(b)}(x, \bar\theta) \,&=& \, C(x) + \bar\theta\,(0) \quad\equiv\quad C(x) + \bar\theta\,(s_b ^{(1)}\,C(x)),\nonumber\\
s_b^{(1)} B = 0\quad\Longrightarrow\quad \tilde B^{(b)}(x, \bar\theta)\, &=& \, B(x) + \bar\theta\,(0) \quad\equiv\quad B(x) 
+ \bar\theta\,(s_b ^{(1)}\, B(x)),\nonumber\\
s_b^{(1)} \tilde\phi = 0\quad\Longrightarrow\quad \tilde\phi^{(b)}(x, \bar\theta)\, &=& \, \tilde\phi(x) + \bar\theta\,(0) \quad\equiv\quad \tilde\phi(x) 
+ \bar\theta\,( s_b ^{(1)}\, \tilde\phi(x)),\nonumber\\
s_b^{(1)} {\cal B} = 0\quad\Longrightarrow\quad \tilde {\cal B}^{(b)}(x, \bar\theta)\, &=& \,{\cal B}(x) + \bar\theta\,(0) \quad\equiv\quad {\cal B}(x) 
+ \bar\theta\,( s_b ^{(1)}\, {\cal B}(x)),
\end{eqnarray} 
where the superscripts $(b)$ denotes the {\it anti-chiral} superfields that have been obtained after the applications of 
the BRST invariant restrictions on the anti-chiral superfields. In other words, we have taken into account 
$F(x, \bar\theta)  = C(x), \tilde B(x, \bar\theta)   = B(x), \tilde\Phi (x, \bar\theta) = \tilde\phi (x),\; \tilde {\cal B} (x) = {\cal B}(x)$
which lead to the precise determination of the secondary fields as: $B_1(x) = 0, f_2 (x) =0, f_3(x) = 0$ and $f_4(x) = 0 $.
As a consequence, we have already determined $s_b^{(1)} C = 0, s_b^{(1)} B = 0, s_b^{(1)} \tilde\phi = 0,\; s_b^{(1)} {\cal B} = 0$
which are nothing but the coefficients of  $\bar\theta$ in the expansions of the {\it anti-chiral} superfields which have been 
obtained after the applications of the BRST-invariant restrictions (61). In other words, we note that 
$\partial_{\bar\theta}\, F^{(b)}(x, \bar\theta) = s_b^{(1)} C, \partial_{\bar\theta}\, \tilde B^{(b)}(x, \bar\theta) = s_b^{(1)} B,
\partial_{\bar\theta}\, \tilde\Phi^{(b)}(x, \bar\theta) = s_b^{(1)} \tilde\phi$ and 
$\partial_{\bar\theta}\, \tilde{\cal B}^{(b)}(x, \bar\theta) = s_b^{(1)} {\cal B}$ which {\it physically} imply that the 
translations of the anti-chiral superfields (with superscripts $(b)$) along $\bar\theta$-direction of the 
(2, 1)-dimensional anti-chiral super-submanifold generates the BRST symmetry transformations for the corresponding 
{\it ordinary} 2D fields (defined on the (1 + 1)-dimensional (2D) {\it ordinary} flat Minkowiskian spacetime manifold).

We discuss a bit more about the determination of {\it secondary} fields  in terms of the basic and auxiliary fields
of our 2D theory described by the Lagrangian density ${\cal L}_{(b_1)}$ (cf. Eq. (25)). It is elementary to check that
 the following equalities
\begin{eqnarray}
&&s_b^{(1)} (\phi \;C) = 0 \qquad\qquad\quad\;\Longrightarrow \quad \Phi (x, \bar\theta)\; F^{(b)} (x, \bar\theta) = \phi (x) \, C(x),\nonumber\\
&&s_b^{(1)} (A^{\mu}\;\partial_\mu C) = 0 \qquad\quad\;\;\Longrightarrow \quad B^{\mu} (x, \bar\theta) \;\partial_\mu F^{(b)} (x, \bar\theta) = A^\mu (x)\;\partial_\mu C(x),
\end{eqnarray} 
lead to the non-trivial solutions $R_\mu  = \kappa_1 \,\partial_\mu C$ and $f_1 = \kappa_2\, C(x)$ where $\kappa_1$ and $\kappa_2$
are some numerical constants. With these as inputs, we now observe the following
\begin{eqnarray}
B_\mu ^{(m)} (x, \bar\theta) = A_{\mu}(x) + \bar\theta\, (\kappa_1\, \partial_\mu C(x)),\quad 
 \Phi ^{(m)} (x, \bar\theta) = \phi (x) + \bar\theta\, (\kappa_2 \,C(x)),
\end{eqnarray}
where the superscript $(m)$ on the {\it anti-chiral} superfields denotes the {\it modified} version  of the anti-chiral superfields $B_\mu (x, \bar\theta)$
and $\Phi (x, \bar\theta)$. At this stage, we utilize 
\begin{eqnarray}
s_b ^{(1)}(m\,A_\mu - \partial_\mu\,\phi) = 0\quad\Longrightarrow \quad m\, B_\mu ^{(m)} (x, \bar\theta) -  \partial_\mu \Phi ^{(m)} (x, \bar\theta)
= m\,A_\mu (x)- \partial_\mu\,\phi (x), 
\end{eqnarray}
which leads to a relationship between $\kappa _1$ and $\kappa _2$ as: $m\,\kappa _1 = \kappa _2$. Finally, the other BRST invariant 
quantities and their generalizations onto the (2, 1)-dimensional {\it anti-chiral} supermanifolds imply the following restrictions on the superfields
\begin{eqnarray*}
s_b^{(1)}[m\, \bar C \, C - i\, B\, \phi] = 0 \quad &\Longrightarrow & \quad m\,\bar F (x, \bar\theta) \;F^{(b)} (x, \bar\theta)\nonumber\\
 - i\,B^{(b)} (x, \bar\theta)\;\Phi^{(m)} (x, \bar\theta) & = &  m\, \bar C(x) \, C(x) - i\, B(x)\, \phi (x),\nonumber\\
s_b^{(1)}[A^{\mu}\;\partial_\mu B + i\;\partial_\mu \bar C\; \partial^\mu C] = 0 \quad &\Longrightarrow & \quad
B_\mu^{(m)} (x, \bar\theta) \;\partial^\mu\tilde B^{(b)}(x, \bar\theta)\nonumber\\
 + i\;\partial_{\mu}\bar F(x, \bar\theta)\;\partial^{\mu}F^{(b)}(x, \bar\theta)
 & = & A^{\mu}(x)\;\partial_{\mu}B(x) + i\; \partial_{\mu}\bar C(x)\;\partial^{\mu}C(x),
 \end{eqnarray*}  
\begin{eqnarray}                     
\tilde B^{(b)} (x, \bar\theta)\;\dot {\bar F} (x, \bar\theta) - \dot {\tilde B}^{(b)} (x, \bar\theta)\;\dot {\bar F} (x) & = &  B(x)\; \dot {\bar C}(x)
 - \dot B (x)\;{\bar C} (x),
\end{eqnarray}
which lead to the derivation of constants  and all the secondary fields in terms of the basic and auxiliary fields 
of the Lagrangian density ${\cal L}_{(b_1)}$ (cf. Eq. (25)) as\footnote{In our Appendix C, we determine the
value of constant $\kappa_1 = +\,1$ in an explicit fashion.}:
\begin{eqnarray}
\kappa_1 = 1,\quad \kappa_2 = m,\quad R_\mu (x) = \partial_\mu C (x), \quad B_2 (x) = B(x), \quad f_1 (x) = m\, C(x).
\end{eqnarray} 
As a consequence, we have the following super expansions
\begin{eqnarray}
B_\mu^{(b)}(x, \bar\theta) &=& \, A_\mu(x) + \bar\theta\,(\partial_\mu C) \;\quad\equiv\quad A_\mu(x) + \bar\theta\,(s_b ^{(1)}\,A_\mu(x)),\nonumber\\
\bar F^{(b)}(x, \bar\theta) \,&=& \, \bar C(x) + \bar\theta\,(i\, B) \;\qquad\equiv\quad \bar C(x) + \bar\theta\,(s_b ^{(1)}\,\bar C(x)),\nonumber\\
\Phi^{(b)}(x, \bar\theta) \,&=& \, \phi (x) + \bar\theta\,(m\,C) \qquad\equiv\quad \phi(x) + \bar\theta\,(s_b ^{(1)}\,\phi (x)),
\end{eqnarray}
in addition to Eq. (62). Thus, we have derived {\it all} the BRST symmetry transformations for ${\cal L}_{(b_1)}$ 
and proven their sanctity within the framework of (anti-)chiral superfield approach.

For the derivation of the off-shell nilpotent anti-BRST symmetry transformations $ s_{ab}^{(1)}$, we generalize the basic and auxiliary fields 
$A_\mu, C,\bar C,\phi,\tilde\phi, B,{\cal B}$ of the theory (onto a (2,1)-dimensional {\it chiral} super-submanifold) as:   
\begin{eqnarray}
 &&A_{\mu} (x) \longrightarrow \; B_\mu (x, \theta) \;= A_{\mu}(x) + \theta\, \bar R_\mu (x),\quad
 C (x) \,\longrightarrow F(x, \theta) = C(x) + i\,\theta\, \bar B_{1}(x),\nonumber\\
&& \bar C (x) \,\;\longrightarrow\; \bar F(x, \theta)~~\, = \bar C_(x) + i\,\theta\; \bar B_{2} (x),\quad
 \phi (x) \; \longrightarrow \Phi  (x, \theta) = \phi (x) + \theta\, \bar f_1 (x),\nonumber\\
&& \tilde\phi (x)  \;\;\longrightarrow \;\tilde\Phi  (x, \theta)~~~ = \tilde\phi (x) + \theta\,\bar f_2 (x),\qquad
 B (x)  \longrightarrow \tilde B (x, \theta) = B(x) + \theta\, \bar f_3(x),\nonumber\\
&& {\cal B} (x) \;\; \longrightarrow \;\tilde{\cal B}  (x, \theta)~~~ = {\cal B} (x) + \theta\, \bar f_4 (x),
 \end{eqnarray}  
where the superspace coordinates $ Z^M = (x^\mu, \theta)$ characterize the (2,1)-dimensional {\it chiral} supermanifold.
Here $x^\mu$ (with $\mu = 0,1)$ are the 2D bosonic coordinates and $\theta$ is a {\it fermionic} (i.e. $\theta^2 = 0)$
Grassmannian variable. The secondary fields $(\bar R_\mu, \bar f_1, \bar f_2, \bar f_3, \bar f_4)$ are 
fermionic in nature whereas $(\bar B_1, \bar B_2)$ are bosonic (due to the fermionic nature of $\theta$).
To determine the {\it secondary} fields in terms of the {\it basic} and {\it auxiliary} fields of the
Lagrangian density ${ \cal L}_{(b_1)}$, we obtain   the following useful  and interesting  anti-BRST invariant quantities:
\begin{eqnarray}
&& s_{ab}^{(1)}\bar C = \; s_{ab}^{(1)}B = \,\; s_{ab}^{(1)} {\cal B} = 0,\,\; s_{ab}^{(1)}(m\,A_\mu - \partial_\mu\,\phi) =
 0,\;\; s_{ab}^{(1)} (\bar C\;\phi) = 0,  \nonumber\\
&& s_{ab}^{(1)}[A^{\mu}\;\partial_\mu B + i\;\partial_\mu \bar C\; \partial^\mu C] = 0,
\quad s_{ab}^{(1)}[m\, \bar C \, C - i\, B\, \phi] = 0,\nonumber\\
 && s_{ab}^{(1)} (B\, \dot C - \dot B\,  C) = 0,\quad\;\; s_{ab}^{(1)} (A^{\mu}\;\partial_\mu \bar C ) = 0,\;\;\;\;\;
 s_{ab}^{(1)} (\tilde\phi\; ) = 0.
\end{eqnarray} 
Following the {\it basic} tenets of (anti-)chiral superfield approach to 
BRST formalism, we note  that the following restrictions have to be imposed on the {\it chiral} superfields:
\begin{eqnarray}
&&\bar F(x, \theta) = \bar C(x), \qquad \tilde B (x, \theta) = B(x),\qquad \tilde{\cal B}(x, \theta) = {\cal B}(x),\nonumber\\
&& m  B_\mu (x, \theta) - \partial_\mu \Phi (x, \theta) = m\, A_\mu(x) -\partial_\mu \phi(x),\quad \tilde\Phi(x, \theta) = \tilde\phi(x),\nonumber\\
&& B^\mu (x, \theta)\, \partial_\mu \bar F (x, \theta) = A^\mu(x)\,\partial_\mu {\bar C}(x),\quad \bar F(x, \theta)\, \Phi(x, \theta)  = \bar C(x),\phi(x),\nonumber\\
&&B^\mu (x, \theta)\,\partial_\mu \tilde B(x, \theta) + i\, \partial_\mu \bar F(x, \theta) \, \partial^\mu C(x, \theta) = A^\mu(x)\, \partial_\mu B(x) + 
i\, \partial_\mu \bar C(x)\, \partial^\mu C(x),\nonumber\\
&& m\,\bar F(x, \theta)\, F(x, \theta) - i\, \tilde B(x, \theta)\,\Phi(x, \theta) = m\, \bar C(x)\, C(x) - i\,  B(x)\,\phi(x),\nonumber\\
&& \tilde B (x, \theta)\, \dot F (x, \theta) - \dot {\tilde B}(x, \theta) F(x,\theta) = B(x)\,\dot C(x) - \dot B(x)\,C(x).
\end{eqnarray}
The above restrictions are {\it physical} because of the fact that any anti-BRST invariant quantity (at the {\it quantum} level) is 
a gauge invariant quantity (at the {\it classical} level). Hence, such quantities should be independent of the {\it mathematical}
quantity $\theta$ (as this Grassmannian variable is {\it not} a physical quantity {\it but} it is a purely  mathematical artifact).

The equalities in (71) lead to the determination  of secondary  fields, in terms of the auxiliary and basic fields
of the Lagrangian density ${\cal L}_{(b_1)}$, as:
\begin{eqnarray}
\bar R_\mu = \partial_\mu \bar C,\qquad \bar B_1 = - B,\quad \bar f_1 = m\, \bar C,\quad \bar f_2 = \bar f_3 = \bar f_4 = \bar B_2 = 0.
\end{eqnarray}
The above deduction has been performed on exactly similar lines of arguments (see, e.g., Appendix C) as we have done for the determination of the BRST symmetry
$(s_b^{(1)}$). The substitutions of {\it all} the secondary fields into the expansion (69) lead to the following
\begin{eqnarray}
&&B_\mu^{(ab)}(x,\theta) = A_\mu(x) + \theta\, (\partial_\mu \bar C) \;\;\equiv\;\;  A_\mu (x) +\theta \;(s_{ab}^{(1)} A_\mu (x)),\nonumber\\
&&F^{(ab)}(x, \theta) = C(x) + \theta \;( -i\,B)  \;\;\equiv\,  C (x) + \theta\; ( s_{ab}^{(1)}\, C(x)),\nonumber\\
&&\bar F^{(ab)}(x, \theta) = \bar C(x) + \theta \;(0)  \qquad\;\equiv\,  \bar C (x) + \theta\; ( s_{ab}^{(1)}\, \bar C(x)),\nonumber\\
&&\Phi^{(ab)}(x, \theta) \,= \, \phi (x) + \theta\;(m\,\bar C) \;\;\; \equiv\;\; \phi(x) + \theta\,(s_{ab} ^{(1)}\,\phi (x)),\nonumber\\
&&\tilde\phi^{(ab)}(x, \theta) = \tilde{\phi}(x) + \theta\;(0)  \qquad\;\;\equiv\;\; \tilde{\phi}(x)+\theta\,(s_{ab} ^{(1)}\,\tilde\phi (x)),\nonumber\\
&&\tilde B^{(ab)}(x, \theta) =  B(x) + \theta\;(0)  \quad\quad\,\equiv\;\;  B(x)+\theta\,(s_{ab} ^{(1)}\,B (x)),\nonumber\\
&&\tilde {\cal B}^{(ab)}(x, \theta) =  {\cal B } (x) + \theta\;(0)  \qquad\;\;\equiv\;\;  {\cal B } (x)+\theta\,(s_{ab} ^{(1)}\, {\cal B }(x),
\end{eqnarray}
where, the anti-BRST symmetry transformations $(s_{ab}^{(1)})$ have been listed in Eq. (26)
and they appear on the r.h.s. of the super  expansions of all the {\it chiral} superfields of our theory as 
the coefficient of $\theta$. Hence, we conclude that we have derived {\it all} the anti-BRST 
symmetry transformations (26) and we have obtained a relationship and a mapping
\begin{eqnarray}
\frac {\partial\Omega}{\partial\theta}^{(ab)} (x, \theta) = s_{ab}^{(1)}\,\omega (x),\qquad \qquad s_{ab}^{(1)} \quad\leftrightarrow\quad \frac {\partial}{\partial\theta},
\end{eqnarray} 
which illustrate that the anti-BRST symmetry transformations for the {\it ordinary} generic field $\omega (x)$  
are nothing but the translation of the generic {\it chiral} superfields ($\Omega^{(ab)}(x, \theta)$), derived after the application
of the anti-BRST invariant restrictions (71), along the 
$\theta$-direction of the (2, 1)-dimensional chiral super-submanifold. Hence, we have established the mapping
$\partial_{\theta}\leftrightarrow   s_{ab}^{(1)}$ which implies that the nilpotency $([s_{ab}^{(1)}]^2 = 0)$
of the anti-BRST symmetry   $(s_{ab}^{(1)})$ is due to the nilpotency ($\partial_{\theta}^2 = 0$) of the translational
generator ($\partial_{\theta}$).

At this stage, we wish to capture the off-shell nilpotency and absolute anticommutativity 
of the conserved (anti-)BRST charges 
$Q_{(a)b}^{(1)}$ that have been expressed in a concise form in Eq. (49).
Taking the helps from the expansions (62), (68), (73), it can be checked that we have the following expressions
\begin{eqnarray*}
Q_b^{(1)}  & = &\frac {\partial}{\partial\bar\theta}\;\int\;d^{D-1}x\;\Big[i\;\dot{\bar F}^{(b)}(x,\bar\theta) F^{(b)}(x,\bar\theta)
- i\; \bar F^{(b)}(x,\bar\theta)\dot F^{(b)}(x,\bar\theta)\Big]\nonumber\\
&\equiv & \int\; d \bar\theta\int\;d^{D-1}x\;\Big[i\;\dot{\bar F}^{(b)}(x,\bar\theta) F^{(b)}(x,\bar\theta)
- i\; \bar F^{(b)}(x,\bar\theta)\dot F^{(b)}(x,\bar\theta)\Big],\nonumber\\
\end{eqnarray*}
\begin{eqnarray}
Q_{b}^{(1)} & = &\frac {\partial}{\partial\theta}\;\int\;d^{D-1}x\;[\,i\, F^{(ab)}(x,\theta)\; \dot { F}^{(ab)}(x,\theta)
\Big]\nonumber\\
&\equiv & \int\; d\theta\;\int\;d^{D-1}x\;[i\,\; F^{(ab)}(x,\theta)\; \dot { F}^{(ab)}(x,\theta)\Big],\nonumber\\
Q_{ab}^{(1)} & = & \frac {\partial}{\partial\theta}\int\; d^{D-1}x\;\Big[ i\;\bar F^{(ab)}(x,\theta)\;\dot F^{(ab)}(x,\theta)  
- i\;\dot {\bar F}^{(ab)}(x,\theta)\;F^{(ab)}(x,\theta)\Big]\nonumber\\
&\equiv & \int\; d\theta\;\int\; d^{D-1}x\;\Big[ i\;\bar F^{(ab)}(x,\theta)\;\dot F^{(ab)}(x,\theta)  
- i\;\dot {\bar F}^{(ab)}(x,\theta)\;F^{(ab)}(x,\theta)\Big],\nonumber\\
 Q_{ab}^{(1)} & = &\frac {\partial}{\partial\bar\theta}\;\int\;d^{D-1}x\;[- \,i\, \bar F^{(b)}(x,\bar\theta)\; \dot {\bar F}^{(b)}(x,\bar\theta)
\Big]\nonumber\\
&\equiv & \int\; d\bar\theta\;\int\;d^{D-1}x\;[- \,i\,\; \bar F^{(b)}(x,\bar\theta)\; \dot {\bar F}^{(b)}(x,\bar\theta)\Big],
\end{eqnarray}
for the (anti-)BRST charges in terms of the {\it (anti-)chiral}  superfields that have been obtained after the
applications of the (anti-)BRST invariant conditions/restrictions. Immediate consequences of the above expressions (due to $\partial_\theta ^2 = 
\partial_{\bar\theta} ^2 = 0 $)  are:
\begin{eqnarray}
\partial_{\theta}\;Q_{ab}^{(1)} = 0, \qquad \partial_{\bar\theta}\;Q_{ab} ^{(1)} = 0,\qquad \partial_{\theta}\;Q_b^{(1)}  = 0,
 \qquad \partial_{\bar\theta}\;Q_b ^{(1)} = 0.
\end{eqnarray}
These equations are very important because they encapsulate  in themselves the off-shell nilpotency and 
absolute anticommutativity properties of the (anti-)BRST charges $(Q_{(a)b}^{(1)})$ corresponding to the
continuous symmetry transformations (26) for the Lagrangian density ${\cal L}_{(b_1)}$.
This claim becomes very clear and  transparent  when we express (76) in the ordinary 2D space  (with $\partial_\theta \leftrightarrow s_{ab}^{(1)}$ and  
$\partial_{\bar\theta} \leftrightarrow s_b^{(1)}$), namely;
\begin{eqnarray}
s_{ab}^{(1)}\;Q_{ab}^{(1)} = 0, \qquad s_{b}^{(1)}\;Q_{ab} ^{(1)} = 0,\qquad s_{ab}^{(1)}\;Q_b^{(1)}  = 0,
 \qquad s_b^{(1)}\;Q_{b} ^{(1)} = 0.
\end{eqnarray}
Taking the help of the  basic principles behind the  definition  of a
generator for the corresponding continuous symmetry transformation (cf. Eq. (6)), we obtain the following:
\begin{eqnarray}
&&s_{ab}^{(1)} \,Q_{ab}^{(1)} = -\,i\,\{Q_{ab}^{(1)},Q_{ab}^{(1)}\} = 0 \quad\; \Longrightarrow  \qquad\qquad [Q_{ab}^{(1)}]^2 = 0,\nonumber\\
&&s_b^{(1)}  Q_b^{(1)} = -\,i\,{\{Q_b^{(1)},Q_b^{(1)}}\} = 0\quad\;\; \Longrightarrow   \quad\qquad [Q_b^{(1)}]^2 = 0,\nonumber\\
&&s_b ^{(1)} Q_{ab}^{(1)}  = -i\,{\{Q_{ab}^{(1)},Q_b^{(1)}}\} = 0 \quad\;\;\,\Longrightarrow    \;\;Q_{ab}^{(1)}\;Q_b^{(1)} + Q_{b}^{(1)}\;Q_{ab}^{(1)} = 0,  \nonumber\\
&&s_{ab}^{(1)} Q_b^{(1)} = -\,i\,{\{Q_b^{(1)},Q_{ab}^{(1)}}\} = 0\quad\;\,\, \Longrightarrow     \;\, Q_b^{(1)}\;Q_{ab}^{(1)} + Q_{ab}^{(1)}\;Q_{b}^{(1)} = 0.
\end{eqnarray} 
Thus, we note that we have captured the off-shell nilpotency and absolute anticommutativity of the conserved charges 
 within the framework of our {\it newly} proposed (see, e.g. [22-25]) (anti-)chiral superfield approach to BRST formalism (cf. Eqs. (77), (78)).

Against the backdrop of the above discussions, we concentrate now on the derivation of  (anti-)BRST
symmetries $s_{(a)b}^{(2)}$ for the Lagrangian density ${\cal L}_{(b_2)}$
within the framework 
of (anti-)chiral superfield approach to BRST formalism [22-25]. For this purpose,
first of all, we take into account the (anti-)chiral superfield expansions given in (69) and (60) with
the following replacements: $B(x)\rightarrow  \bar B(x), {\cal B}(x)\rightarrow  \bar {\cal B}(x),
\tilde B(x, \theta)\rightarrow     \tilde {\bar B}(x, \theta), \tilde B(x, \bar\theta)\rightarrow   \tilde {\bar B}(x, \bar\theta),
\tilde {\cal B}(x, \theta)\rightarrow     {\tilde{\bar {\cal B}}}(x, \theta),
 \tilde {\cal B}(x, \bar\theta)\rightarrow   {\tilde{\bar {\cal B}}}(x, \bar\theta)$.  
We note that the {\it secondary} fields in the expansions (69) and (60) remain the same.
For the derivation of the (anti-)BRST symmetry transformations $s_{(a)b}^{(2)}$, we check that the following are the
(anti-)BRST invariant quantities:
\begin{eqnarray}
&& s_{ab}^{(2)}\bar C = \; s_{ab}^{(2)}\bar B = \,\; s_{ab}^{(2)} \bar {\cal B} = 0,\,\; s_{ab}^{(2)}(m\,A_\mu +
\partial_\mu\,\phi) = 0,\;\; s_{ab}^{(2)} (\bar C\;\phi) = 0,  \nonumber\\
&& s_{ab}^{(2)}\big[A^{\mu}\;\partial_\mu \bar B + i\;\partial_\mu \bar C\; \partial^\mu C \big] = 0,
\quad s_{ab}^{(2)}\big[m\, \bar C \, C + i\, \bar B\, \phi \big] = 0,\nonumber\\
 && s_{ab}^{(2)} (\bar B\, \dot C - \dot {\bar B}\,  C) = 0,\quad\;\; s_{ab}^{(2)} (A^{\mu}\;\partial_\mu \bar C ) = 0,\;\;\;\;\;
 s_{ab}^{(2)} (\tilde\phi\; ) = 0.
\end{eqnarray} 
\begin{eqnarray}
&& s_b^{(2)} C = \; s_b ^{(2)}\bar B = \; s_b^{(2)}\bar {\cal B} = 0,\;\;\; s_b ^{(2)}(m\,A_\mu + \partial_\mu\,\phi) = 0,\;\;\quad s_b^{(2)} (C\;\phi) = 0,  \nonumber\\
&& s_b^{(2)}\big[A^{\mu}\;\partial_\mu \bar B + i\;\partial_\mu \bar C\; \partial^\mu C \big] = 0,\quad
\quad s_b^{(2)}\big[m\, \bar C \, C + i\, \bar B\, \phi\big] = 0,\nonumber\\
&& s_b^{(2)} (\bar B\, \dot{\bar C} - \dot {\bar B}\, \bar C) = 0,\;\;\;\;\; s_b^{(2)} (A^{\mu}\;\partial_\mu C) = 0,\;\;\;
 s_b^{(2)} (\tilde\phi\; ) = 0.
\end{eqnarray} 
According to the basic tenets of the (anti-)chiral superfield  approach, first of all, the anti-BRST invariant
quantities (79) have to be generalized onto the {\it chiral} (2, 1)-dimensional super-submanifold and BRST invariant 
quantities (80) have to be generalized onto (2, 1)-dimensional  {\it anti-chiral}  super-submanifold. After that, we demand the following restrictions
on the {\it chiral} superfields for the derivation of exact $s_{ab}^{(2)}$, namely;
\begin{eqnarray}
&&\bar F(x, \theta) = \bar C(x), \qquad \tilde { \bar B }(x, \theta) = \bar B(x),\qquad {\tilde{ \bar {\cal B}}}(x, \theta) = \bar {\cal B}(x),\nonumber\\
&& m  B_\mu (x, \theta) + \partial_\mu \Phi (x, \theta) = m\, A_\mu(x) + \partial_\mu \phi(x),\quad \tilde\Phi(x, \theta) = \tilde\phi(x),\nonumber\\
&& B^\mu (x, \theta)\, \partial_\mu \bar F (x, \theta) = A^\mu(x)\,\partial_\mu {\bar C}(x),\quad \bar F(x, \theta)\, \Phi(x, \theta)  = \bar C(x)\,\phi(x),\nonumber\\
&&B^\mu (x, \theta)\,\partial_\mu \tilde {\bar B}(x, \theta) + i\, \partial_\mu \bar F(x, \theta) \, \partial^\mu F(x, \theta) = A^\mu(x)\, \partial_\mu {\bar B}(x) + i\, \partial_\mu \bar C(x)\, \partial^\mu C(x),\nonumber\\
&& m\,\bar F(x, \theta)\, F(x, \theta) + i\, \tilde {\bar B}(x, \theta)\,\Phi(x, \theta) = m\, \bar C(x)\, C(x) + i\,  \bar B(x)\,\phi(x),\nonumber\\
&& \tilde {\bar B} (x, \theta)\, \dot F (x, \theta) - \dot {\tilde {\bar B}}(x, \theta) F(x,\theta) = \bar B(x)\,\dot C(x) - \dot {\bar B}(x)\,C(x).
\end{eqnarray}
The arguments for the derivation of the secondary fields, in terms of the basic and auxiliary fields of the  Lagrangian density ${\cal L}_{(b_2)}$, 
go along the similar lines as we have done for the derivations of  $s_{ab}^{(1)}$. We, ultimately, obtain  
the following (see also, e.g., Appendix C):
\begin{eqnarray}
\bar R_\mu = \partial_\mu \bar C,\qquad \bar B_1 = - \bar B,\quad \bar f_1 = -\,m\, \bar C,\quad \bar f_2 = \bar f_3 = \bar f_4 = \bar B_2 = 0.
\end{eqnarray}
The substitutions of these secondary fields into the appropriate super expansions of the {\it chiral} superfields lead to the 
following: 
\begin{eqnarray}
&&B_\mu^{(AB)}(x,\theta) = A_\mu(x) + \theta\, (\partial_\mu \bar C) \;\;\;\,\equiv\;\;  A_\mu (x) +\theta \;(s_{ab}^{(2)} A_\mu (x)),\nonumber\\
&&F^{(AB)}(x, \theta) = C(x) + \theta \;( -\,i\,\bar B)  \;\;\;\equiv\,  C (x) + \theta\; ( s_{ab}^{(2)}\, C(x)),\nonumber\\
&&\bar F^{(AB)}(x, \theta) = \bar C(x) + \theta \;( 0 )  \qquad\;\;\,\equiv\,  \bar C (x) + \theta\; ( s_{ab}^{(2)}\, \bar C(x)),\nonumber\\
&&\Phi^{(AB)}(x, \theta) \,= \, \phi (x) + \theta\;(-\,m\,\bar C) \,\,\equiv\;\; \phi(x) + \theta\,(s_{ab} ^{(2)}\,\phi (x)),\nonumber\\
&&\tilde\phi^{(AB)}(x, \theta) = \tilde{\phi}(x) + \theta\;(0)  \qquad\;\;\;\;\equiv\;\; \tilde{\phi}(x)+\theta\,(s_{ab} ^{(2)}\,\tilde\phi (x)),\nonumber\\
&&\tilde {\bar B}^{(AB)}(x, \theta) =  \bar B(x) + \theta\;(0)  \quad\;\;\quad\,\equiv\;\;  \bar B(x)+\theta\,(s_{ab} ^{(2)}\,\bar B (x)),\nonumber\\
&&{\tilde {\bar {\cal B}}}^{(AB)}(x, \theta) =  \bar {\cal B } (x) + \theta\;(0)  \qquad\;\;\;\equiv\;\;  \bar{\cal B } (x)+\theta\,(s_{ab} ^{(2)}\, 
\bar{\cal B }(x)),
\end{eqnarray}
where, on the r.h.s., we have found the coefficients of $\theta$ as the anti-BRST symmetry transformations
$s_{ab}^{(2)}$ that have been listed in Eq. (27). In other words, we have already derived the anti-BRST
symmetry transformations $s_{ab} ^{(2)}$ for the Lagrangian density ${\cal L}_{(b_2)}$.
We also note that superscript $(AB)$ on the {\it chiral} superfields (cf. l.h.s. of (83)) denotes the 
superfields that have been obtained after the applications of the restrictions (81).

For the derivation of the BRST-symmetry transformations  $s_b ^{(2)}$, we generalize the BRST invariant
quantities (80) onto (2, 1)-dimensional {\it anti-chiral} super-submanifold and invoke the following restrictions
on the {\it anti-chiral} superfields:
\begin{eqnarray}
&& F(x, \bar\theta) =  C(x), \qquad \tilde {\bar B} (x, \bar\theta) = \bar B(x),\qquad {\tilde{\bar{\cal B}}}(x, \bar\theta) = \bar{\cal B}(x),\nonumber\\
&& m  B_\mu (x, \bar\theta) + \partial_\mu \Phi (x, \bar\theta) = m\, A_\mu(x) + \partial_\mu \phi(x),\quad \tilde\Phi(x, \bar\theta) = \tilde\phi(x),\nonumber\\
&& B^\mu (x, \bar\theta)\, \partial_\mu  F (x, \bar\theta) = A^\mu(x)\,\partial_\mu  C(x),\quad F(x, \bar\theta)\, \Phi(x, \bar\theta)  
= C(x)\,\phi(x),\nonumber\\
&&B^\mu (x, \bar\theta)\,\partial_\mu \tilde {\bar B}(x, \bar\theta) + i\, \partial_\mu \bar F(x, \bar\theta) \, \partial^\mu F(x, \bar\theta) =
 A^\mu(x)\, \partial_\mu \bar B(x) + 
i\, \partial_\mu \bar C(x)\, \partial^\mu C(x),\nonumber\\
&& m\,\bar F(x, \bar\theta)\, F(x, \bar\theta) + i\, \tilde {\bar B}(x, \bar\theta)\,\Phi(x, \bar\theta) = m\, \bar C(x)\, C(x) + i\,  \bar B(x)\,\phi(x),\nonumber\\
&& {\tilde {\bar B}} (x, \bar\theta)\, \dot{\bar F} (x, \bar\theta) - \dot {\tilde{\bar B}}(x, \bar\theta) \bar F(x,\bar\theta) = \bar B(x)\,\dot {\bar C}(x) - \dot{\bar B}(x)\,\bar C(x).
\end{eqnarray}
The above restrictions lead to the determination of the secondary fields of the appropriate {\it anti-chiral} superfields (cf. Eq. (60)), in the terms 
of the basic and auxiliary fields of the Lagrangian density ${\cal L}_{(b_2)}$, as follows:
\begin{eqnarray}
 R_\mu = \partial_\mu C,\qquad B_2 = \bar B,\quad f_1 = -\,m\, C,\quad  f_2 =  f_3 =  f_4 =  B_1 = 0.
\end{eqnarray}
In the derivation of (85), the arguments and discussions have been taken on the similar lines as {\it that}
in the context of the derivation of $s_{b} ^{(1)}$ (cf. Appendix C, too). The substitutions of (85), into the appropriate super expansions 
of the {\it anti-chiral} superfields, leads to 
\begin{eqnarray}
&&B_\mu^{(B)}(x,\bar\theta) = A_\mu(x) + \bar\theta\, (\partial_\mu C) \;\;\;\equiv\;\;  A_\mu (x) +\bar\theta \;(s_b^{(2)} A_\mu (x)),\nonumber\\
&&F^{(B)}(x, \bar\theta) = C(x) + \bar\theta \;(0)  \;\;\qquad\equiv\,  C (x) + \bar\theta\; ( s_b^{(2)}\, C(x)),\nonumber\\
&&\bar F^{(B)}(x, \bar\theta) = \bar C(x) + \bar\theta \;(i\,\bar B )  \;\;\;\;\;\;\equiv\,  \bar C (x) + \bar\theta\; ( s_b^{(2)}\, \bar C(x)),\nonumber\\
&&\Phi^{(B)}(x, \bar\theta) \,= \, \phi (x) + \bar\theta\;(-\,m\, C) \, \equiv\;\; \phi(x) + \bar\theta\,(s_b ^{(2)}\,\phi (x)),\nonumber\\
&&\tilde\phi^{(B)}(x, \bar\theta)~ = \tilde{\phi}(x) + \bar\theta\;(0)  \;\qquad\;\equiv\;\; \tilde{\phi}(x)+\bar\theta\,(s_b^{(2)}\,\tilde\phi (x)),\nonumber\\
&&\tilde {\bar B}^{(B)}(x, \bar\theta) =  \bar B(x) + \bar\theta\;(0)  \qquad\;\,\equiv\;\;  \bar B(x)+\bar\theta\,(s_b ^{(2)}\,\bar B (x)),\nonumber\\
&&{\tilde {\bar {\cal B}}}^{(B)}(x, \bar\theta) =  \bar {\cal B } (x) + \bar\theta\;(0)  \qquad\;\;\equiv\;\;  \bar{\cal B } (x)+\bar\theta\,(s_b^{(2)}\, 
\bar{\cal B }(x)),
\end{eqnarray}
where, on the r.h.s. of (86),  we have obtained the BRST symmetry transformation $s_b ^{(2)}$ as the coefficients of $\bar\theta$
(which have been quoted in Eq. (27)). The superscript $(B)$ on the {\it anti-chiral} superfields denotes the superfields 
that have been obtained after the applications of the BRST invariant restrictions (84) and which lead to the determination of 
the BRST symmetry transformation  $s_b^{(2)}$ as the coefficients of $\bar\theta$ in {\it their} super expansions.

Against the backdrop of the super expansions (83) and (86), we capture the off-shell nilpotency
and absolute anticommutativity of the conserved charges $Q_{(a)b}^{(2)}$ which are associated
with the Lagrangian density ${\cal L}_{(b_2)}$. For this purpose, we take into account 
the concise forms of the nilpotent and conserved (anti-)BRST charges $Q_{(a)b}^{(2)}$  that are listed in Eq. (57).
It can be checked  that we have the following expressions for $Q_{(a)b}^{(2)}$ (cf. Eq. (57))
in terms of the superfields (derived in Eqs.(83) and (86)), Grassmannian differentials $(d\,\theta, d\,\bar\theta)$
and corresponding partial derivatives $(\partial_\theta, \partial_{\bar\theta})$, namely;
\begin{eqnarray}
Q_b^{(2)}  & = &\frac {\partial}{\partial\bar\theta}\;\int\;d^{D-1}x\;\Big[i\;\dot{\bar F}^{(B)}(x,\bar\theta) F^{(B)}(x,\bar\theta)
- i\; \bar F^{(B)}(x,\bar\theta)\dot F^{(B)}(x,\bar\theta)\Big]\nonumber\\
&\equiv & \int\; d \bar\theta\int\;d^{D-1}x\;\Big[i\;\dot{\bar F}^{(B)}(x,\bar\theta) F^{(B)}(x,\bar\theta)
- i\; \bar F^{(B)}(x,\bar\theta)\dot F^{(B)}(x,\bar\theta)\Big],\nonumber\\
Q_b^{(2)}  & = & \frac {\partial}{\partial\theta}\int\; d^{D-1}x\;\Big[i\;F^{(AB)}(x,\theta)\dot F^{(AB)}(x,\theta)\Big]\nonumber\\
& \equiv & \int\; d\theta\;\int\; d^{D-1}x\;\Big[i\;F^{(AB)}(x,\theta)\dot F^{(AB)}(x,\theta)\Big],\nonumber\\
 Q_{ab}^{(2)} & = & \frac {\partial}{\partial\theta}\int\; d^{D-1}x\;\Big[ i\;\bar F^{(AB)}(x,\theta)\;\dot F^{(AB)}(x,\theta)  
- i\;\dot {\bar F}^{(AB)}(x,\theta)\;F^{(AB)}(x,\theta)\Big]\nonumber\\
&\equiv & \int\; d\theta\;\int\; d^{D-1}x\;\Big[ i\;\bar F^{(AB)}(x,\theta)\;\dot F^{(AB)}(x,\theta)  
- i\;\dot {\bar F}^{(AB)}(x,\theta)\;F^{(AB)}(x,\theta)\Big],\nonumber\\
 Q_{ab}^{(2)} & = &\frac {\partial}{\partial\bar\theta}\;\int\;d^{D-1}x\;\Big[-\,i\, \bar F^{(B)}(x,\bar\theta)\; \dot {\bar F}^{(B)}(x,\bar\theta)
\Big]\nonumber\\
&\equiv & \int\; d\bar\theta\;\int\;d^{D-1}x\;\Big[- \,i\, \bar F^{(B)}(x,\bar\theta)\; \dot {\bar F}^{(B)}(x,\bar\theta)\Big],
\end{eqnarray}
where the superfields with superscript $(B)$ and $(AB)$ have already been explained earlier. A close look
at (87) implies that we have {\it already} the following (due to  $\partial_\theta^2 = \partial_{\bar\theta}^2 = 0$): 
\begin{eqnarray}
\partial_{\bar\theta}\,Q_b^{(2)} = 0,\qquad \partial_{\theta}\,Q_b^{(2)} = 0, \qquad \partial_{\theta}\,Q_{ab}^{(2)} = 0,
\qquad \partial_{\bar\theta}\,Q_{ab}^{(2)} = 0.
\end{eqnarray}
These relations are crucial for capturing the off-shell nilpotency and absolute anticommutativity
of the charges $Q_{(a)b}^{(2)}$ in view of the observations that $\partial_\theta\leftrightarrow  
s_{ab}^{(2)}, \partial_{\bar\theta}\leftrightarrow s_b^{(2)}$. To be more precise, it can be checked
that the relationships of (88) can be expressed,
in the ordinary 2D spacetime in terms of the (anti-)BRST symmetry transformations $(s_{(a)b}^{(2)})$ , as:
\begin{eqnarray}
&& s_{b}^{(2)}\,Q_{b}^{(2)}\, = -\,i\,\lbrace{Q_{b}^{(2)},\,Q_{b}^{(2)}}\rbrace\, = \,0  \qquad 
 \Longrightarrow\quad [Q_{b}^{(2)}]^2 = 0,\nonumber\\
&& s_{ab}^{(2)}\,Q_{ab}^{(2)}\, = -\,i\,\lbrace{Q_{ab}^{(2)},\,Q_{ab}^{(2)}}\rbrace\, = \,0 \quad\quad 
\Longrightarrow\quad [Q_{ab}^{(2)}]^2 = 0,\nonumber\\
 && s_{ab}^{(2)}\,Q_{b}^{(2)}\, = -\,i\,\lbrace{Q_{b}^{(2)},\,Q_{ab}^{(2)}}\rbrace\, = \,0 
\qquad\;\Longrightarrow\quad Q_{b}^{(2)}\,Q_{ab}^{(2)} + Q_{ab}^{(2)}\,Q_{b}^{(2)} = 0,\nonumber\\
&& s_{b}^{(2)}\,Q_{ab}^{(2)}\, = -\,i\,\lbrace{Q_{ab}^{(2)},\,Q_{b}^{(2)}}\rbrace\, = \,0 
\qquad\;\Longrightarrow\quad Q_{ab}^{(2)}\,Q_{b}^{(2)} + Q_{b}^{(2)}\,Q_{ab}^{(2)} = 0.
\end{eqnarray}
The above relationships, in a very explicit fashion, demonstrate the nilpotency and absolute anticommutativity
of the (anti-)BRST charges $Q_{(a)b}^{(2)}$ where we have exploited the key ideas behind the intimate connection
between the continuous symmetry transformations and their generators (cf. Eq. (6)). An interesting  result is 
the observation that the nilpotency of the BRST charge $Q_{(b)}^{(2)}$ is connected with the nilpotency $(\partial_{\bar\theta}^2 = 0)$
of the translational generator $\partial_{\bar\theta}$ {\it but} its absolute anticommutativity,
with the anti-BRST charge, is deeply related with the nilpotency $(\partial_{\theta}^2 = 0)$ of the 
translational generator $\partial_{\theta}$ . Geometrically, the translation of BRST charge $Q_{b}^{(2)}$ along $\bar\theta$-direction
of the (2, 1)-dimensional {\it anti-chiral} super-submanifold is related with its nilpotency. However, the translation of the 
{\it same} charge along $\theta$-direction of the {\it chiral} super-submanifold leads to the observation of absolute
anticommutativity of the BRST charge {\it with} anti-BRST charge (i.e.  $Q_{b}^{(2)}\,Q_{ab}^{(2)} + Q_{ab}^{(2)}\,Q_{b}^{(2)} = 0 $). 
Similar kinds of statements can be made for the anti-BRST charge $Q_{ab}^{(2)}$ as well.

We end this subsection with the remarks that the nilpotency $(\partial_\theta ^2 = 0, \partial_{\bar \theta} ^2 = 0)$ 
properties of the translational generators $(\partial_\theta , \partial_{\bar \theta} )$ are deeply 
connected with the off-shell nilpotency $([s_{(a)b} ^{(1,2)}]^ 2 = 0,\; [Q_{(a)b} ^{(1,2)}]^ 2 = 0)$ of the (anti-)BRST symmetry transformations 
$s_{(a)b} ^{(1, 2)}$ and corresponding conserved charges 
 $Q_{(a)b}^{(1, 2)}$ for the Lagrangian densities  ${\cal L}_{(b_1, b_2)}$ which have been considered  for our present discussions
on the {\it modified} 2D Proca theory.

\vskip 1cm

\subsection{Off-Shell Nilpotent (Anti-)co-BRST Symmetries and Conserved Charges: (Anti-)Chiral Superfield Approach}

We exploit the (anti-)chiral super expansions of (69) and (60) to derive, first of all, the nilpotent (anti-)co-BRST symmetry transformations
 $(s_{(a)d} ^{(1)}) $ for the Lagrangian density ${\cal L}_{(b_1)}$. Towards this goal in mind, we note that the following 
 (anti-)co-BRST invariant quantities
\begin{eqnarray}
&&s_{ad}^{(1)} C = 0, \quad s_{ad}^{(1)} (\partial\cdot A + m\, \phi) = 0,\quad s_{ad}^{(1)} B = s_{ad}^{(1)} {\cal B} = 0,
\quad s_{ad}^{(1)}(\tilde\phi\; C) = 0,\nonumber\\
&&s_{ad}^{(1)}[m\,A_\mu - \varepsilon_{\mu\nu}\,\partial^\nu\, \tilde\phi\,] = 0,\quad
s_{ad}^{(1)} [\varepsilon^{\mu\nu}\, A_\nu\, \partial_\mu {\cal B} - i\,\partial_\mu \bar C\quad \partial^\mu C] = 0,\quad s_{ad}^{(1)}(\partial\cdot A) = 0,\nonumber\\
&& s_{ad}^{(1)} (\varepsilon ^{\mu\nu} A_\mu\;\partial_\nu  C) = 0,\qquad s_{ad}^{(1)}[{\cal B}\, \dot {\bar C} - \dot{\cal B}\,  \bar C] = 0,\nonumber\\
&&s_{d}^{(1)} \bar C = 0, \quad s_{d}^{(1)} (\partial\cdot A + m\, \phi) = 0,\quad s_{d}^{(1)} B = s_{d}^{(1)} {\cal B} = 0,\quad
s_{d}^{(1)}(\tilde\phi\;\bar C) = 0,  \nonumber\\
&&s_{d}^{(1)}[m\,A_\mu - \varepsilon_{\mu\nu}\partial^\nu\, \tilde\phi] = 0,\quad
s_d^{(1)} [\varepsilon^{\mu\nu}\, A_\nu\, \partial_\mu {\cal B} - i\,\partial_\mu \bar C\, \partial^\mu C] = 0,\quad s_{d}^{(1)}(\partial\cdot A) = 0,
\nonumber\\
&& s_d^{(1)} (\varepsilon ^{\mu\nu} A_\mu\;\partial_\nu \bar C) = 0, \quad s_d^{(1)}[{\cal B}\, \dot C - \dot{\cal B}\,  C] = 0,
 \end{eqnarray}
are to be generalized onto a (2 ,1)-dimensional (anti-)chiral super-submanifold and we have to demand specific restrictions on the 
(anti-)chiral superfields to obtain the secondary fields of (69) and (60) in terms of the basic and auxiliary fields of the Lagrangian density
${\cal L}_{(b_1)}$.

We concentrate on the derivation of the anti-co-BRST symmetry transformation $ s_{ad}^{(1)}$ by imposing the following 
restrictions on the {\it anti-chiral} superfields  
\begin{eqnarray}
 &&F(x, \bar\theta)  =  C(x), \quad \tilde{\cal B}(x, \bar\theta) = {\cal B}(x), \quad {\tilde B}(x, \bar\theta) =  B(x),\quad 
\partial_\mu B^\mu (x, \bar\theta) = \partial _\mu A^\mu (x),\nonumber\\
 &&\Phi (x, \bar\theta) = \phi (x),\quad m\, B_\mu (x, \bar\theta) - \varepsilon_{\mu\nu}\,\partial^\nu \tilde\Phi(x, \bar\theta) = m\,A_\mu(x) - \varepsilon_{\mu\nu}\,\partial^\nu\, \tilde\phi\,(x),  \nonumber\\
&& \tilde{\cal B}(x, \bar\theta)\,
\dot{\bar F}(x, \bar\theta) - {\dot{\tilde {\cal B}}} (x, \bar\theta) \bar F(x, \bar\theta) = {\cal B}(x)\dot {\bar C}(x) - \dot {\cal B}(x)\,\bar C(x),\nonumber\\
&&\tilde\Phi(x, \bar\theta) F (x, \bar\theta) = \tilde\phi(x)\,C(x),\quad
\varepsilon^{\mu\nu}\,B_\nu(x, \bar\theta)\,\partial_\mu \tilde{\cal B}(x, \bar\theta)
 - i \, \partial_\mu \bar F(x, \bar\theta)\,\partial^\mu F(x, \bar\theta)
\nonumber\\
&& = \varepsilon^{\mu\nu}\, A_\nu(x)\, \partial_\mu {\cal B}(x) - i\,\partial_\mu \bar C(x)\, \partial^\mu C(x),
\end{eqnarray}
which lead to the determination of some of the {\it trivial} expressions for the secondary fields in the super expansions (60)  as follows:
\begin{eqnarray}
\partial_\mu\,R^\mu = 0,\quad\quad B_1 = 0,\quad \quad f_1 = f_3 = f_4 = 0.  
\end{eqnarray}
The substitution of $B_1 =  f_1 = f_3 = f_4 = 0$, in the expansions (60), leads to:
\begin{eqnarray}
&&F^{(ad)}(x, \bar\theta)  =  C(x) + \bar\theta\;(0)\quad\equiv \quad C(x) + \bar\theta\;(s_{ad}^{(1)}\; C(x)) ,\nonumber\\
&&\tilde{\cal B}^{(ad)}(x, \bar\theta) = {\cal B}(x) + \bar\theta\;(0)\;\quad \equiv \quad{\cal B}(x) + \bar\theta\;(s_{ad}^{(1)}\; {\cal B}(x)) , \nonumber\\
 &&{\tilde B}^{(ad)}(x, \bar\theta) =  B(x) + \bar\theta\;(0) \quad \equiv \quad B(x) + \bar\theta\;(s_{ad}^{(1)}\; {B}(x))  ,\nonumber\\
&&\Phi^{(ad)}(x, \bar\theta)  =  \phi(x) + \bar\theta\;(0)\quad \;\,\equiv \quad \phi(x) + \bar\theta\;(s_{ad}^{(1)}\; \phi (x)),
\end{eqnarray}
which shows that we have already derived the transformations $ s_{ad}^{(1)}\; C(x) = s_{ad}^{(1)}\; {\cal B}(x) = s_{ad}^{(1)}\; {B}(x)
 = s_{ad}^{(1)}\; \phi (x) = 0 $ as the coefficients of $\bar\theta$ in the expansions for the superfields with the superscript $(ad)$. 
 The latter symbol denotes that the {\it anti-chiral} superfields, on the l.h.s. of (93), have been obtained after the applications
 of the restrictions (91).

 The arguments and discussions for the determination of the secondary fields in terms of the basic and auxiliary fields
 of the Lagrangian density ${\cal L}_{(b_1)}$ go along similar lines as we have done  in the  previous subsection 5.1 (see also, Appendix C for details). Ultimately, 
we obtain the following expressions for the secondary fields: 
\begin{eqnarray}
R_\mu (x) =  -\, \varepsilon_{\mu\nu} \,\partial^\nu\,C (x), \qquad  f_2 (x) = -\,m\,C,\qquad  B_2 (x) =  {\cal B}(x).
\end{eqnarray} 
The substitutions of the above values into the appropriate expansions for the (2, 1)-dimensional {\it anti-chiral} superfields lead to:
\begin{eqnarray}
&& B_\mu^ {(ad)}(x, \bar\theta) ~~~ = \; A_\mu + \bar\theta\;(-\, \varepsilon_{\mu\nu}\,\partial^\nu\,C)\;\;\equiv \quad A_\mu (x)
 + \bar\theta\;(s_{ad}^{(1)}\;A_\mu (x)) ,\nonumber\\
&& \bar F^{(ad)}(x, \bar\theta)\quad = \; \bar C(x) + \bar\theta\;(i\,{\cal B})\;\qquad\;\; \equiv \quad \bar C(x) 
+ \bar\theta\;(s_{ad}^{(1)}\; \bar C(x)) , \nonumber\\
&& \tilde{\Phi}^{(ad)}(x, \bar\theta)\quad  = \; \tilde\phi(x) + \bar\theta\;(-\,m\,C)\quad\;\;\equiv \quad \tilde\phi(x) + 
\bar\theta\;(s_{ad}^{(1)}\; \tilde\phi (x)).
\end{eqnarray}
From Eqs. (93) and (95), it is crystal clear that we have computed  {\it all} the anti-co-BRST symmetry transformations
$s_{ad}^{(1)}$ for {\it all} the fields of the Lagrangian density ${\cal L}_{(b_1)} $.

To determine all the secondary fields of (69) in terms of the basic and auxiliary fields 
of the Lagrangian density ${\cal L}_{(b_1)} $, we have to invoke the co-BRST (i.e dual-BRST) invariant
quantities  of Eq. (90) and generalize {\it them} onto the (2, 1)-dimensional {\it chiral}
super-submanifold with the following restrictions:
\begin{eqnarray}
&&\bar F(x, \theta)  =  \bar C(x), \quad \tilde{\cal B}(x, \theta) = {\cal B}(x), \quad {\tilde B}(x, \theta) =  B(x),\quad 
\partial_\mu B^\mu (x, \theta) = \partial _\mu A^\mu (x),\nonumber\\
 &&\Phi (x, \theta) = \phi (x),\quad m\, B_\mu (x, \theta) - \varepsilon_{\mu\nu}\,\partial^\nu \tilde\Phi(x, \theta) = m\,A_\mu(x) - \varepsilon_{\mu\nu}\,\partial^\nu\, \tilde\phi\,(x),  \nonumber\\
&& \tilde{\cal B}(x, \theta)\,
\dot F(x, \theta) - {\dot {\tilde {\cal B}}} (x, \theta)  F(x, \theta) = {\cal B}(x)\dot  C(x) - \dot {\cal B}(x)\, C(x),\nonumber\\
&&\tilde\Phi(x, \theta) \bar F (x, \theta) = \tilde\phi(x)\,\bar C(x),\quad
\varepsilon^{\mu\nu}\,B_\nu(x, \theta)\,\partial_\mu \tilde{\cal B}(x, \theta)
 - i \, \partial_\mu \bar F(x, \theta)\,\partial^\mu F(x, \theta)
\nonumber\\
&& = \varepsilon^{\mu\nu}\, A_\nu(x)\, \partial_\mu {\cal B}(x) - i\,\partial_\mu \bar C(x)\, \partial^\mu C(x).
\end{eqnarray}  
We demand that the {\it chiral} superfields (and their useful combinations) on the l.h.s.
of the above equations {\it must} be independent of the Grassmannian variable $\theta$
because the co-BRST invariant quantities (for a model of a Hodge theory) are a set of {\it physical}
quantities at the {\it quantum} level. The above restrictions lead to the following relationships
between the secondary fields of the expansions (69) and the basic and auxiliary fields of ${\cal L}_{(b_1)}$,
namely:
 \begin{eqnarray}
 \bar R_\mu = -\,\varepsilon_{\mu\nu}\partial^\nu \bar C, \quad \bar B_1 = - \,{\cal B},\quad 
\bar f_2 = -\,m\, \bar C,\quad  \bar f_1 =  \bar f_3 =  \bar f_4 =  \bar B_2 = 0.
\end{eqnarray}
The substitutions of the above {\it secondary} fields into the expansions (69) lead to  
\begin{eqnarray}
&& B_\mu^ {(d)}(x, \theta)  = \; A_\mu + \theta\;(-\, \varepsilon_{\mu\nu}\,\partial^\nu\,\bar C)\;\equiv \quad A_\mu (x)
 + \theta\;(s_{d}^{(1)}\;A_\mu (x)) ,\nonumber\\
&& \bar F^{(d)}(x, \theta) = \; \bar C(x) + \theta\;(0)\;\;\;\qquad\;\; \equiv \quad \bar C(x) 
+ \theta\;(s_{d}^{(1)}\; \bar C(x)) , \nonumber\\
&&F^{(d)}(x, \theta)  =  C(x) + \theta\;(-\,i\,{\cal B})\qquad\equiv \quad C(x) + \theta\;(s_{d}^{(1)}\; C(x)) ,\nonumber\\
&&\Phi^{(d)}(x, \theta)  =  \phi(x) + \theta\;(0)\qquad\quad\;\; \;\,\equiv \quad \phi(x) + \theta\;(s_{d}^{(1)}\; \phi (x)),\nonumber\\
&& \tilde{\Phi}^{(d)}(x, \theta)  = \; \tilde\phi(x) + \theta\;(-\,m\,\bar C)\quad\;\,\equiv \quad \tilde\phi(x) + 
\theta\;(s_{d}^{(1)}\; \tilde\phi (x)),\nonumber\\
&&{\tilde B}^{(d)}(x, \theta) =  B(x) + \theta\;(0) \qquad\quad\;\;\; \equiv \quad B(x) + \theta\;(s_{d}^{(1)}\; {B}(x)),\nonumber\\
&&\tilde{\cal B}^{(d)}(x, \theta) = {\cal B}(x) + \theta\;(0)\;\qquad\quad\;\;\; \equiv \quad{\cal B}(x) + \theta\;(s_{d}^{(1)}\; {\cal B}(x)), 
\end{eqnarray}
where the coefficients of $\theta$, on the r.h.s., of the above expansions are nothing but the 
co-BRST symmetry transformations (35) and the superscript $(d)$ on the {\it chiral} superfields, on the l.h.s., denotes the
superfields that have been obtained after the applications of the restrictions (96) {\it and} which lead to the determination
 of the co-BRST symmetry transformations $s_d^{(1)}$ for the Lagrangian density ${\cal L}_{(b_1)}$ 
of our 2D {\it modified} Proca theory.

Taking the helps of expansions in (93), (95) and (98), we can now express the co-BRST and anti-co-BRST charges in the following 
explicit forms:
\begin{eqnarray}
Q_d^{(1)} &=& \int dx \,\Bigl[\frac{\partial}{\partial{\theta}}\,\Big(i\, \bar F^{(d)}(x, {\theta})\,
\dot F^{(d)}(x, {\theta})
- i\, \dot{\bar F}^{(d)}(x, {\theta})\,F^{(d)}(x, {\theta})\Big)\,\Bigr]\nonumber\\
&\equiv & \int dx \Bigl[\int\, d\,\theta\,\Big(i \bar F^{(d)}(x, {\theta})\dot F^{(d)}(x, {\theta}) 
- i\, \dot{\bar F}^{(d)}(x, {\theta})\,F^{(d)}(x, {\theta})\Big)\,\Bigr],\nonumber\\
Q_{d}^{(1)} &=& \int dx \,\Bigl[\frac{\partial}{\partial{\bar\theta}}\,\Big(i\, 
\dot{\bar F}^{(ad)}(x, \bar{\theta})\,{\bar F}^{(ad)}(x, \bar{\theta})\Big)\,\Bigr] \nonumber \\
&\equiv &  \int dx \Bigl[\int d\,\bar\theta\,\Big(i\, \dot{\bar F}^{(ad)}(x, \bar{\theta})\,{\bar F}^{(ad)}
(x, \bar{\theta})\Big)\,\Bigr],\nonumber\\
Q_{ad}^{(1)} &=& \int dx \Bigl[\frac{\partial}{\partial\bar\theta}\, \Big(i\,\dot{\bar F} ^{(ad)}(x, \bar\theta)\,
 F^{(ad)}(x, \bar\theta) \, - \,i \bar F^{(ad)}(x, \bar\theta)\, 
\dot F^{(ad)}(x, \bar\theta)\Big) \Bigr] \nonumber\\
&\equiv & \int dx \,\Bigl[\int\, d\,\bar\theta\,\Big(i\,\dot{\bar F} ^{(ad)}(x, \bar\theta)\, F^{(ad)}(x, \bar\theta) 
\, - \,i \bar F^{(ad)}(x, \bar\theta)\, 
\dot F^{(ad)}(x, \bar\theta)\Big)\Bigr],\nonumber\\
Q_{ad}^{(1)} &=& -\,i\,\int dx \Bigl[\frac{\partial}{\partial\theta}\,
\Big(\dot F^{(d)}(x, {\theta})\, F^{(d)}(x, {\theta})\Big)\, \Bigr] \nonumber \\
&\equiv &  -\,i\,\int dx \,\Bigl[\int\,d\,\theta\,\Big( \dot F^{(d)}(x, {\theta})\,
F^{(d)}(x, {\theta})\Big)\, \Bigr].
\end{eqnarray}   
A close and clear observation of the above expressions for the (anti-)co-BRST
charges $(Q_{(a)d}^{(1)})$ immediately  implies the following (due to $\partial_\theta^2 = \partial_{\bar\theta}^2 = 0 $),
namely;
\begin{eqnarray}
\partial_{\bar\theta}\,Q_d^{(1)} = 0,\qquad \partial_{\theta}\,Q_d^{(1)} = 0, \qquad \partial_{\theta}\,Q_{ad}^{(1)} = 0,
\qquad \partial_{\bar\theta}\,Q_{ad}^{(1)} = 0,
\end{eqnarray}
which encompass, in their  folds, the off-shell nilpotency and absolute anticommutativity of the (anti-)co-BRST
charges $Q_{(a)d}^{(1)}$. The above statement becomes transparent when we express (100) in the 2D {\it ordinary} spacetime 
(with the identifications: $\partial_{\bar\theta} \leftrightarrow s_{ad}^{(1)}, \,\partial_{\theta} \leftrightarrow s_{d}^{(1)}$),
in the language of the continuous symmetry transformations  $s_{(a)d}^{(1)}$ (and corresponding conserved charges $Q_{(a)d}^{(1)}$) for the Lagrangian density
${\cal L}_{(b_1)}$, namely; 
\begin{eqnarray}
&& s_{d}^{(1)}\,Q_{d}^{(1)}\, = -\,i\,\lbrace{Q_{d}^{(1)},\,Q_{d}^{(1)}}\rbrace\, = \,0  \qquad 
 \Longrightarrow\quad [Q_{d}^{(1)}]^2 = 0,\nonumber\\
&& s_{ad}^{(1)}\,Q_{ad}^{(1)}\, = -\,i\,\lbrace{Q_{ad}^{(1)},\,Q_{ad}^{(1)}}\rbrace\, = \,0 \quad\quad 
\Longrightarrow\quad [Q_{ad}^{(1)}]^2 = 0,\nonumber\\
 && s_{ad}^{(1)}\,Q_{d}^{(1)}\, = -\,i\,\lbrace{Q_{d}^{(1)},\,Q_{ad}^{(1)}}\rbrace\, = \,0 
\qquad\;\Longrightarrow\quad Q_{d}^{(1)}\,Q_{ad}^{(1)} + Q_{ad}^{(1)}\,Q_{d}^{(1)} = 0,\nonumber\\
&& s_{d}^{(1)}\,Q_{ad}^{(1)}\, = -\,i\,\lbrace{Q_{ad}^{(1)},\,Q_{d}^{(1)}}\rbrace\, = \,0 
\qquad\;\Longrightarrow\quad Q_{ad}^{(1)}\,Q_{d}^{(1)} + Q_{d}^{(1)}\,Q_{ad}^{(1)} = 0,
\end{eqnarray}
which demonstrate the off-shell nilpotency and absolute anticommutativity of the  conserved (anti-)co-BRST 
charges  $Q_{(a)d}^{(1)}$ for the Lagrangian density ${\cal L}_{(b_1)}$.

 As we have discussed various aspects of (anti-)co-BRST symmetries $(s_{(a)d}^{(1)})$ and corresponding charges $(Q_{(a)d}^{(1)})$
 for the Lagrangian density ${\cal L}_{(b_1)}$, we can do the {\it same} for the Lagrangian density ${\cal L}_{(b_2)}$. Towards this objective in mind,
 first of all, we note that the following (anti-)co-BRST invariant quantities w.r.t. $s_{(a)d}^{(2)}$, namely:
 \begin{eqnarray}
&&s_{ad}^{(2)} C = 0, \quad s_{ad}^{(2)} (\partial\cdot A - m\, \phi) = 0,\quad s_{ad}^{(2)} \bar B = s_{ad}^{(2)} \bar {\cal B} = 0,
\quad s_{ad}^{(2)}(\tilde\phi\; C) = 0,\nonumber\\
&&s_{ad}^{(2)}[m\,A_\mu + \varepsilon_{\mu\nu}\,\partial^\nu\, \tilde\phi\,] = 0,\quad
s_{ad}^{(2)} [\varepsilon^{\mu\nu}\, A_\nu\, \partial_\mu \bar {\cal B} - i\,\partial_\mu \bar C\; \partial^\mu C] = 0,\quad s_{ad}^{(2)}(\partial\cdot A) = 0,
\nonumber\\
&& s_{ad}^{(2)} (\varepsilon ^{\mu\nu} A_\mu\;\partial_\nu  C) = 0,\qquad s_{ad}^{(2)}[\bar {\cal B}\, \dot {\bar C} - {\dot{\bar{\cal B}}}\,  \bar C] = 0,\nonumber\\
&&s_{d}^{(2)} \bar C = 0, \quad s_{d}^{(2)} (\partial\cdot A - m\, \phi) = 0,\quad s_{d}^{(2)} \bar B = s_{d}^{(2)} \bar {\cal B} = 0,\quad
s_{d}^{(2)}(\tilde\phi\;\bar C) = 0,  \nonumber\\
&&s_{d}^{(2)}[m\,A_\mu + \varepsilon_{\mu\nu}\partial^\nu\, \tilde\phi] = 0,\quad
s_d^{(2)} [\varepsilon^{\mu\nu}\, A_\nu\, \partial_\mu \bar{\cal B} - i\,\partial_\mu \bar C\, \partial^\mu C] = 0,\quad s_{d}^{(2)}(\partial\cdot A) = 0,
\nonumber\\
&& s_d^{(2)} (\varepsilon ^{\mu\nu} A_\mu\;\partial_\nu \bar C) = 0, \quad s_d^{(2)}[\bar {\cal B}\, \dot C - {\dot{\bar {\cal B}}}\,  C] = 0,
 \end{eqnarray}
 are to be generalized onto (2, 1)-dimensional (anti-)chiral super-submanifolds (of the {\it general} (2, 2)-dimensional supermanifold) and we have to invoke 
 specific restrictions on {\it them} so that we could derive the (anti-)co-BRST symmetry transformation $(s_{(a)d}^{(2)})$ for the Lagrangian density 
${\cal L}_{(b_2)}$ within the framework of (anti-)chiral superfield formalism.

First and foremost, we concentrate on the derivation of anti-co-BRST symmetry transformations $(s_{ad}^{(2)})$. In this context, the following 
restrictions on the {\it anti-chiral} superfields (emerging from a close look at (102)), namely;
\begin{eqnarray}
 &&F(x, \bar\theta)  =  C(x), \quad {\tilde{\bar{\cal B}}}(x, \bar\theta) = \bar{\cal B}(x), \quad {\tilde {\bar B}}(x, \bar\theta) =  \bar B(x),\quad 
\partial_\mu B^\mu (x, \bar\theta) = \partial _\mu A^\mu (x),\nonumber\\
 &&\Phi (x, \bar\theta) = \phi (x),\quad m\, B_\mu (x, \bar\theta) + \varepsilon_{\mu\nu}\,\partial^\nu \tilde\Phi(x, \bar\theta) = m\,A_\mu(x) + \varepsilon_{\mu\nu}\,\partial^\nu\, \tilde\phi\,(x),  \nonumber\\
&& {\tilde{\bar{\cal B}}}(x, \bar\theta)\,
\dot{\bar F}(x, \bar\theta) - {\dot{\tilde{\bar{\cal B}}}} (x, \bar\theta) \bar F(x, \bar\theta) = \bar{\cal B}(x)\dot {\bar C}(x) - {\dot {\bar{\cal B}}}
(x)\,\bar C(x),\nonumber\\
&&\tilde\Phi(x, \bar\theta) F (x, \bar\theta) = \tilde\phi(x)\,C(x),\quad
\varepsilon^{\mu\nu}\,B_\nu(x, \bar\theta)\,\partial_\mu {\tilde{\bar{\cal B}}}(x, \bar\theta)
 - i \, \partial_\mu \bar F(x, \bar\theta)\,\partial^\mu F(x, \bar\theta)
\nonumber\\
&& = \varepsilon^{\mu\nu}\, A_\nu(x)\, \partial_\mu {\bar{\cal B}}(x) - i\,\partial_{\mu} {\bar C}(x)\, \partial^{\mu} C(x),
\end{eqnarray}
lead to the derivation of secondary fields of super expansions (60) in terms of the auxiliary and basic fields of ${\cal L}_{(b_2)}$ as:
 \begin{eqnarray}
 R_\mu = -\,\varepsilon_{\mu\nu}\partial^\nu  C, \quad B_2 = i\,\bar{\cal B},\quad f_2 = m\,C,\quad B_1 =  f_1 = f_3 = f_4 = 0.
\end{eqnarray}
Substitutions of these secondary fields into the super expansions (60)  leads to the following:
\begin{eqnarray}
&& B_\mu^ {(AD)}(x, \bar\theta)  = \; A_\mu + \bar\theta\;(-\, \varepsilon_{\mu\nu}\,\partial^\nu\, C)\;\equiv \quad A_\mu (x)
 + \bar\theta\;(s_{ad}^{(2)}\;A_\mu (x)) ,\nonumber\\
&& \bar F^{(AD)}(x, \bar\theta) = \; \bar C(x) + \bar\theta\;(\,i\,\bar{\cal B})\;\;\;\quad\;\; \equiv \quad \bar C(x) 
+ \bar\theta\;(s_{ad}^{(2)}\; \bar C(x)) , \nonumber\\
&&F^{(AD)}(x, \bar\theta)  =  C(x) + \bar\theta\;(0)\;\;\;\;\;\;\qquad\equiv \quad C(x) + \bar\theta\;(s_{ad}^{(2)}\; C(x)),\nonumber\\
&&\Phi^{(AD)}(x, \bar\theta)  =  \phi(x) + \bar\theta\;(0)\qquad\quad\;\; \;\,\equiv \quad \phi(x) + \bar\theta\;(s_{ad}^{(2)}\; \phi (x)),\nonumber\\
&& \tilde{\Phi}^{(AD)}(x, \bar\theta)  = \; \tilde\phi(x) + \bar\theta\;(m\,C)\quad\;\;\;\;\,\,\equiv \quad \tilde\phi(x) + 
\bar\theta\;(s_{ad}^{(2)}\; \tilde\phi (x)),\nonumber\\
&&{\tilde{\bar B}}^{(AD)}(x, \bar\theta) =  \bar B(x) + \bar\theta\;(0) \qquad\quad\;\;\; \equiv \quad \bar B(x) + \bar\theta\;(s_{ad}^{(2)}\; \bar{B}(x)),\nonumber\\
&&{\tilde{\bar {\cal B}}}^{(AD)}(x, \bar\theta) = \bar {\cal B}(x) + \bar\theta\;(0)\;\qquad\quad\;\;\; \equiv \quad \bar {\cal B}(x) + \bar\theta\;(s_{ad}^{(2)}\; \bar{\cal B}(x)). 
\end{eqnarray}
where the superscript $(AD)$ stands for the super expansions of the {\it chiral} superfields that have been obtained after the
 applications of restrictions from Eq. (102). It should be noted that we have derived {\it all} the anti-co-BRST symmetry transformations $s_{ad} ^{(2)}$ as the coefficients of $\bar \theta$ in the final super expansions (105).

Taking into account the co-BRST invariant quantities from Eq. (102) and generalizing them onto the (2, 1)-dimensional {\it{chiral}} super-submanifold, we demand the following restrictions on these specific combinations of {\it{chiral}} superfields
\begin{eqnarray}
&&\bar F(x, \theta)  =  \bar C(x), \quad {\tilde{\bar{\cal B}}}(x, \theta) = \bar{\cal B}(x), \quad {\tilde {\bar B}}(x, \theta) =  \bar B(x),\quad 
\partial_\mu B^\mu (x, \theta) = \partial _\mu A^\mu (x),\nonumber\\
 &&\Phi (x, \theta) = \phi (x),\quad m\, B_\mu (x, \theta) + \varepsilon_{\mu\nu}\,\partial^\nu \tilde\Phi(x, \theta) = m\,A_\mu(x) + \varepsilon_{\mu\nu}\,\partial^\nu\, \tilde\phi\,(x), \nonumber\\
&& {\tilde{\bar{\cal B}}}(x, \theta)\,
\dot F(x, \theta) - {\dot {\tilde {\bar{\cal B}}}} (x, \theta)  F(x, \theta) = \bar {\cal B}(x)\dot  C(x) - {\dot {\bar{\cal B}}}(x)\, C(x),\nonumber\\
&&\tilde\Phi(x, \theta) \bar F (x, \theta) = \tilde\phi(x)\,\bar C(x),\quad
\varepsilon^{\mu\nu}\,B_\nu(x, \theta)\,\partial_\mu {\tilde{\bar{\cal B}}}(x, \theta)
 - i \, \partial_\mu \bar F(x, \theta)\,\partial^\mu F(x, \theta)
\nonumber\\
&& = \varepsilon^{\mu\nu}\, A_\nu(x)\, \partial_\mu {\bar{\cal B}}(x) - i\,\partial_\mu \bar C(x)\, \partial^\mu C(x).
\end{eqnarray} 
due to the basic tenets of (anti-)chiral superfield approach to BRST formalism where we demand  that {\it all} the co-BRST invariant 
quantities must be independent of the Grassmannian variables $\theta$. As a consequence of the restrictions in (106), we obtain 
the following expressions for the {\it secondary} fields (cf. Eq. (69)) in terms of the {\it basic} and {\it auxiliary} fields of the Lagrangian density ${\cal L}_{(b_2)}$, namely;
 \begin{eqnarray}
 \bar R_\mu = -\,\varepsilon_{\mu\nu}\partial^\nu \bar C, \quad \bar B_1 = - \,\bar {\cal B},\quad 
f_2 = m\, \bar C,\quad  \bar f_1 =  \bar f_3 =  \bar f_4 =  \bar B_2 = 0.
\end{eqnarray} 
The substitutions of these secondary fields into the super expansions (69) lead to the following super expansions for the {\it chiral} superfields
\begin{eqnarray}
&& B_\mu^ {(D)}(x, \theta)  = \; A_\mu + \theta\;(-\, \varepsilon_{\mu\nu}\,\partial^\nu\,\bar C)\;\equiv \quad A_\mu (x)
 + \theta\;(s_{d}^{(2)}\;A_\mu (x)) ,\nonumber\\
&& \bar F^{(D)}(x, \theta) = \; \bar C(x) + \theta\;(0)\;\;\;\qquad\;\; \equiv \quad \bar C(x) 
+ \theta\;(s_{d}^{(2)}\; \bar C(x)) , \nonumber\\
&&F^{(D)}(x, \theta)  =  C(x) + \theta\;(-\,i\,\bar{\cal B})\qquad\equiv \quad C(x) + \theta\;(s_{d}^{(2)}\; C(x)) ,\nonumber\\
&&\Phi^{(D)}(x, \theta)  =  \phi(x) + \theta\;(0)\qquad\quad\;\; \;\,\equiv \quad \phi(x) + \theta\;(s_{d}^{(2)}\; \phi (x)),\nonumber\\
&& \tilde{\Phi}^{(D)}(x, \theta)  = \; \tilde\phi(x) + \theta\;(m\,\bar C)\quad\;\;\;\;\,\,\equiv \quad \tilde\phi(x) + 
\theta\;(s_{d}^{(2)}\; \tilde\phi (x)),\nonumber\\
&&{\tilde {\bar B}}^{(D)}(x, \theta) = \bar B(x) + \theta\;(0) \qquad\quad\;\;\; \equiv \quad \bar B(x) + \theta\;(s_{d}^{(2)}\; \bar{B}(x)),\nonumber\\
&&{\tilde{\bar {\cal B}}}^{(D)}(x, \theta) = \bar{\cal B}(x) + \theta\;(0)\;\qquad\quad\;\;\; \equiv \quad\bar{\cal B}(x) +
 \theta\;(s_{d}^{(2)}\; \bar{\cal B}(x)), 
\end{eqnarray}
where the {\it chiral} superfields (with superscript $(D)$) denote the superfields that have been obtained after the applications of
 the restrictions quoted in (106). A close look at (108) shows that we have {\it{already}} derived  the co-BRST symmetry transformations $s_d^{(2)}$ 
(for ${\cal L}_{(b_2)}$) as the coefficients of $\theta$ in the above {\it{chiral}} super expansions.

At this stage, we can use the super expansions $(105)$ and $(108)$ to express  the (anti-)co-BRST charges $Q_{(a)d}^{(2)}$ connected with the nilpotent transformations $s_{(a)d}^{(2)}$ (for the Lagrangian density ${\cal L}_{(b_2)}$) as: 
\begin{eqnarray}
Q_d^{(2)} &=& \int dx \,\Bigl[\frac{\partial}{\partial{\theta}}\,\Big(i\, \bar F^{(D)}(x, {\theta})\,
\dot F^{(D)}(x, {\theta})
- i\, \dot{\bar F}^{(D)}(x, {\theta})\,F^{(D)}(x, {\theta})\Big)\,\Bigr]\nonumber\\
&\equiv & \int dx \Bigl[\int\, d\,\theta\,\Big(i \bar F^{(D)}(x, {\theta})\dot F^{(D)}(x, {\theta}) 
- i\, \dot{\bar F}^{(D)}(x, {\theta})\,F^{(D)}(x, {\theta})\Big)\,\Bigr],\nonumber\\
Q_{d}^{(2)} &=& \int dx \,\Bigl[\frac{\partial}{\partial{\bar\theta}}\,\Big(i\, 
\dot{\bar F}^{(AD)}(x, \bar{\theta})\,{\bar F}^{(AD)}(x, \bar{\theta})\Big)\,\Bigr] \nonumber \\
&\equiv &  \int dx \Bigl[\int d\,\bar\theta\,\Big(i\, \dot{\bar F}^{(AD)}(x, \bar{\theta})\,{\bar F}^{(AD)}
(x, \bar{\theta})\Big)\,\Bigr],\nonumber\\
Q_{ad}^{(2)} &=& \int dx \Bigl[\frac{\partial}{\partial\bar\theta}\, \Big(i\,\dot{\bar F} ^{(AD)}(x, \bar\theta)\,
 F^{(AD)}(x, \bar\theta) \, - \,i \bar F^{(AD)}(x, \bar\theta)\, 
\dot F^{(AD)}(x, \bar\theta)\Big) \Bigr] \nonumber\\
&\equiv & \int dx \,\Bigl[\int\, d\,\bar\theta\,\Big(i\,\dot{\bar F} ^{(AD)}(x, \bar\theta)\, F^{(AD)}(x, \bar\theta) 
\, - \,i \bar F^{(AD)}(x, \bar\theta)\, 
\dot F^{(AD)}(x, \bar\theta)\Big)\Bigr],\nonumber\\
Q_{ad}^{(2)} &=& -\,i\,\int dx \Bigl[\frac{\partial}{\partial\theta}\,
\Big(\dot F^{(D)}(x, {\theta})\, F^{(D)}(x, {\theta})\Big)\, \Bigr] \nonumber \\
&\equiv &  -\,i\,\int dx \,\Bigl[\int\,d\,\theta\,\Big( \dot F^{(D)}(x, {\theta})\,
F^{(D)}(x, {\theta})\Big)\, \Bigr].
\end{eqnarray}   
It is now very clear that we have the following  very interesting and informative relationships (due to the nilpotency $\partial_{\theta}^2 = \partial_{\bar{\theta}}^2$ properties of $(\partial_{\theta}, \partial_{\bar{\theta}})$), namely;
\begin{eqnarray}
\partial_{\bar\theta}\,Q_d^{(2)} = 0,\qquad \partial_{\theta}\,Q_d^{(2)} = 0, \qquad \partial_{\theta}\,Q_{ad}^{(2)} = 0,
\qquad \partial_{\bar\theta}\,Q_{ad}^{(2)} = 0.
\end{eqnarray}
The above equation {\it{actually}} captures the off-shell nilpotency and absolute anticommutativity of the conserved charges $Q_{(a)d}^{(2)}$. This statement becomes very transparent when we express $(110)$ in the {\it{ordinary}} space (with the mappings: $s_d \leftrightarrow \partial_{\theta}, s_{ad} \leftrightarrow \partial_{\bar{\theta}}$) and exploit  the idea behind the continuous symmetry transformations and their relationships with thier generators (cf. Eq. (6)), namely;
\begin{eqnarray}
&& s_{d}^{(2)}\,Q_{d}^{(2)}\, = -\,i\,\lbrace{Q_{d}^{(2)},\,Q_{d}^{(2)}}\rbrace\, = \,0  \qquad 
 \Longrightarrow\quad [Q_{d}^{(2)}]^2 = 0,\nonumber\\
&& s_{ad}^{(2)}\,Q_{ad}^{(2)}\, = -\,i\,\lbrace{Q_{ad}^{(2)},\,Q_{ad}^{(2)}}\rbrace\, = \,0 \quad\quad 
\Longrightarrow\quad [Q_{ad}^{(2)}]^2 = 0,\nonumber\\
 && s_{ad}^{(2)}\,Q_{d}^{(2)}\, = -\,i\,\lbrace{Q_{d}^{(2)},\,Q_{ad}^{(2)}}\rbrace\, = \,0 
\qquad\;\Longrightarrow\quad Q_{d}^{(2)}\,Q_{ad}^{(2)} + Q_{ad}^{(2)}\,Q_{d}^{(2)} = 0,\nonumber\\
&& s_{d}^{(2)}\,Q_{ad}^{(2)}\, = -\,i\,\lbrace{Q_{ad}^{(2)},\,Q_{d}^{(2)}}\rbrace\, = \,0 
\qquad\;\Longrightarrow\quad Q_{ad}^{(2)}\,Q_{d}^{(2)} + Q_{d}^{(2)}\,Q_{ad}^{(2)} = 0.
\end{eqnarray}
Thus, we have captured   the off-shell nilpotency and absolute anticommutativity of the conserved charges $Q_{(a)d}^{(2)}$ within the framework of (anti-)chiral superfield approach to BRST formalism which are primarily connected with the nilpotency $(\partial_{\theta} ^2 = \partial_{\bar\theta} ^2= 0)$ of
the translational generators $(\partial_{\theta}, \partial_{\bar\theta})$ along the {\it chiral} and {\it anti-chiral} super-submanifolds.

\section{Invariance of the Lagrangian Densities: Chiral and Anti-Chiral Superfield Approach}

In this section, first of all, we capture the (anti-)BRST invariance(s) of the Lagrangian densities 
 ${\cal L}_{(b_1)}$ and ${\cal L}_{(b_2)}$  within the framework of our (anti-)chiral
 superfield approach to BRST formalism by using the super expansions (62), (68), (73), (83) and (86)
 which have been obtained after the (anti-)BRST invariant restrictions on the (anti-)chiral superfields 
(defined on the (2, 1)-dimensional {\it (anti-)chiral} super-submanifolds of the {\it general} (2, 2)-dimensional supermanifold).
We note, in this connection, that the ordinary Lagrangian density  ${\cal L}_{(b_1)}$ 
can be generalized onto the (2, 1)-dimensional (anti-)chiral super-submanifolds (of the {\it general} (2, 2)-dimensional supermanifold) as:    
\begin{eqnarray}
{\cal L}_{(b_1)}\longrightarrow \tilde {\cal L}^{(ac)}_{(b_1)}& = & \, \tilde {\cal B}^{(b)}(x, \bar\theta)\,\big(\tilde E^{( b)}(x, \bar\theta) -
m\,\tilde \Phi^{( b)}(x, \bar\theta)\big) 
- \frac {1}{2}\,  \tilde {\cal B}^{( b)}(x, \bar\theta)\,\tilde {\cal B}^{( b)}(x, \bar\theta)  \nonumber\\
& + &  m \tilde E^{( b)}(x, \bar\theta)\;\tilde \Phi^{( b)}(x, \bar\theta)- \frac {1}{2} \;\partial_\mu {\tilde \Phi^{( b)}}(x, \bar\theta)\,
\partial^\mu \tilde \Phi^{( b)}(x, \bar\theta)\nonumber\\
& + & \frac {m^2}{2} B_\mu ^{( b)}(x, \bar\theta)\, B^{\mu( b)}(x, \bar\theta)  + 
\frac {1}{2} \;\partial_\mu { \Phi^{( b)}}(x, \bar\theta)\,
\partial^\mu  \Phi^{( b)}(x, \bar\theta)\nonumber\\
& - & m\, B_\mu ^{( b)}(x, \bar\theta)\; \partial^\mu  \Phi^{( b)}(x, \bar\theta) +  \tilde B^{( b)}(x, \bar\theta)\big[\partial_\mu B^{\mu( b)}(x, \bar\theta) \nonumber\\
&+& m\,  \Phi^{(b)}(x, \bar\theta)\big] + \frac {1}{2}\; \tilde B^{(b)}(x, \bar\theta) \;\tilde B^{(b)}(x, \bar\theta)\nonumber\\  
& - & \; i\; \partial_{\mu}\bar F^{(b)}(x, \bar\theta)\;\partial^{\mu} F^{(b)}(x, \bar\theta) + i\, m^2\, {\bar F} ^{(b)} ( x, \bar\theta)\, 
F ^{(b)} ( x, \bar\theta),\nonumber\\
{\cal L}_{(b_1)}\longrightarrow \tilde {\cal L}^{(c)}_{(b_1)}& = & \, \tilde {\cal B}^{(ab)}(x, \theta)\,\big(\tilde E^{( ab)}(x, \theta) -
m\,\tilde \Phi^{( ab)}(x, \theta)\big) 
- \frac {1}{2}\,  \tilde {\cal B}^{( ab)}(x, \theta)\;\tilde {\cal B}^{( ab)}(x, \theta) \nonumber\\
& + &  m \tilde E^{( ab)}(x, \theta)\;\tilde \Phi^{( ab)}(x, \theta)- \frac {1}{2} \;\partial_\mu {\tilde \Phi^{( ab)}}(x, \theta)\;
\partial^\mu \tilde \Phi^{( ab)}(x, \theta)\nonumber\\
& + & \frac {m^2}{2} B_\mu ^{( ab)}(x, \theta)\; B^{\mu( ab)}(x, \theta) + 
\frac {1}{2} \;\partial_\mu { \Phi^{( ab)}}(x, \theta)\;
\partial^\mu  \Phi^{(a b)}(x, \theta)\nonumber\\
& - & m\, B_\mu ^{( ab)}(x, \theta) \;\partial^\mu  \Phi^{( ab)}(x, \theta)+  \tilde B^{( ab)}(x, \theta)\big[\partial_\mu B^{\mu( ab)}(x, \theta) \nonumber\\
& + & m\, \Phi^{( ab)}(x, \theta)\big] + \frac {1}{2}\; \tilde B^{(ab)}(x, \theta) \;\tilde B^{(ab)}(x, \theta)\nonumber\\  
& - & \; i\; \partial_{\mu}\bar F^{(ab)}(x, \theta)\;\partial^{\mu} F^{(ab)}(x, \theta)+ i\, m^2\, {\bar F} ^{(ab)} (x, \theta)\, 
F ^{(ab)} (x, \theta),
\end{eqnarray}
where the superscripts $(c)$ and $(ac)$ on the {\it super} Lagrangian densities denote the
 generalizations of the {\it ordinary} Lagrangian densities 
to their {\it chiral} and {\it anti-chiral} counterparts. Furthermore, 
we note that superscripts $(b)$ and $(ab)$ on the (anti-)chiral superfields 
denote that these superfields have been obtained after the applications of the (anti-)BRST 
invariant restrictions (cf. Eqs. (62), (68), (73)). In addition, we have to use the following
\begin{eqnarray}
&&\tilde B^{(b)}(x, \bar\theta) =  \tilde B^{(ab)}(x, \theta)  = B(x),\qquad \tilde E^{(b)}(x, \bar\theta) =  \tilde E^{(ab)}(x, \theta)  = E(x),\nonumber\\
&&F^{(b)}(x, \bar\theta) = C(x),\qquad \tilde \Phi^{(b)}(x, \bar\theta) = \tilde \Phi^{(ab)}(x, \theta) = \tilde \phi(x),\nonumber\\
&&  \tilde {\cal B}^{(b)}(x, \bar\theta)  =  \tilde {\cal B}^{(ab)}(x, \theta)  = {\cal B}(x),\quad \bar F^{(ab)}(x, \theta) = \bar C(x),  
 \end{eqnarray}
 due to the fact that
 we have: $s_b (E, B, {\cal B}, C, \tilde\phi) = 0$ and  $s_{ab} (E, B, {\cal B}, \bar C, \tilde\phi) = 0$.
 In other words, we have some (anti-)chiral superfields with superscripts $(b)$ and $(ab)$ which are {\it actually} ordinary fields. 
 In view of the mappings $s_b^{(1)}\leftrightarrow \partial_{\bar\theta}$
and $s_{ab}^{(1)}\leftrightarrow \partial_{\theta}$, we observe the following (as far as the (super)Lagrangian densities are concerned), namely;
\begin{eqnarray}
 \frac {\partial}{\partial\theta}\Big[\tilde {\cal L}_{(b_1)}^{(c)}\Big] = \partial_{\mu}\big(B\;\partial^{\mu}\bar C\big)\;
\equiv s_{ab}^{(1)}\,{\cal L}_{(b_1)},\qquad
\frac {\partial}{\partial\bar\theta}\Big[\tilde {\cal L}_{(b_1)}^{(ac)}\Big] = \partial_{\mu}\big(B\;\partial^{\mu} C\big)\;\equiv s_{b}^{(1)}\,{\cal L}_{(b_1)},
\end{eqnarray}
which show the (anti-)BRST invariance of the ordinary action integral $S = \int\, d^2 x\, {\cal L}_{(b_1)}$ 
(that can be {\it also} written as {\it super} action
 integrals $\tilde S = \int\,d\,\bar\theta\int\, d^2 x\, \tilde {\cal L}^{(ac)}_{(b_1)}\equiv \int\,d\,\theta\int\, d^2 x\, \tilde {\cal L}^{(c)}_{(b_1)}$).

As we have captured the (anti-)BRST invariance of the Lagrangian density  ${\cal L}_{(b_1)}$, we can also express
the (anti-)BRST invariance of the Lagrangian density ${\cal L}_{(b_2)}$. For this purpose, we have to generalize 
the {\it ordinary} 2D Lagrangian density to its counterparts {\it super} Lagrangian densities on the (anti-)chiral super-submanifolds as:
\begin{eqnarray}
{\cal L}_{(b_2)}\longrightarrow \tilde {\cal L}^{(AC)}_{(b_2)}& = & \, {\tilde {\bar {\cal B}}}^{(B)}(x, \bar\theta)\,\big(\tilde E^{(B)}(x, \bar\theta) +
m\,\tilde \Phi^{(B)}(x, \bar\theta)\big) 
- \frac {1}{2}\,  {\tilde {\bar {\cal B}}}^{(B)}(x, \bar\theta)\,{\tilde {\bar {\cal B}}}^{(B)}(x, \bar\theta)  \nonumber\\
& - &  m \,\tilde E^{(B)}(x, \bar\theta)\;\tilde \Phi^{(B)}(x, \bar\theta)- \frac {1}{2} \;\partial_\mu {\tilde \Phi^{(B)}}(x, \bar\theta)\;
\partial^\mu \tilde \Phi^{(B)}(x, \bar\theta)\nonumber\\
& + & \frac {m^2}{2} B_\mu ^{(B)}(x, \bar\theta)\, B^{\mu(B)}(x, \bar\theta)  + 
\frac {1}{2} \;\partial_\mu { \Phi^{( B)}}(x, \bar\theta)\;
\partial^\mu  \Phi^{( B)}(x, \bar\theta)\nonumber\\
& + & m\, B_\mu ^{( B)}(x, \bar\theta)\, \partial^\mu  \Phi^{( B)}(x, \bar\theta) +  \tilde {\bar B}^{( B)}(x, \bar\theta)\big[\partial_\mu B^{\mu( B)}(x, \bar\theta) \nonumber\\
&-& m\,  \Phi^{( B)}(x, \bar\theta)\big] + \frac {1}{2}\;  \tilde {\bar B}^{(B)}(x, \bar\theta) \; \tilde {\bar B}^{(B)}(x, \bar\theta)\nonumber\\  
& - & \; i\; \partial_{\mu}\bar F^{(B)}(x, \bar\theta)\;\partial^{\mu} F^{(B)}(x, \bar\theta) + i\, m^2\, {\bar F} ^{(B)} ( x, \bar\theta)\, 
F ^{(B)} ( x, \bar\theta),\nonumber\\
{\cal L}_{(b_2)}\longrightarrow \tilde {\cal L}^{(C)}_{(b_2)}& = & \, {\tilde {\bar {\cal B}}}^{(AB)}(x, \theta)\,\big(\tilde E^{( AB)}(x, \theta) +
m\,\tilde \Phi^{( AB)}(x, \theta)\big) 
- \frac {1}{2}\,  {\tilde {\bar {\cal B}}}^{( AB)}(x, \theta)\;{\tilde {\bar {\cal B}}}^{( AB)}(x, \theta) \nonumber\\
& - &  m \,\tilde E^{( AB)}(x, \theta)\;\tilde \Phi^{( AB)}(x, \theta)- \frac {1}{2} \;\partial_\mu {\tilde \Phi^{( AB)}}(x, \theta)\;
\partial^\mu \tilde \Phi^{( AB)}(x, \theta)\nonumber\\
& + & \frac {m^2}{2} B_\mu ^{( AB)}(x, \theta)\, B^{\mu( AB)}(x, \theta) + 
\frac {1}{2} \;\partial_\mu { \Phi^{( AB)}}(x, \theta)\;
\partial^\mu  \Phi^{(AB)}(x, \theta)\nonumber\\
& + & m\, B_\mu ^{( AB)}(x, \theta) \;\partial^\mu  \Phi^{( AB)}(x, \theta)+  \tilde {\bar B}^{( AB)}(x, \theta)\big[\partial_\mu B^{\mu( AB)}(x, \theta) \nonumber\\
&-& m\,  \Phi^{( AB)}(x, \theta)\big] + \frac {1}{2}\; \tilde {\bar B}^{(AB)}(x, \theta) \;\tilde {\bar B}^{(AB)}(x, \theta)\nonumber\\  
& - & \; i\; \partial_{\mu}\bar F^{(AB)}(x, \theta)\;\partial^{\mu} F^{(AB)}(x, \theta)+ i\, m^2\, {\bar F} ^{(AB)} (x, \theta)\, 
F ^{(AB)} (x, \theta),
\end{eqnarray}
where the superscripts $(B)$ and $(AB)$, on the superfields, denote the (anti-)chiral superfields that have been 
obtained after the applications of the (anti-)BRST invariant restrictions for the Lagrangian density ${\cal L}_{(b_2)}$
and, on the l.h.s., the superscripts $(C)$ and $(AC)$ on the super Lagrangian densities stand for the {\it chiral} and {\it anti-chiral}
versions of the ordinary Lagrangian density ${\cal L}_{(b_2)}$ which contain {\it chiral} and {\it anti-chiral} superfields. It should be noted that 
some of the (anti-)chiral superfields, with superscripts $(B)$ and $(AB)$, are {\it actually} ordinary 2D fields. For instance,
we observe that the following are true, namely; 
\begin{eqnarray}
&&{\tilde  {\bar B}}^{(B)}(x, \bar\theta) =  {\tilde {\bar B}}^{(AB)}(x, \theta)  = \bar B(x),\qquad \tilde E^{(B)}(x, \bar\theta) =  
\tilde E^{(AB)}(x, \theta)  = E(x),\nonumber\\
&&F^{(B)}(x, \bar\theta) = C(x),\quad \tilde \Phi^{(B)}(x, \bar\theta) = \tilde \Phi^{(AB)}(x, \theta)
 = \tilde \phi(x),\nonumber\\
&&  {\tilde {\bar {\cal B}}}^{(B)}(x, \bar\theta)  =  {\tilde {\bar {\cal B}}}^{(AB)}(x, \theta)  = {\bar {\cal B}}(x),\quad \bar F^{(AB)}(x, \theta) = \bar C(x). 
 \end{eqnarray}
 In view of the mappings $s_b^{(2)}\leftrightarrow \partial_{\bar\theta}$
and $s_{ab}^{(2)}\leftrightarrow \partial_{\theta}$, it is elementary to check that the following are true in the context of 
(super)Lagrangian densities, namely;
\begin{eqnarray}
 \frac {\partial}{\partial\theta}\Big[\tilde {\cal L}_{(b_2)}^{(C)}\Big] = \partial_{\mu}\big(\bar B\;\partial^{\mu}\bar C\big)\;
\equiv s_{ab}^{(2)}\,{\cal L}_{(b_2)},\qquad
\frac {\partial}{\partial\bar\theta}\Big[\tilde {\cal L}_{(b_2)}^{(AC)}\Big] = \partial_{\mu}\big(\bar B\;\partial^{\mu} C\big)\;\equiv s_{b}^{(2)}\,{\cal L}_{(b_2)},
\end{eqnarray}
Hence, we have captured the (anti-)BRST invariance of the Lagrangian density ${\cal L}_{(b_2)}$ because the corresponding  action integral(s)
\begin{eqnarray}
S = \int d^2x \,{\cal L}_{(b_2)} \Longleftrightarrow \int \,d\theta\,\int\,d^2x\,\tilde{\cal L}_{(b_2)}^{(C)} \equiv \int\,d\,\bar{\theta}\int\,d^2x\;\tilde {\cal L}_{(b_2)}^{(AC)},
\end{eqnarray}
vanish due to Gauss's divergence theorem where {\it all} the physical fields vanish off at $x = \pm \infty $.

We now explain the existence of the  restriction (i.e. $ B+\bar B+ 2\,(\partial\cdot A) = 0$) 
within the framework of the (anti-)chiral superfield approach. Towards this goal in mind, first of all, 
we derive the non-trivial (anti-)BRST symmetry transformations (30)
 for the auxiliary fields $\bar B(x), B(x)$ (i.e. $s_b^{(1)}\bar{B} = 
-\,2\,\Box\;{C} \equiv s_b^{(2)}\,B$ and $s_{(ab)}^{(1)}\bar{B} = -\,2\,\Box\,{\bar{C}} 
\equiv s_{(ab)}^{(2)}\,\ B $). In this context, it can be seen that 
$s_{(a)b}^{(1)}(B+\bar{B}+ 2\,(\partial\cdot A)) = 0$ and $s_{(a)b}^{(2)}(B+\bar{B}+2\;(\partial\cdot A)) = 0$ 
imply the following restrictions on the (anti-)chiral superfields:
\begin{eqnarray}  
&&\tilde B^{(b)}(x, \bar{\theta}) + {\tilde {\bar B}}(x, \bar{\theta}) + 2\;\partial^{\mu}B_{\mu}^{(b)}(x, \bar{\theta}) = B(x) + \bar{B}(x) + 2\,(\partial\cdot A)(x),\nonumber\\
&&\tilde B^{(ab)}(x, {\theta}) + {\tilde {\bar B}}(x, {\theta}) + 2\;\partial^{\mu}B_{\mu}^{(ab)}(x, {\theta}) = B(x) + \bar{B}(x) + 2\;(\partial\cdot A)(x),\nonumber\\ 
&&\tilde B(x, \bar{\theta}) + {\tilde {\bar B}}^{(B)}(x, \bar{\theta}) + 2\;\partial^{\mu}B_{\mu}^{(B)}(x, \bar{\theta}) = B(x) + \bar{B}(x) + 2\;(\partial\cdot A)(x),\nonumber\\
&&\tilde B(x, {\theta}) + {\tilde {\bar B}}^{(AB)}(x, {\theta}) + 2\;\partial^{\mu}B_{\mu}^{(AB)}(x, {\theta}) = B(x) + \bar{B}(x) + 2\;(\partial\cdot A)(x).
\end{eqnarray}
In the above, the auxiliary fields $B(x)$ and $\bar B(x)$ have been generalized to the (anti-)chiral super-submanifolds with the following super expansions:
\begin{eqnarray}
&&{B}(x) \longrightarrow \tilde{{B}}(x, \theta) = {B}(x) + \theta\,\bar {f}_3(x),\nonumber\\
&&{B}(x) \longrightarrow \tilde{{B}}(x, \bar{\theta}) = {B}(x) + \bar{\theta}\,f_3(x),\nonumber\\
&&\bar{B}(x) \longrightarrow \tilde{\bar{B}}(x, \theta) = \bar{B}(x) + \theta\,\bar {f}_5(x),\nonumber\\
&&\bar{B}(x) \longrightarrow \tilde{\bar{B}}(x, \bar{\theta}) = \bar{B}(x) + \bar{\theta}\,f_5(x).
\end{eqnarray}
We also note that the super expansions for the superfields $\tilde B (x, {\bar\theta}) $ and $\tilde B (x, {\theta})$
have been {\it also} given in (60) and (69), respectively.
In equation (119), it is very clear that we have the trivial equalities: $\tilde B^{(b)}(x, \bar{\theta}) = \tilde B^{(ab)}(x, \theta) = B(x)$ due to $s_{(a)b}^{(1)}B(x) = 0$ as well as ${\tilde{\bar B}}^{(B)}(x, \bar{\theta}) = {\tilde{\bar B}}^{(AB)}(x, {\theta}) = \bar B(x)$ due to our observations: $s_{(a)b}^{(2)}\bar B(x) = 0$. Plugging in these values {\it and} (120) into (119) yields the following expressions for the secondary fields:
\begin{eqnarray}
f_3 (x) \equiv f_5(x) = -2\,\Box{C},\quad\quad \bar f_3 (x) \equiv\bar{f}_5(x) = -2\,\Box{\bar{C}}.
\end{eqnarray}
It is straightforward  to note that the secondary fields $(f_5 (x),  \bar f_5 (x))$ in (120) are fermionic in nature because of the fermionic nature of $(\theta, \bar\theta)$.
Thus, we have obtained $s_b^{(1)}\bar{B}(x) = -2\,\Box{C} \equiv s_b^{(2)}B(x)$ and $s_{ab}^{(1)}{\bar B}(x) = -2\,\Box{\bar{C}} = s_{ab}^{(2)}B(x)$. In other words, we have the following (anti-)chiral super expansions for the (anti-)chiral superfields: 
\begin{eqnarray} 
&&\tilde B^{(AB)}(x, \theta) \,\,= B(x) + \theta\,(-\,2\,\Box\,{\bar{C}}) \,\equiv B(x) + \theta\,(s_{ab}^{(2)}B(x)), \nonumber \\
&&\tilde B^{(B)}(x, \bar\theta) = B(x) + \bar\theta\,(-\,2\,\Box\,{{C}}) \equiv B(x) + \bar\theta\,(s_{b}^{(2)}B(x)).
\end{eqnarray}
We note that the coefficients of $\theta$ and $\bar\theta$, in the above super expansions, are nothing but the (anti-)BRST symmetry transformations $s_{(a)b}^{(2)}$. We further point out that the following emerges from the restrictions (119), namely;
\begin{eqnarray} 
&&\tilde{\bar{B}}^{(ab)}(x, {\theta}) = \bar{B}(x) + {\theta}\,(-\,2\,\Box\,{{\bar C}}) \equiv \bar{B}(x) + {\theta}\,(s_{(ab)}^{(1)}\bar{B}(x)),\nonumber \\
&&\tilde{\bar{B}}^{(b)}(x, \bar{\theta}) \,\,= \bar{B}(x) + \bar{\theta}\,(-\,2\,\Box\,{{C}}) \,\equiv \bar{B}(x) + \bar{\theta}\,(s_{b}^{(1)}\bar{B}(x)).
\end{eqnarray}
which show that we have already derived the (anti-)BRST symmetry transformations $s_{ab}^{(1)}\bar{B}(x) = -2\,\Box\,{\bar{C}}$ and $s_b^{(1)}\bar{B}(x) = -2\,\Box\,{C}$ as the coefficients of $\theta$ and $\bar\theta$.

Taking into account the inputs from (122) and (123), we can generalize the ordinary Lagrangian densities ${\cal L}_{(b_1)}$  in the following forms:
\begin{eqnarray}
{\cal L}_{(b_1)} \longrightarrow \tilde{{\cal L}}_{(b_1)}^{(AC)} & = &\tilde{\cal B}^{(B)}(x, \bar{\theta})\,\big(\tilde E^{( B)}(x, \bar\theta) -
m\,\tilde \Phi^{( B)}(x, \bar\theta)\big) 
- \frac {1}{2}\,  \tilde {\cal B}^{( B)}(x, \bar\theta)\,\tilde {\cal B}^{( B)}(x, \bar\theta)  \nonumber\\
& + &  m \tilde E^{( B)}(x, \bar\theta)\;\tilde \Phi^{( B)}(x, \bar\theta)- \frac {1}{2} \;\partial_\mu {\tilde \Phi^{( B)}}(x, \bar\theta)\;
\partial^\mu \tilde \Phi^{(B)}(x, \bar\theta)\nonumber\\
& + & \frac {m^2}{2} B_\mu ^{(B)}(x, \bar\theta)\, B^{\mu(B)}(x, \bar\theta)  + 
\frac {1}{2} \;\partial_\mu { \Phi^{(B)}}(x, \bar\theta)\;
\partial^\mu  \Phi^{(B)}(x, \bar\theta)\nonumber\\
& - & m\, B_\mu ^{(B)}(x, \bar\theta)\, \partial^\mu  \Phi^{(B)}(x, \bar\theta) +  \tilde B^{(B)}(x, \bar\theta)\big[\partial_\mu B^{\mu(B)}(x, \bar\theta) \nonumber\\
&+& m\,  \Phi^{(B)}(x, \bar\theta)\big] + \frac {1}{2}\; \tilde B^{(B)}(x, \bar\theta) \;\tilde B^{(B)}(x, \bar\theta)\nonumber\\  
& - & \; i\; \partial_{\mu}\bar F^{(B)}(x, \bar\theta)\;\partial^{\mu} F^{(B)}(x, \bar\theta) + i\, m^2\, {\bar F} ^{(B)} ( x, \bar\theta)\, 
F ^{(B)} ( x, \bar\theta),\nonumber\\
{\cal L}_{(b_1)} \longrightarrow \tilde{{\cal L}}_{(b_1)}^{(C)} \,\,\, &=& \tilde{\cal B}^{(AB)}(x, \theta)\,\big(\tilde E^{(AB)}(x, \theta) -
m\,\tilde \Phi^{(AB)}(x, \theta)\big) 
- \frac {1}{2}\,  \tilde {\cal B}^{(AB)}(x, \theta)\;\tilde {\cal B}^{(AB)}(x, \theta) \nonumber\\
& + &  m \tilde E^{(AB)}(x, \theta)\;\tilde \Phi^{(AB)}(x, \theta)- \frac {1}{2} \;\partial_\mu {\tilde \Phi^{(AB)}}(x, \theta)\;
\partial^\mu \tilde \Phi^{(AB)}(x, \theta)\nonumber\\
& + & \frac {m^2}{2} B_\mu ^{(AB)}(x, \theta)\, B^{\mu(AB)}(x, \theta) + 
\frac {1}{2} \;\partial_\mu { \Phi^{(AB)}}(x, \theta)\;
\partial^\mu  \Phi^{(AB)}(x, \theta)\nonumber\\
& - & m\, B_\mu ^{(AB)}(x, \theta)\; \partial^\mu  \Phi^{(AB)}(x, \theta)+  \tilde B^{(AB)}(x, \theta)\big[\partial_\mu B^{\mu(AB)}(x, \theta) \nonumber\\
&+& m\, \Phi^{(AB)}(x, \theta)\big] + \frac {1}{2}\; \tilde B^{(AB)}(x, \theta) \;\tilde B^{(AB)}(x, \theta)\nonumber\\  
& - & \; i\; \partial_{\mu}\bar F^{(AB)}(x, \theta)\;\partial^{\mu} F^{(AB)}(x, \theta)\nonumber\\
 &+& i\, m^2\, {\bar F} ^{(AB)} (x, \theta)\, 
F ^{(AB)} (x, \theta),
\end{eqnarray}
where the (anti-)chiral superfields with superscripts $(B)$ and $(AB)$ have already been explained earlier. It should be noted that we {\it also} have the following
\begin{eqnarray}
&&{\tilde{{\bar B}}}^{(B)}(x, {\bar\theta}) = {\tilde{{\bar  B}}}^{(AB)}(x, {\theta}) = {\bar B}(x),\quad {\tilde{{\cal B}}}^{(B)}(x, {\bar\theta}) = {\tilde{{\cal B}}}^{(AB)}(x, {\theta}) = {\cal B}(x),\nonumber\\
&& {\tilde{\bar{\cal B}}}^{(B)}(x, {\bar\theta}) = {\tilde{\bar {\cal B}}}^{(AB)}(x, {\theta}) = \bar{\cal B}(x),\quad
 \tilde{E}^{(B)}(x, \bar{\theta}) = \tilde{E}^{(AB)}(x, \theta) = E(x), \nonumber\\
&&\tilde{\Phi}^{(B)}(x, \bar{\theta}) = \tilde{\Phi}^{(AB)}(x, {\theta}) = \tilde{\phi}(x),\quad F^{(B)}(x, \bar{\theta}) = C(x),\,\,\,\, \bar{F}^{(AB)}(x, {\theta}) = \bar{C}(x),  
\end{eqnarray} 
which show that there are (anti-)chiral superfields in the super Lagrangian density (124) that are, in reality, the {\it{ordinary}} fields (defined on the 2D Minkowskian spacetime manifold). In exactly similar fashion, we can generalize the ordinary Lagrangian density ${\cal L}_{(b_2)}$ onto the (2, 1)-dimensional (anti-)chiral super-submanifolds (of the {\it{general}} (2, 2)-dimensional supermanifold) as \footnote{It is an elementary exercise to note that ${\tilde{{B}}}^{(b)}(x, {\bar\theta}) = {\tilde{{ B}}}^{(ab)}(x, {\theta}) = {B}(x),\quad {\tilde{{\cal B}}}^{(b)}(x, {\bar\theta}) = {\tilde{ {\cal B}}}^{(ab)}(x, {\theta}) = {\cal B}(x),\quad \tilde{E}^{(b)}(x, \bar{\theta}) = \tilde{E}^{(ab)}(x, \theta) = E(x),\; 
\tilde{\Phi}^{(b)}(x, \bar{\theta}) = \tilde{\Phi}^{(ab)}(x, {\theta}) = \tilde{\phi}(x),\quad F^{(b)}(x, \bar{\theta}) = C(x),\,\,\,\, \bar{F}^{(ab)}(x, {\theta}) = \bar{C}(x)$, etc. Thus, there are a set of {\it ordinary} fields in (126), too.}
\begin{eqnarray}
{\cal L}_{(b_2)} \longrightarrow {\tilde{\cal L}}_{(b_2)}^{(ac)} & = & {\tilde{\bar{\cal B}}}^{(b)}(x, \bar{\theta})\,\big(\tilde E^{(b)}(x, \bar\theta) +
m\,\tilde \Phi^{(b)}(x, \bar\theta)\big) 
- \frac {1}{2}\,  {\tilde {\bar {\cal B}}}^{(b)}(x, \bar\theta)\,{\tilde {\bar {\cal B}}}^{(b)}(x, \bar\theta)  \nonumber\\
& - &  m \,\tilde E^{(b)}(x, \bar\theta)\;\tilde \Phi^{(b)}(x, \bar\theta)- \frac {1}{2} \;\partial_\mu {\tilde \Phi^{(b)}}(x, \bar\theta)\;
\partial^\mu \tilde \Phi^{(b)}(x, \bar\theta)\nonumber\\
& + & \frac {m^2}{2} B_\mu ^{(b)}(x, \bar\theta)\, B^{\mu(b)}(x, \bar\theta)  + 
\frac {1}{2} \;\partial_\mu { \Phi^{(b)}}(x, \bar\theta)\;
\partial^\mu  \Phi^{(b)}(x, \bar\theta)\nonumber\\
& + & m\, B_\mu ^{(b)}(x, \bar\theta)\, \partial^\mu  \Phi^{(b)}(x, \bar\theta) + 
 \tilde {\bar B}^{(b)}(x, \bar\theta)\big[\partial_\mu B^{\mu (b)}(x, \bar\theta) \nonumber\\
&-& m\, \Phi^{(b)}(x, \bar\theta)\big] + \frac {1}{2}\;  \tilde {\bar B}^{(b)}(x, \bar\theta) \; \tilde {\bar B}^{(b)}(x, \bar\theta)\nonumber\\  
& - & \; i\; \partial_{\mu}\bar F^{(b)}(x, \bar\theta)\;\partial^{\mu} F^{(b)}(x, \bar\theta) + i\, m^2\, {\bar F} ^{(b)} ( x, \bar\theta)\; 
F ^{(b)} ( x, \bar\theta),\nonumber\\
{\cal L}_{(b_2)} \longrightarrow {\tilde {\cal L}}_{(b_2)}^{(c)} \,\,\, & = & {\tilde{\bar{\cal B}}}^{(ab)}(x, \theta)\,\big(\tilde E^{(ab)}(x, \theta) +
m\, \tilde\Phi^{(ab)}(x, \theta)\big) 
- \frac {1}{2}\,  {\tilde {\bar {\cal B}}}^{(ab)}(x, \theta)\;{\tilde {\bar {\cal B}}}^{(ab)}(x, \theta) \nonumber\\
& - &  m \,\tilde E^{(ab)}(x, \theta)\;\tilde \Phi^{(ab)}(x, \theta)- \frac {1}{2} \;\partial_\mu {\tilde \Phi^{(ab)}}(x, \theta)\;
\partial^\mu \tilde \Phi^{(ab)}(x, \theta)\nonumber\\
& + & \frac {m^2}{2} B_\mu ^{(ab)}(x, \theta)\, B^{\mu(ab)}(x, \theta) + 
\frac {1}{2} \;\partial_\mu { \Phi^{(ab)}}(x, \theta)\;
\partial^\mu  \Phi^{(ab)}(x, \theta)\nonumber\\
& + & m\, B_\mu ^{(ab)}(x, \theta)\; \partial^\mu  \Phi^{(ab)}(x, \theta)+  \tilde {\bar B}^{(ab)}(x, \theta)\big[\partial_\mu B^{\mu (ab)}(x, \theta) \nonumber\\
&-& m\,  \Phi^{(ab)}(x, \theta)\big] + \frac {1}{2}\; \tilde {\bar B}^{(ab)}(x, \theta) \;\tilde {\bar B}^{(ab)}(x, \theta)\nonumber\\  
& - & \; i\; \partial_{\mu}\bar F^{(ab)}(x, \theta)\;\partial^{\mu} F^{(ab)}(x, \theta)+ i\, m^2\, {\bar F} ^{(ab)} (x, \theta)\, 
F ^{(ab)} (x, \theta),
\end{eqnarray}
where all the superscripts $(b)$ and $(ab)$ (on the r.h.s.) have been explained earlier and the superscripts $(C)$ and $(AC)$ on the Lagrangian densities (on the l.h.s.) denote the generalizations of the {\it ordinary } Lagrangian densities to the corresponding super Lagrangian densities so that we can study the variation of Lagrangian density ${\cal L}_{(b_1)}$ 
w.r.t. $s_{(a)b}^{(2)}$ {\it {and}} Lagrangian density ${\cal L}_{(b_2)}$ with respect to $s_{(a)b}^{(1)}$. Keeping in our mind the mappings: $s_b^{(1, 2)} \leftrightarrow \partial_{\bar{\theta}}$ and $s_{ab}^{(1, 2)} \leftrightarrow \partial_{\theta}$, we note the following
\begin{eqnarray*} 
\frac{\partial}{\partial\theta}{\tilde{\cal L}}^{(C)}_{(b_1)}   & = & 
  \partial_\mu  \Bigl [ \bar B \,\partial^\mu \, \bar C 
+  2 \,m^2\, A^\mu\, \bar  C - 2\, m \,\phi \,\partial^\mu \, \bar C \, \Bigr ]\nonumber\\ 
& - & \,\Bigl [ B + \bar B + 2 \,\,(\partial \cdot A) \Bigr ]\, ( \Box + m^2)\bar C \quad\equiv \; s_{ab}^{(2)}{\cal L}_{(b_1)}, \\
\frac{\partial}{\partial\bar\theta}{\tilde{\cal L}}^{(AC)}_{(b_1)}   & = & 
\partial_\mu  \Bigl [ \bar B \,\partial^\mu \, C 
+ 2 \,m^2\, A^\mu\,  C - 2\, m \,\phi \,\partial^\mu \, C \, \Bigr ]\nonumber\\
& - & \,\Bigl [ B + \bar B + 2 \,\,(\partial \cdot A) \Bigr ]\, ( \Box + m^2)C\quad \equiv \;  s_{b}^{(2)}{\cal L}_{(b_1)},\nonumber\\
\end{eqnarray*}
\begin{eqnarray} 
\frac{\partial}{\partial\theta}{\tilde{\cal L}}^{(c)}_{(b_2)}   & = &  
 \partial_\mu \, \Bigl [ B \,\partial^\mu \,\bar C 
+ 2\, m^2\, A^\mu \,\bar C + 2 \,m \,\phi \,\partial^\mu \,\bar C \, \Bigr ] \nonumber\\
&-& \,\Bigl [ B + \bar B + 2 \,\,(\partial \cdot A) \Bigr ] \,( \Box + m^2)\, \bar C\quad\equiv \; s_{ab}^{(1)}{\cal L}_{(b_2)}, \nonumber\\
\frac{\partial}{\partial\bar\theta}{\tilde{\cal L}}^{(ac)}_{(b_2)}   & = & 
 \partial_\mu \, \Bigl [ B \,\partial^\mu \,C 
+ 2\, m^2\, A^\mu \,C + 2 \,m \,\phi \,\partial^\mu \,C \, \Bigr ] \nonumber\\
&-& \,\Bigl [ B + \bar B + 2 \,\,(\partial \cdot A) \Bigr ] \,( \Box + m^2)\, C\quad\equiv \;  s_{b}^{(1)}{\cal L}_{(b_2)},
\end{eqnarray}
which show that we have captured the existence of restrictions within the framework of (anti-)chiral superfield 
approach to BRST formalism because we can have the symmetry invariance on the r.h.s. of (127) {\it iff} $B + \bar B + 2\,(\partial\cdot A) = 0$
(provided we do {\it not} impose the mass-shell conditions: $(\Box + m^2)\,C = (\Box + m^2)\,\bar C = 0$ from {\it outside}).

We now concentrate on encapsulating the (anti-)co-BRST invariance of the Lagrangian densities ${\cal L}_{(b_1)}$ and ${\cal L}_{(b_2)}$
within the framework (anti-)chiral superfield approach to BRST formalism. Towards this goal in mind, we generalize the {\it ordinary} 
Lagrangian densities onto (2, 1)-dimensional (anti-)chiral super-submanifolds as \footnote{We point out that $s_d ^{(1)}[\bar C, (\partial\cdot A), {\cal B},
\phi] = 0$ implies that we have: $\bar F ^{(d)}(x, \theta) = \bar C (x),\; \partial^\mu  B_{\mu}^{(d)}x, \theta) =  (\partial\cdot A)(x),\; 
\tilde {\cal B}^{(d)}(x, \theta) = {\cal B} (x),\; {\Phi}^{(d)}(x, \theta) = {\Phi} (x). $ Similarly, we observe that $s_{ad} ^{(1)}( C, \,(\partial\cdot A),\, {\cal B},
\phi ) = 0$  implies that we have: $ F ^{(ad)}(x, \bar\theta) =  C (x),\; \partial^\mu B_{\mu}^{(ad)}x, \bar\theta) =  (\partial\cdot A)(x),\; 
\tilde {\cal B}^{(ad)}(x, \theta) = {\cal B} (x),\; {\Phi}^{(ad)}(x, \theta) = {\Phi} (x)$, etc.} 
\begin{eqnarray}
{\cal L}_{(b_1)}\longrightarrow \tilde {\cal L}^{(ac,  ad)}_{(b_1)}& = & \, \tilde {\cal B}^{(ad)}(x, \bar\theta)\,\big(\tilde E^{(ad)}(x, \bar\theta) -
m\,\tilde \Phi^{(ad)}(x, \bar\theta)\big) 
- \frac {1}{2}\,  \tilde {\cal B}^{(ad)}(x, \bar\theta)\,\tilde {\cal B}^{(ad)}(x, \bar\theta)  \nonumber\\
& + &  m \tilde E^{(ad)}(x, \bar\theta)\;\tilde \Phi^{(ad)}(x, \bar\theta)- \frac {1}{2} \;\partial_\mu {\tilde \Phi^{(ad)}}(x, \bar\theta)\;
\partial^\mu \tilde \Phi^{(ad)}(x, \bar\theta)\nonumber\\
& + & \frac {m^2}{2} B_\mu ^{(ad)}(x, \bar\theta)\, B^{\mu (ad)}(x, \bar\theta)  + 
\frac {1}{2} \;\partial_\mu { \Phi^{(ad)}}(x, \bar\theta)\;
\partial^\mu  \Phi^{(ad)}(x, \bar\theta)\nonumber\\
& - & m\, B_\mu ^{(ad)}(x, \bar\theta)\, \partial^\mu  \Phi^{(ad)}(x, \bar\theta) +  \tilde B^{(ad)}(x, \bar\theta)\big[\partial_\mu B^{\mu(ad)}(x, \bar\theta) \nonumber\\
&+& m\,  \Phi^{(ad)}(x, \bar\theta)\big] + \frac {1}{2}\; \tilde B^{(ad)}(x, \bar\theta) \;\tilde B^{(ad)}(x, \bar\theta)\nonumber\\  
& - & \; i\; \partial_{\mu}\bar F^{(ad)}(x, \bar\theta)\;\partial^{\mu} F^{(ad)}(x, \bar\theta) + i\, m^2\, {\bar F} ^{(ad)} ( x, \bar\theta)\, 
F ^{(ad)} ( x, \bar\theta),\nonumber\\
{\cal L}_{(b_1)}\longrightarrow \tilde {\cal L}^{(c, d)}_{(b_1)}& = & \, \tilde {\cal B}^{(d)}(x, \theta)\,\big(\tilde E^{(d)}(x, \theta) -
m\,\tilde \Phi^{(d)}(x, \theta)\big) 
- \frac {1}{2}\,  \tilde {\cal B}^{(d)}(x, \theta)\;\tilde {\cal B}^{(d)}(x, \theta) \nonumber\\
& + &  m \tilde E^{(d)}(x, \theta)\;\tilde \Phi^{(d)}(x, \theta)- \frac {1}{2} \;\partial_\mu {\tilde \Phi^{(d)}}(x, \theta)\;
\partial^\mu \tilde \Phi^{(d)}(x, \theta)\nonumber\\
& + & \frac {m^2}{2} B_\mu ^{(d)}(x, \theta)\, B^{\mu (d)}(x, \theta) + 
\frac {1}{2} \;\partial_\mu { \Phi^{(d)}}(x, \theta)\;
\partial^\mu  \Phi^{(d)}(x, \theta)\nonumber\\
& - & m\, B_\mu ^{(d)}(x, \theta)\; \partial^\mu  \Phi^{(d)}(x, \theta)+  \tilde B^{(d)}(x, \theta)\big[\partial_\mu B^{\mu (d)}(x, \theta) \nonumber\\
& + & m\, \Phi^{(d)}(x, \theta)\big] + \frac {1}{2}\; \tilde B^{(d)}(x, \theta) \;\tilde B^{(d)}(x, \theta)\nonumber\\  
& - & \; i\; \partial_{\mu}\bar F^{(d)}(x, \theta)\;\partial^{\mu} F^{(d)}(x, \theta)+ i\, m^2\, {\bar F} ^{(d)} (x, \theta)\, 
F ^{(d)} (x, \theta),
\end{eqnarray}
where the superscript  $(ac,  ad)$ on the super Lagrangian density denotes that the superfields, contained in {\it it}, are {\it anti-chiral}
which have been obtained after the anti-co-BRST invariant restrictions (cf. Eqs. (93), (95)). In exactly similar fashion, we note that 
the super Lagrangian density, with superscript $(c, d)$, incorporates {\it chiral} superfields that have been obtained after the 
applications of the co-BRST invariant restrictions (cf. Eq. (98)). It is straightforward now to check that:
\begin{eqnarray}
 \frac {\partial}{\partial\theta}\Big[\tilde {\cal L}_{(b_1)}^{(c, d)} \Big] = \partial_\mu \,\Bigl [ \,{\cal B}\, \partial^\mu \,\bar C 
+ m \, \varepsilon^{\mu\nu}\, \bigl (m\, A_\nu \,\bar C 
+ \phi \,\partial_\nu \,\bar C \bigr )
+ m \, \tilde \phi\, \partial^\mu\, \bar C \Bigr ]\;
\equiv s_{d}^{(1)}\,{\cal L}_{(b_1)},\nonumber\\
\frac {\partial}{\partial\bar\theta}\Big[\tilde {\cal L}_{(b_1)}^{(ac, ad)}\Big] = \partial_\mu \,\Bigl [ \,{\cal B} \,\partial^\mu \,C 
+ m \, \varepsilon^{\mu\nu}\, \bigl (m \,A_\nu \,C + \phi\, \partial_\nu \,C \bigr )
+ m \,\tilde \phi \,\partial^\mu \,C \Bigr ]\;\equiv s_{ad}^{(1)}\,{\cal L}_{(b_1)}.
\end{eqnarray}
The above equation captures the (anti-)co-BRST invariance of the action integral $S= \int d^2 x\, {\cal L}_{(b_1)}$ as it matches precisely 
with our earlier observation in   equation (36).

We would like to capture {\it now} the (anti-)co-BRST invariance of the Lagrangian density ${\cal L}_{(b_2)}$ within the framework 
of (anti-)chiral superfield approach to BRST formalism. In this connection, first of all, we generalize the {\it ordinary } Lagrangian
density ${\cal L}_{(b_2)}$ onto the suitably chosen (2, 1)-dimensional (anti-)chiral super-submanifolds (of the {\it general} (2, 2)-dimensional supermanifold) as follows:
\begin{eqnarray}
{\cal L}_{(b_2)}\longrightarrow \tilde {\cal L}^{(AC,  AD)}_{(b_2)}& = & \, {\tilde {\bar {\cal B}}}^{(AD)}(x, \bar\theta)\,\big(\tilde E^{(AD)}(x, \bar\theta) +
m\,\tilde \Phi^{(AD)}(x, \bar\theta)\big) 
- \frac {1}{2}\,  {\tilde {\bar {\cal B}}}^{(AD)}(x, \bar\theta)\,{\tilde {\bar {\cal B}}}^{(AD)}(x, \bar\theta)  \nonumber\\
& - &  m \,\tilde E^{(AD)}(x, \bar\theta)\;\tilde \Phi^{(AD)}(x, \bar\theta)- \frac {1}{2} \;\partial_\mu {\tilde \Phi^{(AD)}}(x, \bar\theta)\;
\partial^\mu \tilde \Phi^{(AD)}(x, \bar\theta)\nonumber\\
& + & \frac {m^2}{2} B_\mu ^{(AD)}(x, \bar\theta)\, B^{\mu (AD)}(x, \bar\theta)  + 
\frac {1}{2} \;\partial_\mu { \Phi^{(AD)}}(x, \bar\theta)\;
\partial^\mu  \Phi^{(AD)}(x, \bar\theta)\nonumber\\
& + & m\, B_\mu ^{(AD)}(x, \bar\theta)\, \partial^\mu  \Phi^{(AD)}(x, \bar\theta) +  \tilde {\bar B}^{(AD)}(x, \bar\theta)\big[\partial_\mu B^{\mu (AD)}(x, \bar\theta) \nonumber\\
&-& m\,  \Phi^{(AD)}(x, \bar\theta)\big] + \frac {1}{2}\;  \tilde {\bar B}^{(AD)}(x, \bar\theta) \; \tilde {\bar B}^{(AD)}(x, \bar\theta)\nonumber\\  
& - & \; i\; \partial_{\mu}\bar F^{(AD)}(x, \bar\theta)\;\partial^{\mu} F^{(AD)}(x, \bar\theta) + i\, m^2\, {\bar F} ^{(AD)} ( x, \bar\theta)\, 
F ^{(AD)} ( x, \bar\theta),\nonumber\\
{\cal L}_{(b_2)}\longrightarrow \tilde {\cal L}^{(C, D)}_{(b_2)}& = & \, {\tilde {\bar {\cal B}}}^{(D)}(x, \theta)\,\big(\tilde E^{(D)}(x, \theta) +
m\,\tilde \Phi^{(D)}(x, \theta)\big) 
- \frac {1}{2}\,  {\tilde {\bar {\cal B}}}^{(D)}(x, \theta)\;{\tilde {\bar {\cal B}}}^{(D)}(x, \theta) \nonumber\\
& - &  m \,\tilde E^{(D)}(x, \theta)\;\tilde \Phi^{(D)}(x, \theta)- \frac {1}{2} \;\partial_\mu {\tilde \Phi^{(D)}}(x, \theta)\;
\partial^\mu \tilde \Phi^{(D)}(x, \theta)\nonumber\\
& + & \frac {m^2}{2} B_\mu ^{(D)}(x, \theta)\, B^{\mu (D)}(x, \theta) + 
\frac {1}{2} \;\partial_\mu { \Phi^{(D)}}(x, \theta)\;
\partial^\mu  \Phi^{(D)}(x, \theta)\nonumber\\
& + & m\, B_\mu ^{(D)}(x, \theta)\; \partial^\mu  \Phi^{(D)}(x, \theta)+  \tilde {\bar B}^{(D)}(x, \theta)\big[\partial_\mu B^{\mu (D)}(x, \theta) \nonumber\\
&-& m\,  \Phi^{(D)}(x, \theta)\big] + \frac {1}{2}\; \tilde {\bar B}^{(D)}(x, \theta) \;\tilde {\bar B}^{(D)}(x, \theta)\nonumber\\  
& - & \; i\; \partial_{\mu}\bar F^{(D)}(x, \theta)\;\partial^{\mu} F^{(D)}(x, \theta)+ i\, m^2\, {\bar F} ^{(D)} (x, \theta)\, 
F ^{(D)} (x, \theta),
\end{eqnarray}
where the super Lagrangian density, with superscript $(AC, AD)$, contains the {\it anti-chiral} superfields that have been obtained after the applications of anti-co-BRST invariant restrictions (cf. Eqs. (102)). In exactly similar fashion, we note that the super Lagrangian density ${\cal L}_{(b_2)}^{(C, \,D)}$ incorporates the {\it chiral} superfields that have been obtained after the applications of the co-BRST invariant restrictions (cf. Eq. (102)). At this stage, taking the helps of the mappings: $s_d^{(2)} \leftrightarrow \partial_{\theta}, s_{ad}^{(2)} \leftrightarrow \partial_{\bar{\theta}}$, we observe the following
\begin{eqnarray}
&&\frac{\partial}{\partial\bar{\theta}}[\tilde{{\cal L}}_{(b_2)}^{(AC, AD)}] = \partial_{\mu}\,\Bigl [ \,\bar {\cal B}\, \partial^\mu \,C 
+ m \, \varepsilon^{\mu\nu}\, \bigl (m \,A_\nu \,C - \phi \,\partial_\nu\, C \bigr )
- m\, \tilde \phi\, \partial^\mu\, C \Bigr ] \equiv s_{ad}^{(2)}{{\cal L}}_{b_2}, \nonumber\\
&& \frac{\partial}{\partial{\theta}}[\tilde{{\cal L}}_{(b_2)}^{(C, D)}] = \partial_{\mu}\,\Bigl [ \,\bar {\cal B}\, \partial^\mu \,\bar C 
+ m \, \varepsilon^{\mu\nu}\, \bigl ( m \,A_\nu \,\bar C - \phi\, \partial_\nu\, \bar C \bigr )
- m \,\tilde \phi \,\partial^\mu \,\bar C \Bigr ] \equiv s_{d}^{(2)}{{\cal L}}_{b_2}.
\end{eqnarray}
Thus, we note that we have captured the (anti-)co-BRST {\it invariance} of the action integral $ S = \int d^2 x\, {\cal L}_{(b_2)}$
because we observe that Eq. (131) matches with Eq. (38). It goes without saying that there are some superfields, with superscripts 
$(C, D)$ and $(AC, AD)$, which are {\it actually} ordinary fields on the 2D Minkowskian spacetime manifold as is evident from the observations 
$s_{(a)d}^{(2)}\;[\bar {\cal B}, (\partial\cdot A), \phi]= 0$ as well as $s_d^{(2)} \bar C = 0, s_{ad}^{(2)} C = 0$, etc.

We concentrate, at this stage, on capturing the restriction: ${\cal B} + \bar {\cal B} - 2\, E = 0$
within the framework of the (anti-)chiral superfield approach to BRST formalism. In this context, we generalize the Lagrangian 
density ${\cal L}_{(b_1)}$ to the (2, 1)-dimensional (anti-)chiral super-submanifolds as follows 
\begin{eqnarray}
{\cal L}_{(b_1)}\longrightarrow \tilde {\cal L}^{(AC,  AD)}_{(b_1)}& = & \, \tilde {\cal B}^{(AD)}(x, \bar\theta)\,\big(\tilde E^{(AD)}(x, \bar\theta) -
m\,\tilde \Phi^{(AD)}(x, \bar\theta)\big) 
- \frac {1}{2}\,  \tilde {\cal B}^{(AD)}(x, \bar\theta)\,\tilde {\cal B}^{(AD)}(x, \bar\theta)  \nonumber\\
& + &  m \tilde E^{(AD)}(x, \bar\theta)\;\tilde \Phi^{(AD)}(x, \bar\theta)- \frac {1}{2} \;\partial_\mu {\tilde \Phi^{(AD)}}(x, \bar\theta)\;
\partial^\mu \tilde \Phi^{(AD)}(x, \bar\theta)\nonumber\\
& + & \frac {m^2}{2} B_\mu ^{(AD)}(x, \bar\theta)\, B^{\mu (AD)}(x, \bar\theta)  + 
\frac {1}{2} \;\partial_\mu { \Phi^{(AD)}}(x, \bar\theta)\;
\partial^\mu  \Phi^{(AD)}(x, \bar\theta)\nonumber\\
& - & m\, B_\mu ^{(AD)}(x, \bar\theta)\, \partial^\mu  \Phi^{(AD)}(x, \bar\theta) +  \tilde B^{(AD)}(x, \bar\theta)\big[\partial_\mu B^{\mu (AD)}(x, \bar\theta) \nonumber\\
&+& m\,  \Phi^{(AD)}(x, \bar\theta)\big] + \frac {1}{2}\; \tilde B^{(AD)}(x, \bar\theta) \;\tilde B^{(AD)}(x, \bar\theta)\nonumber\\  
& - & \; i\; \partial_{\mu}\bar F^{(AD)}(x, \bar\theta)\;\partial^{\mu} F^{(AD)}(x, \bar\theta) + i\, m^2\, {\bar F} ^{(AD)} ( x, \bar\theta)\, 
F ^{(AD)} ( x, \bar\theta),\nonumber\\
{\cal L}_{(b_1)}\longrightarrow \tilde {\cal L}^{(C, D)}_{(b_1)}& = & \, \tilde {\cal B}^{(D)}(x, \theta)\,\big(\tilde E^{(D)}(x, \theta) -
m\,\tilde \Phi^{(D)}(x, \theta)\big) 
- \frac {1}{2}\,  \tilde {\cal B}^{(D)}(x, \theta)\;\tilde {\cal B}^{(D)}(x, \theta) \nonumber\\
& + &  m \tilde E^{(D)}(x, \theta)\;\tilde \Phi^{(D)}(x, \theta)- \frac {1}{2} \;\partial_\mu {\tilde \Phi^{(D)}}(x, \theta)\;
\partial^\mu \tilde \Phi^{(D)}(x, \theta)\nonumber\\
& + & \frac {m^2}{2} B_\mu ^{(D)}(x, \theta)\, B^{\mu (D)}(x, \theta) + 
\frac {1}{2} \;\partial_\mu { \Phi^{(D)}}(x, \theta)\;
\partial^\mu  \Phi^{(D)}(x, \theta)\nonumber\\
& - & m\, B_\mu ^{(D)}(x, \theta)\; \partial^\mu  \Phi^{(D)}(x, \theta)+  \tilde B^{(D)}(x, \theta)\big[\partial_\mu B^{\mu (D)}(x, \theta) \nonumber\\
& + & m\, \Phi^{(D)}(x, \theta)\big] + \frac {1}{2}\; \tilde B^{(D)}(x, \theta) \;\tilde B^{(D)}(x, \theta)\nonumber\\  
& - & \; i\; \partial_{\mu}\bar F^{(D)}(x, \theta)\;\partial^{\mu} F^{(D)}(x, \theta)+ i\, m^2\, {\bar F} ^{(D)} (x, \theta)\, 
F ^{(D)} (x, \theta),
\end{eqnarray}
where all the superscripts and their implicants have been clarified earlier. It is now straightforward to check that the following are true, namely;  
\begin{eqnarray*}
 \frac {\partial}{\partial\bar\theta}\Big[\tilde {\cal L}_{(b_1)}^{(AC, AD)}\Big] &=& \partial_\mu \Bigl [ \bar {\cal B}\, \partial^\mu\,  C
+ m \,\varepsilon^{\mu\nu}\,  \bigl (m\, A_\nu \, C +  \phi\, \partial_\nu \, C \bigr ) 
- m \,\tilde \phi\, \partial^\mu \,  C \Bigr ]\nonumber\\ 
&-& \Bigl [ {\cal B} + \bar {\cal B} - 2 E \Bigr ] \,
(\Box + m^2 )\; C \;\equiv s_{ad}^{(2)}\,{\cal L}_{(b_1)},
\end{eqnarray*}
\begin{eqnarray}
\frac {\partial}{\partial\theta}\Big[\tilde {\cal L}_{(b_1)}^{(C, D)} \Big] &=& \partial_\mu \Bigl [  \bar {\cal B}\, \partial^\mu\, \bar C
+ m \,\varepsilon^{\mu\nu}\,  \bigl (m\, A_\nu \,\bar C +  \phi\, \partial_\nu \,\bar C \bigr ) 
- m \,\tilde \phi\, \partial^\mu \, \bar C \Bigr ]\nonumber\\
&-& \Bigl [ {\cal B} + \bar {\cal B} - 2 E \Bigr ] \,
(\Box + m^2 ) \;\bar C
\;\equiv s_{d}^{(2)}\,{\cal L}_{(b_1)},
\end{eqnarray}
where we have taken into account the mappings: $s_d^{(2)} \leftrightarrow \partial_{\theta}, s_{ad}^{(2)} \leftrightarrow \partial_{\bar{\theta}}$.
Thus, we note that we have captured the restriction: ${\cal B} + \bar {\cal B} - 2\, E = 0$ which has appeared in Eq. (40) in connection with 
the applications of the (anti-)co-BRST symmetry transformations $(s_{(a)d}^{(2)})$ on Lagrangian density  ${\cal L}_{(b_1)}$.
It goes without saying that there are some superfields, with superscripts $(C, D)$ and $(AC, CD)$ that are {\it primarily ordinary} fields because of the observations: $s_{(a)d}^{(2)}\;[{\cal B}, (\partial\cdot A), \phi]= 0$ as well as $s_d^{(2)}\bar C = 0, s_{ad}^{(2)} C = 0$, etc.

As we have captured the  restriction in connection with the applications of $s_{(a)d}^{(2)}$ on the Lagrangian density ${\cal L}_{(b_1)}$, in exactly similar fashion, we {\it now} explain the existence of the {\it above}  restriction in the context of the applications of $s_{(a)d}^{(1)}$ on the Lagrangian density ${\cal L}_{(b_2)}$. Towards this goal in mind, we generalize the Lagrangian density ${\cal L}_{(b_2)}$ onto (2, 1)-dimensional (anti-)chiral super-submanifolds as   
\begin{eqnarray}
{\cal L}_{(b_2)}\longrightarrow \tilde {\cal L}^{(ac,  ad)}_{(b_2)}& = & \, {\tilde {\bar {\cal B}}}^{(ad)}(x, \bar\theta)\,\big(\tilde E^{(ad)}(x, \bar\theta) +
m\,\tilde \Phi^{(ad)}(x, \bar\theta)\big) 
- \frac {1}{2}\,  {\tilde {\bar {\cal B}}}^{(ad)}(x, \bar\theta)\,{\tilde {\bar {\cal B}}}^{(ad)}(x, \bar\theta)  \nonumber\\
& - &  m \,\tilde E^{(ad)}(x, \bar\theta)\;\tilde \Phi^{(ad)}(x, \bar\theta)- \frac {1}{2} \;\partial_\mu {\tilde \Phi^{(ad)}}(x, \bar\theta)\;
\partial^\mu \tilde \Phi^{(ad)}(x, \bar\theta)\nonumber\\
& + & \frac {m^2}{2} B_\mu ^{(ad)}(x, \bar\theta)\, B^{\mu (ad)}(x, \bar\theta)  + 
\frac {1}{2} \;\partial_\mu { \Phi^{(ad)}}(x, \bar\theta)\;
\partial^\mu  \Phi^{(ad)}(x, \bar\theta)\nonumber\\
& + & m\, B_\mu ^{(ad)}(x, \bar\theta)\, \partial^\mu  \Phi^{(ad)}(x, \bar\theta) +  \tilde {\bar B}^{(ad)}(x, \bar\theta)\big[\partial_\mu B^{\mu (ad)}(x, \bar\theta) \nonumber\\
&-& m\,  \Phi^{(ad)}(x, \bar\theta)\big] + \frac {1}{2}\;  \tilde {\bar B}^{(ad)}(x, \bar\theta) \; \tilde {\bar B}^{(ad)}(x, \bar\theta)\nonumber\\  
& - & \; i\; \partial_{\mu}\bar F^{(ad)}(x, \bar\theta)\;\partial^{\mu} F^{(ad)}(x, \bar\theta) + i\, m^2\, {\bar F} ^{(ad)} ( x, \bar\theta)\; 
F ^{(ad)} ( x, \bar\theta),\nonumber\\
{\cal L}_{(b_2)}\longrightarrow \tilde {\cal L}^{(c, d)}_{(b_2)}& = & \, {\tilde {\bar {\cal B}}}^{(d)}(x, \theta)\,\big(\tilde E^{(d)}(x, \theta) +
m\,\tilde \Phi^{(d)}(x, \theta)\big) 
- \frac {1}{2}\,  {\tilde {\bar {\cal B}}}^{(d)}(x, \theta)\;{\tilde {\bar {\cal B}}}^{(d)}(x, \theta) \nonumber\\
& - &  m \,\tilde E^{(d)}(x, \theta)\;\tilde \Phi^{(d)}(x, \theta)- \frac {1}{2} \;\partial_\mu {\tilde \Phi^{(d)}}(x, \theta)\;
\partial^\mu \tilde \Phi^{(d)}(x, \theta)\nonumber\\
& + & \frac {m^2}{2} B_\mu ^{(d)}(x, \theta)\, B^{\mu (d)}(x, \theta) + 
\frac {1}{2} \;\partial_\mu { \Phi^{(d)}}(x, \theta)\;
\partial^\mu  \Phi^{(d)}(x, \theta)\nonumber\\
& + & m\, B_\mu ^{(d)}(x, \theta)\; \partial^\mu  \Phi^{(d)}(x, \theta)+  \tilde {\bar B}^{(d)}(x, \theta)\big[\partial_\mu B^{\mu (d)}(x, \theta) \nonumber\\
&-& m\,  \Phi^{(d)}(x, \theta)\big] + \frac {1}{2}\; \tilde {\bar B}^{(d)}(x, \theta) \;\tilde {\bar B}^{(d)}(x, \theta)\nonumber\\  
& - & \; i\; \partial_{\mu}\bar F^{(d)}(x, \theta)\;\partial^{\mu} F^{(d)}(x, \theta)+ i\, m^2\, {\bar F} ^{(d)} (x, \theta)\, 
F ^{(d)} (x, \theta),
\end{eqnarray}
where all the symbols and superscripts have been clarified earlier. It is now elementary exercise to note that the following are true, namely;
\begin{eqnarray*}  
\frac{\partial}{\partial\bar{\theta}}\Big[\tilde{\cal L}_{(b_2)}^{(ac, ad)}\Big] &=& \partial_\mu \,\Bigl [  {\cal B}\, \partial^\mu \, C
+ m \,\varepsilon^{\mu\nu} \, \bigl (m \,A_\nu \, C -  \phi\, \partial_\nu \, C \bigr ) 
+ m \,\tilde \phi\, \partial^\mu \, C \Bigr ] 
\nonumber\\ &-& \Bigl [ {\cal B} + \bar {\cal B} - 2 E \Bigr ] \,
(\Box + m^2 ) \; C \;\equiv s_{ad}^{(1)}{\cal L}_{(b_2)}
\end{eqnarray*}
\begin{eqnarray}  
\frac{\partial}{\partial{\theta}}\Big[{\cal L}_{(b_2)}^{(c, d)}\Big] &=& \partial_\mu \,\Bigl [ {\cal B}\, \partial^\mu \,\bar C
+ m \,\varepsilon^{\mu\nu} \, \bigl (m \,A_\nu \,\bar C -  \phi\, \partial_\nu \,\bar C \bigr ) 
+ m \,\tilde \phi\, \partial^\mu \bar C \Bigr ] 
\nonumber\\ &-& \Bigl [ {\cal B} + \bar {\cal B} - 2 E \Bigr ] \,
(\Box + m^2 )\; \bar C\; \equiv s_{d}^{(1)}{\cal L}_{(b_2)}.
\end{eqnarray}
The above equation shows that we have proven the sanctity of the restriction within the framework of (anti-)chiral superfield approach to BRST formalism (provided we do {\it not} take into account the mass-shell conditions: $(\Box + m^2)\,C = 0,\;(\Box + m^2)\,\bar C = 0 $ for the (anti-)ghost fields. It should be pointed that some of the superfields, with superscripts $(ac, ad)$ and $(c, d)$ are {\it actually} ordinary fields. For instance, we have: $\bar F ^{(d)} (x, \theta) = \bar C (x),
F ^{(ad)} (x, \bar\theta) =  C (x), \Phi^{(d)} (x, \theta) = \Phi^{(ad)} (x, \bar\theta) = \phi (x),$ etc.

\section{CF-Type Restrictions and Pseudo-Scalar Field with Negative Kinetic Term: A Few Comments}

In this section, we dwell a bit on the existence of the (anti-)BRST and (anti-)co-BRST invariant  restrictions 
(e.g. $B + \bar B + 2\, (\partial\cdot A) = 0,\; {\cal B } + \bar {\cal B}
 - 2\, E = 0 $) that have appeared (cf. Eq. (33)) in our BRST analysis of the 2D modified version of the Proca theory 
{\it and} discuss their drastic differences and some similarities  {\it vis-{\`a}-vis} the 
 {\it usual} CF-condition [21] that exists in the BRST analysis of the {\it non-Abelian} 1-form gauge theory. In the case of the {\it latter} theory
(defined in any arbitrary dimension of spacetime), the coupled [34] Lagrangian densities ${\cal L}_{(B)}$  and ${\cal L}_{(\bar B)}$, in the Curci-Ferrari gauge [35, 36], are  [34, 37]
\begin{eqnarray}
{\cal L}_{B}  & = & -\,\frac{1}{4}F^{\mu\nu}\cdot F_{\mu\nu} + B\cdot (\partial_\mu A^\mu)
 +  \frac{1}{2}(B\cdot B + \bar B\cdot \bar B) - i\,\partial_{\mu}\bar C\cdot D^{\mu} C,\nonumber\\
{\cal L}_{\bar B}  & = & -\,\frac{1}{4}F^{\mu\nu}\cdot F_{\mu\nu}  - \bar B\cdot (\partial_\mu A^\mu)
 +  \frac{1}{2}(B\cdot B + \bar B\cdot \bar B)
 -  i\,D_{\mu}\bar C\cdot \partial^{\mu} C, 
\end{eqnarray}
where $F_{\mu\nu}  = \partial_\mu A_\nu - \partial_\nu A_\mu + i\, (A_\mu\times A_\nu)$ is the field strength tensor which is derived from the 2-form 
$F^{(2)} = d\,A^{(1)} + i\,(A^{(1)}\wedge  A^{(1)})$ with the 1-form $A^{(1)} = d\, x^\mu\, A_\mu\cdot T \equiv d\, x^\mu\, A_\mu$.
The $SU(N)$ non-Abelian symmetry transformations are expressed in terms of the generator $T^a$ which obey the $SU(N)$ Lie algebra $[T^a, T^b] = f^{abc}\;T^c$ where $a, b, c = 1, 2...N^2 - 1$
are the group indices in the $SU(N)$  Lie algebraic space where the cross and dot products between two non-null vectors  $(P^a, Q^a)$ are defined as $(P\times Q)^a = f^{abc}P^b Q^c$ and $P\cdot Q = P^a Q^a$. The covariant derivatives $D_\mu C = \partial_\mu C  + i\,(A_\mu\times C)$ and 
$D_\mu \bar C = \partial_\mu \bar C  + i\,(A_\mu\times \bar C)$ are in the adjoint representation of the $SU(N)$ Lie algebra. For the  $SU(N)$ algebra,
the structure constants $ f^{abc}$ can be chosen to be {\it totally} antisymmetric in {\it all} the indices\footnote{To be precise, for a specific 
representation of $T^a$, the structure constants $f^{abc}$ become totally  antisymmetric for any arbitrary Lie algebra 
(see, e.g. [38] for details).} (see, e.g. [38] for details). In the above coupled Lagrangian densities 
$ B= B\cdot T = B^a T^a $ and $\bar{B}= \bar{B}\cdot T = \bar{B}^a T^a $ are the Nakanishi-Lautrup type auxiliary fields which obey the Curci-Ferrari 
(CF) condition [21] as follows
\begin{eqnarray}
B + \bar B + (C\times\bar C) = 0,  
\end{eqnarray}
where $C^a$ and ${\bar C}^a$ are the {\it fermionic} [i.e. $(C^a)^2 = ({\bar C}^a)^2 = 0, C^a C^b + C^b C^a = 0,
 \bar C^a \bar C^b + \bar C^b \bar C^a = 0,  C^a \bar C^b +  C^b \bar C^a = 0,$ etc.] ghost and anti-ghost fields  which are needed in the theory to maintain the unitarty at any arbitrary 
 order of perturbative  computations.

 The CF-condition $(137)$ emerges out when we equate ${\cal L}_{B}$ and ${\cal L}_{\bar B}$ and demand their equivalence 
(modulo a total spacetime derivative term). To be precise, we have:
\begin{eqnarray}
{\cal L}_B \equiv {\cal L}_{\bar{B}}\quad \Longrightarrow \quad B + \bar{B} + (C \times \bar{C}) = 0.
\end{eqnarray}
In other words, the very existence of the coupled Lagrangian densities ${{\cal L}}_B$ and ${{\cal L}}_{\bar{B}}$ depends crucially on the CF-condition. Furthermore, the absolute anticommutativity of the nilpotent (anti-)BRST symmetry transformations (i.e. $s_b\,s_{ab} + s_{ab}\,s_b = 0$) is valid {\it{only}} when the CF-condition is satisfied in the non-Abelian 1-form gauge theory. This also gets reflected at the level of the conserved and nilpotent charges $Q_{(a)b}$ because the absolute anticommutativity of these charges (i.e. ($Q_b\, Q_{ab} +  Q_{ab}\,Q_b = 0$)), once again, crucially depends on the existence of CF-condition ($B + \bar B + (C\times\bar C) = 0$).
In addition, we note that the applications of the off-shell nilpotent and absolutely anticommuting (anti-)BRST symmetry transformations $s_{(a)b}$ on the Lagrangian densities ${\cal L}_B$ and ${\cal L}_{\bar{B}}$ of the non-Abelian 1-form gauge theory,
lead to the following\footnote{ The off-shell nilpotent (anti-)BRST symmetry transformations for the coupled Lagrangian densities are: $s_{ab}\; A_\mu= D_\mu\bar C,\qquad  s_{ab}\; \bar C= -\frac{i}{2}\,(\bar C\times\bar C),
\; s_{ab}\;C   = i{\bar B},\quad  s_{ab}\;\bar B  = 0,\;
s_{ab}\; F_{\mu\nu} = i \,(F_{\mu\nu}\times\bar C),\quad s_{ab} (\partial_\mu A^\mu) = \partial_\mu D^\mu\bar C,
\quad s_{ab}\; B = i\,(B \times \bar C),$ and
$s_b\; A_\mu = D_\mu C,\quad s_b \;C =  - \frac{i}{2}\; (C\times C),\quad  s_b\;\bar C \;= i\,B ,\;
 \quad s_b\; B = 0, \;
 s_b\;\bar B = i\,(\bar B\times C),\quad s_b\; (\partial_\mu A^\mu) = \partial_\mu D^\mu C,
\quad s_b \;F_{\mu\nu} = i\,(F_{\mu\nu}\times C)$ (see, e.g. [34, 37] for details).} (see, e.g. [39]):
\begin{eqnarray}
&&s_b{\cal L}_B   =  \partial_\mu[B \cdot D^\mu C],  \qquad s_{ab}{\cal L}_{\bar B}= - \;\partial_\mu{[\bar B \cdot D^\mu \bar C]},\nonumber\\
&&s_b{\cal L }_{\bar B}\;  =  \partial_\mu\,[ {\{ B + ( C \times \bar C )\}}\cdot
\partial^\mu C \,]-{\{B + \bar B + ( C\times\bar C )\}}\cdot D_\mu\partial^\mu C,\nonumber\\
&&s_{ab}{\cal L}_B  = -\;\partial_\mu\,[{\{\bar B + ( C\times\bar C)\} \cdot \partial^\mu \bar C}\,] 
 +  \;\{(B+\bar B + ( C \times {\bar C})\} \cdot D_\mu \partial^\mu \bar C.
\end{eqnarray}
Thus, we observe that {\it both} the Lagrangian densities (i.e. ${\cal L}_B$ and ${\cal L}_{\bar{B}}$) respect {\it both} the off-shell nilpotent (anti-)BRST
symmetries provided we confine ourselves on a hypersurface in the $D$-dimensional Minkowskian spacetime manifold  where the CF-condition ($B + \bar B + (C\times\bar C) = 0$) is satisfied. In other words,
we have the following:
\begin{eqnarray}
&&s_b{\cal L}_B   =  \partial_\mu[B \cdot D^\mu C],  \;\;\qquad\qquad s_{ab}{\cal L}_{\bar B}= - \;\partial_\mu{[\bar B \cdot D^\mu \bar C]},\nonumber\\
&&s_b{\cal L}_{\bar B}  = -\, \partial_\mu[\bar B \cdot \partial^\mu C],  \qquad\qquad s_{ab}{\cal L}_{B}= \partial_\mu{[B \cdot \partial^\mu \bar C]}.
\end{eqnarray} 
The above equation establishes  that the action integrals $S_1 = \int\,d^D x\,{\cal L}_{B}$ and $S_2 = \int\,d^D x\,{\cal L}_{\bar B}$
respect {\it both} the (anti-)BRST symmetries on the hypersurafce in the $D$-dimensional flat Minkowskian spacetime manifold where (i) the CF-condition is satisfied, and (ii) the off-shell nilpotent and absolutely anticommuting (anti-)BRST symmetries are defined.

Against the backdrop of the above statements, we now concentrate on the discussion of our restrictions (cf. Eq. (33)). It can be checked that the requirement of equivalence between ${\cal L}_{(b_1)}$ and ${\cal L}_{(b_2)}$ (i.e. ${\cal L}_{(b_1)} \equiv  {\cal L}_{(b_2)}$) for our theory leads 
to the following: 
\begin{eqnarray}
(B + \bar B +2\,(\partial\cdot A))\,(B - \bar B +2\,m\,\phi) - ({\cal B} + \bar{\cal B} - 2\,E)\,({\cal B} - \bar{\cal B} + 2\,m\,\tilde\phi) = 0.
\end{eqnarray}
Thus, we find that {\it all} the {\it four}  restrictions (that have been pointed  out in (33)) appear very naturally in the equality:
 ${\cal L}_{(b_1)} =  {\cal L}_{(b_2)}$ (modulo some {\it total} spacetime derivatives). Therefore, we conclude that {\it one }
 of the solutions of (141) is nothing but the restrictions derived in (33). This observation is exactly similar to our observation in the context of non-Abelian 1-form gauge theory (cf. Eq. (138). We now focus on the symmetry properties of the Lagrangian densities 
 ${\cal L}_{(b_1)}$ and ${\cal L}_{(b_2)}$
 which have been illustrated in Eq. (34). We observe that {\it both} the Lagrangian densities respect {\it both} the (anti-)BRST symmetry 
 transformations $s_{(a)b}^{(1, 2)}$ on the constrained hypersurface where $B + \bar B +2\,(\partial\cdot A) = 0$ is satisfied.
 Thus, there is, once again, similarity between our 2D modified Proca  (i.e. {\it{massive}} Abelian 1-form gauge) theory  and the non-Abelian 1-form gauge theory (cf. Eq. (140)). We point out that the restriction: ${\cal B} + \bar{\cal B} - 2\,E = 0$ {\it also} appears in 
(141) which is, once again, similar to the observation in 2D non-Abelian theory in the context of the existence of the off-shell nilpotent and absolutely anticommuting (anti-)co-BRST symmetries [39].

We would like to point out here that {\it both} the factorized terms in Eq. (141) are {\it zero} {\it separately} and {\it independently} because both of them 
owe their origins to mathematically {\it independent} cohomological operators of differential geometry. For instance, as pointed out earlier, the restriction (${\cal B} + \bar{\cal B} - 2\,E = 0$) owes its origin to the exterior derivative $d = d x^\mu \partial_\mu $\;(with $d^2 = 0)$  because the 2-form 
$F^{(2)} = d \, A^{(1)}\equiv \frac {1}{2}(dx^\mu\wedge dx^\nu)\,F_{\mu\nu}$ defines the field strength tensor $F_{\mu\nu}$ which possesses only {\it one}
non-zero component in 2D (that is nothing but the electric field $E$). In exactly  similar fashion, we note that the restriction:
$B+ \bar B + 2\,(\partial\cdot A) = 0$ owes its origin to the co-exterior derivative $\delta  = -\,\ast d\,\ast$ because we observe that 
$\delta A^{(1)} = -\,\ast d\,\ast \,A^{(1)} = (\partial\cdot A)$ which  defines the gauge-fixing term of the Lagrangian densities 
${\cal L}_{(b_1)}$ and ${\cal L}_{(b_2)}$ (where we generalize {\it it} to $[(\partial\cdot A \pm m\,\phi)]$ on the dimensional ground). Since the cohomological
 operators $(d)$ and  $(\delta)$ are linearly independent of each-other, we argue  that both the terms of Eq. (141) would vanish off on their {\it own}.
At present  level of our understanding, we do {\it not} know as to why the  restrictions ${\cal B} + \bar{\cal B} - 2\,E = 0$ and 
$B+ \bar B + 2\,(\partial\cdot A) = 0$ are picked out, from Eq. (33), in the discussions of the (anti-)co-BRST  and (anti-)BRST symmetry transformations 
of the Lagrangian densities  ${\cal L}_{(b_1)}$ and ${\cal L}_{(b_2)}$ {\it but } the other constraints $ B - \bar B + 2\,m\,\phi = 0$  and    
${\cal B}- \bar{\cal B} + 2\,m\,\tilde\phi = 0$ are {\it not } utilized by the (anti-)co-BRST  and (anti-)BRST symmetries of our 2D 
Proca theory.

We would like to mention a few things connected with the 2D {\it non-Abelian} 1-form gauge theory which we have discussed  in our earlier work
[39]  where we have shown the existence of the (anti-)co-BRST symmetries (in addition to (anti-)BRST symmetries). In fact, we have derived the off-shell nilpotent and absolutely anticommuting  (anti-)co-BRST symmetry transformation for the 2D non-Abelian 1-form gauge theory under which the Lagrangian
 densities and, {\it specifically},  the gauge-fixing term remain {\it invariant}. To be precise, we have considered the following coupled Lagrangian densities for the 2D non-Abelian gauge theory [34, 37] in the Curci-Ferrari gauge [35, 36] for our discussions, namely;
\begin{eqnarray}
&&{\cal L}_B = {\cal B} {\cdot E} - \frac {1}{2}\,{\cal B} \cdot {\cal B} +\, B\cdot (\partial_{\mu}A^{\mu}) 
+ \frac{1}{2}(B\cdot B + \bar B \cdot \bar B) - i\,\partial_{\mu}\bar C \cdot D^{\mu}C, \nonumber\\
&&{\cal L}_{\bar B} = {\cal B} {\cdot E} - \frac {1}{2}\,{\cal B} \cdot {{\cal B}} - \bar B\cdot (\partial_{\mu}A^{\mu}) 
+ \frac{1}{2}(B\cdot B + \bar B \cdot \bar B) - i\, D_{\mu}\bar C \cdot \partial^{\mu}C,
\end{eqnarray}
where ${\cal B} = {\cal B}\cdot T$ is the Nakanishi-Lautrup type auxiliary field which has been invoked to linearize the kinetic term
 $(\frac {1}{2} \,{E}\cdot {E} = -\,\frac {1}{4} \,F_{\mu\nu}\cdot F^{\mu\nu})$ for the 2D non-Abelian theory. It is clear that, in 2D spacetime, 
we have only one existing component of  $F_{\mu\nu}$ (i.e. $F_{01} = \partial_0\,A_1-\partial_1\,A_0 +i\,(A_0\times A_1)\equiv E$). We have found out that the following  are true \footnote{The off-shell nilpotent and absolutely anticommuting (anti-)co-BRST symmetries 
for the 2D coupled Lagrangian densities are: $s_{ad} A_\mu = - \varepsilon_{\mu\nu}\partial^\nu C,\quad\, s_{ad} C = 0,\quad
s_{ad} \bar C =  i \,{\cal B},\quad\,s_{ad} {\cal B} = 0,\quad
s_{ad} E =D_\mu\partial^\mu C,\quad\,\,\, s_{ad} B = 0,\quad
s_{ad}\bar B = 0,\quad\, s_{ad}({\partial_\mu A^\mu})= 0,$ and 
$s_d A_\mu = - \varepsilon_{\mu\nu}\partial^\nu \bar C,
\quad s_d C = - i\, {\cal B},\quad s_d \bar C = 0,\quad s_d{\cal B} = 0,\quad
s_d E = D_\mu\partial^\mu\bar C,\quad s_d B = 0,
\quad s_d\bar B = 0, \quad s_d({\partial_\mu A^\mu})= 0$ (see, e.g. [39] for details).}, namely;
\begin{eqnarray}
&&s_{ad}\,{\cal L}_{\bar B} = \partial_\mu[{\cal B}\cdot \partial^\mu C],  \qquad\qquad\quad s_d {\cal L}_B = \partial_\mu[{\cal B}
\cdot\partial^{\mu} \bar C],\nonumber\\
&& s_d{\cal L}_{\bar B} = \partial_\mu[{\cal B}\cdot D^\mu\bar C- \varepsilon^{\mu\nu} (\partial_\nu\bar C\times  
\bar C)\cdot C] + i\; (\partial_\mu A^\mu)\cdot({\cal B}\times\bar C),\nonumber\\
&& s_{ad}{\cal L}_B = \partial_\mu[{\cal B}\cdot D^\mu C+ \varepsilon^{\mu\nu}\bar C\cdot(\partial_\nu C\times C)] 
+ i\; (\partial_\mu A^\mu)\cdot({\cal B}\times C),
\end{eqnarray}
which demonstrate that, for the {\it both} the (anti-)co-BRST symmetries to be respected by  {\it both} Lagrangian densities, we need 
to invoke the following restrictions: 
\begin{eqnarray}
{\cal B} \times C = 0,\quad\quad\quad {\cal B} \times \bar C = 0.
\end{eqnarray}
It should be pointed out, at this stage, that the restriction (cf. Eq. (40)) that emerges out in our discussions on the 2D modified Proca theory 
(i.e. ${\cal B} + \bar {\cal B} - 2\, E = 0$) is analogous to the  restrictions  ${\cal B} \times C = 0$ and ${\cal B} \times \bar C = 0$
 that are essential for the BRST analysis  of the 2D non-Abelian theory. Hence, there is some kind of similarity.

Now we pin-point a few distinct differences between the restrictions of our 2D modified Proca theory and standard 
 {\it non-Abelian} 1-form gauge  theory. In the context of the {\it{latter}}, we know that the (anti-)BRST symmetry transformations $s_{(a)b}$ absolutely anticommute (i.e. $s_b\,s_{ab} + s_{ab}\,s_b = 0$) {\it{only}} when we impose the CF-condition $B + \bar{B} + (C \times \bar{C}) = 0$. This observation is {\it{not}} true in the context of our 2D modified Proca theory because we observe that {\it{only}} the pairs $(s_b^{(1)}, s_{ab}^{(1)})$ and $(s_b^{(2)}, s_{ab}^{(2)})$ absolutely anticommute {\it{without}} any recourse to the restrictions $(33)$. Except the above pairs, we point out that the rest of the {\it{fermionic}} symmetry transformations $s_{(a)b}^{(1)}$ and $s_{(a)b}^{(2)}$ do {\it{not}} absolutely anticommute ({\it{even}} if the restrictions $(33)$ are imposed from {\it{outside}}). These kinds of results are {\it also} true in the case of (anti-)co-BRST symmetry transformations $s_{(a)d}^{(1, 2)}$. We have collected {\it{all}} such possible anticommutators in our Appendix A. The above observations, at the symmetry level, are {\it{also}} reflected at the level of conserved charges because we find that the pairs $(Q_b^{(1)}, Q_{ab}^{(1)}), (Q_b^{(2)}, Q_{ab}^{(2)}), (Q_d^{(1)}, Q_{ad}^{(1)})$ and $(Q_d^{(2)}, Q_{ad}^{(2)})$ absolutely anticommute {\it{but}} other possible combinations do not absolutely anticommute {\it{even}} if we impose the restrictions $(33)$ from {\it{outside}}. We have collected {\it{all}} these results in our Appendix B. Furthermore, we note that, in the CF-condition (137),
there is {\it no} gauge field. However, we find that in the restriction $B + \bar B + 2\, (\partial\cdot A) = 0$ (connected with the (anti-)BRST
symmetries), the gauge field  appears in the 
form of Lorentz gauge (i.e. $(\partial\cdot A)$) and electric field appears in ${\cal B} + \bar {\cal B} - 2\, E = 0$ which is the restriction 
in the context of (anti-)co-BRST symmetry transformations for our 2D modified version of Proca theory.

At this juncture, we comment on the appearance of a pseudo-scaler field  $(\tilde{\phi})$ in our theory which is endowed with  the {\it negative} kinetic term but {\it it} possesses a  {\it{properly}} well defined mass (as it satisfies the Klein-Gordon equation $(\Box + m^2)\,\tilde{\phi} = 0$). 
In fact, we observe  that this pseudo-scalar field is essential for our discussion because we have shown the existence  of a set of appropriate {\it discrete}
symmetries (cf. Eq. (43)) which provide the physical realizations of the Hodge duality $\ast$ operation of differential geometry (cf. Eq. (45)). Thus,
the appearance of such kind of term is very {\it natural} in our whole discussion. We would like to point out that such kinds of fields have become very popular 
in the realm of cosmology where these kinds of fields have been christened as the ``ghost" fields (which are distinctly  different from the {\it fermionic} Faddeev-Popov ghost terms) (see, e.g. [40-48]). Such kinds of fields have  also been proposed as the candidates  for the dark matter and dark energy in modern literature (see, e.g. [49, 50]). In the context of the dark energy, these fields have {\it no} mass (which is the {\it massless } limit of the {\it massive} field theory {\it with} {\it only} the  negative kinetic term(s) for the field(s) {\it but} without any explicit mass term).

We end this section with the remark that we have generalized our present discussion to the 4D {\it massive} theory of Abelian 2-form gauge theory [51] where we have discussed the physical implications  of the existence of such kinds of
fields (which are endowed  with negative kinetic terms but properly defined mass) in the context of bouncing, cyclic and self-accelerated models of Universe [52-57].

\vskip 1cm

\section {Conclusions}

\vskip 1cm

In our present investigation, we have considered the St\"{u}ckelberg-modified version of the 2D Proca theory and shown that 
there are {\it two} Lagrangian densities for this theory which respect the {\it off-shell} nilpotent (anti-)BRST and (anti-)co-BRST symmetry
transformations besides respecting the ghost-scale and bosonic {\it continuous} symmetries. There exists a couple of discrete symmetries,
too, in our theory which make {\it both} the above Lagrangian densities represent a couple of field-theoretic  examples of  Hodge theory
because {\it all} the above symmetries, taken {\it together}, provide  the physical realizations of the de Rham cohomological operators [7-11]
of differential geometry at the {\it algebraic} level.

 We have applied the (anti-)chiral superfield approach to derive the {\it fermionic} (anti-) BRST and (anti-)co-BRST symmetry
transformations where we have defined the superfields on the (2, 1)-dimensional (anti-)chiral super-submanifolds of the {\it general} 
(2, 2)-dimensional supermanifold on which our 2D theory has been generalized. One of the key results of our present endeavor  has been the proof of
the off-shell nilpotency and absolute anticommutativity of the (anti-)BRST and (anti-)co-BRST charges. For instance, we have shown that the off-shell
nilpotency ($Q_b^2 = 0 $) of the BRST charge is deeply connected with nilpotency ($\partial_{\bar \theta}^2 = 0$)
of the translational generator ($\partial_{\bar\theta}$) along the $\bar\theta$-direction of the {\it anti-chiral} super-submanifold. However, the absolute anticommutativity of the BRST charge $with$ the anti-BRST charge has been found to be encoded in the nilpotency  ($\partial_{\theta}^2 = 0$) of the
translational generator  ($\partial_{\theta}$) along the $\theta$-direction of the $chiral$ super-submanifold of the general (2, 2)-dimensional
supermanifold. Similar kinds of statements could be made in connection with the other fermionic charges
(e.g. anti-BRST and (anti-)co-BRST charges) of our present theory. These observations are completely novel results within the framework of the
superfield approach to BRST formalism (cf. Sec. 5 for details).

The novel observations, in our discussions on 2D modified Proca theory, are (i) the introduction of a pseudo-scalar field (on {\it symmetry} ground) which 
is endowed with the negative kinetic term, and (ii) the existence of the restrictions which are found to have some kinds of similarities and a 
few distinct  differences with the {\it standard} CF-condition [21] that exists in the realm of BRST approach to non-Abelian 1-form gauge theory
in any arbitrary dimension of spacetime. Thus, we note that  our restrictions (cf. (33)) exist {\it only} in  2D modified model of Proca theory {\it but} the 
standard CF-condition [21] exists for the  non-Abelian 1-form theory in any arbitrary $D$-dimension of spacetime.
Our restrictions do {\it not} play any role in the proof of absolute anticommutativity property (cf. Sec. 7). The 
existence of a pseudo-scalar field with negative kinetic term is {\it important} because it is a precursor  to the existence of such kinds of fields 
in 4D theory where it is expected to play {\it important} role in the cosmological models of Universe and it might provide a clue to the ideas behind the dark matter/dark energy (see, Sec. 7 for details).

In our present endeavor, we have concentrated $only$ on the 2D {\it massive} Abelian (i.e. St\"{u}ckelberg-modified) 1-form gauge  theory.
However, we expect that this analysis could  be generalized to the physical four (3+1)-dimensions of spacetime.
In this context, we would like to mention our recent work [51] on the 4D {\it massive} Abelian 2-form gauge theory where we have shown the
existence of a pseudo-scalar and  an axial-vector fields with negative kinetic terms. Both the above models (i.e. the 2D modified Proca and 4D Abelian 2-form theories)  are the {\it massive}  field-theoretic 
examples of Hodge theory. We plan to apply our ideas to the {\it  massive} 6D  Abelian 3-form gauge theory and find out the consequences 
therein. The $massless$ version of the $latter$ theory has already been proven to be a tractable field-theoretic example  of Hodge theory 
in our earlier work (see, e.g. [14]).

We speculate that the {\it massive } models of Hodge theory would solve the problem of dark matter/dark energy   from the point of view of {\it symmetries} as these field-theoretic models would invoke some new kinds of fields endowed  with a few exotic physical properties. These theories might
turn out to be useful in the context of cosmology, too, where one requires the existence of ``ghost'' fields (i.e. scalar fields
with negative kinetic terms) [40-48]. We are, at present, very much involved with the {\it massive} version of gauge theories and we plan to prove 
$these$ theories to be the models for the Hodge theory. In the process, we shall be discussing about the fields/particles with exotic properties [58]
which might turn out to be useful in the context of various kinds of cosmological models. There is yet another direction that could be explored in the future 
{\it even} in the case of an interacting 2D Proca theory where there exists a coupling between the massive 
Abelian 1-form gauge field and the Dirac fields (see, e.g. [4] for details).
\\

\vskip 1cm

 \noindent
{\bf Acknowledgements}\\

\vskip 1cm

\noindent
The present investigation has been carried out under the BHU-fellowships received  by S. Kumar and A. Tripathi as well as the
DST-INSPIRE fellowship (Govt. of India) awarded to B. Chauhan. All these authors express their deep  sense of gratefulness to
the above funding agencies for  financial supports. Fruitful and enlightening comments by our esteemed Reviewer is thankfully 
acknowledged.\\

\vskip 1cm

\begin{center}
{\bf Appendix A: On the Absolute Anticommutativity of Nilpotent Symmetries}\\
\end{center}

\vskip 1cm

\noindent
We have already seen that the pairs ($s_b^{(1)}, s_{ab}^{(1)}), \;(s_b^{(2)}, s_{ab}^{(2)}),\; (s_d^{(1)}, s_{ad}^{(1)})$ and $(s_d^{(2)}, s_{ad}^{(2)})$
absolutely anticommute (separately and independently) without any recourse to the restrictions that have been listed in Eq. (33).
However, we show here that the other combinations of the above fermionic symmetries {\it  do not} absolutely anticommute.
In this context, we note that the following combinations of the (anti-)BRST symmetries $s_{(a)b}^{(1, 2)}$
\[\{ s_b^{(1)}, s_b^{(2)}\},\;\;\;\;\{ s_b^{(1)}, s_{ab}^{(2)}\},\;\;\;\;\{ s_{ab}^{(1)}, s_b^{(2)}\},\;\;\;\;\{ s_{ab}^{(1)}, s_{ab}^{(2)}\}, \eqno (A.1)\]
are the non-trivial anticommutators which we have to apply  on {\it all} the fields of our theory (that has been described by the Lagrangian densities 
${\cal L}_{(b_1)}$ and ${\cal L}_{(b_2)}$). It is straightforward, in this connection, that the following is true, namely;
\[\{ s_b^{(1)}, s_b^{(2)}\}\,\Psi  = 0, \qquad \Psi = A_\mu, C, \phi, \tilde\phi, B, \bar B, {\cal B}, \bar {\cal B},\eqno (A.2)\]
except the non-zero anticommutator:
\[\{ s_b^{(1)}, s_b^{(2)}\}\,\bar C = -\,4\,i\,\Box C.\eqno (A.3)\]
Thus, we conclude that the BRST symmetry transformations $s_{b}^{(1)}$ and $s_{b}^{(2)}$ do {\it not} absolutely anticommute with each-other.
Hence, their corresponding charges $Q_b^{(1)}$ and $Q_b^{(2)}$ would {\it also} not absolutely anticommute with each-other (cf. Appendix B below).

We focus now on the computation of the anticommutator $\{s_{b}^{(1)}, s_{ab}^{(2)}\}$ for our theory. In this connection, we observe the following:
\[\{ s_b^{(1)}, s_{ab}^{(2)}\}\,\Psi = 0,\qquad \Psi = \tilde \phi, {\cal B}, \bar {\cal B}. \eqno (A.4)\]
However, the other fields (e.g. $A_\mu, C, \bar C, \phi, B, \bar B, {\cal B}, \bar {\cal B})$ obviously do {\it not} satisfy (A.4).
We list here, the non-trivial anticommutators (acting on these fields), as
\[\{s_b^{(1)}, s_{ab}^{(2)}\}A_{\mu} = i\,\partial_{\mu} (B - \bar{B}) = -\,2\,i\,m\,\partial_{\mu}{\phi},\quad \{s_b^{(1)}, s_{ab}^{(2)}\}C = 2\,i\,\Box{C}, \]
\[\{s_b^{(1)}, s_{ab}^{(2)}\}\bar{C} = -\,2\,i\,\Box{\bar{C}},\quad \{s_b^{(1)}, s_{ab}^{(2)}\}\phi = -\,i\,m\,(B+\bar{B}) \equiv  2\,i\,m\,(\partial\cdot A),\] 
\[\{s_b^{(1)}, s_{ab}^{(2)}\} B = -\,2\,i\,\Box{B},\quad \{s_b^{(1)}, s_{ab}^{(2)}\} \bar{B} = 2\,i\,\Box{\bar{B}},\eqno (A.5)\]
where we have also exploited the restrictions (33) to demonstrate that {\it even if } we impose them from {\it outside},
the above anticommutators are {\it not} zero. We proceed ahead  and compute the anticommutator $\{s_{ab}^{(1)}, s_{ab}^{(2)}\} $.
In this context, we observe  the following:
\[\{ s_{ab}^{(1)}, s_{ab}^{(2)}\}\,\Psi  = 0, \qquad \Psi = A_\mu, \bar C, \phi, \tilde\phi, B, \bar B, {\cal B}, \bar {\cal B}.\eqno (A.6)\]
However, we note that the following is true, namely;
\[\{ s_{ab}^{(1)}, s_{ab}^{(2)}\}\, C = 4\,i\,\Box {\bar C},\eqno (A.7)\]
which demonstrates that the absolute anticommutativity between $s_{ab}^{(1)}$ and  $s_{ab}^{(2)}$ is {\it not} satisfied for our 2D theory because one of the
 fields (i.e. $C$) does not respect it.

We concentrate now on the {\it last} non-trivial anticommutator amongst the (anti-)BRST symmetry transformations $ s_{(a)b}^{(1, 2)}$.
In this connection, we note the following
\[\{ s_b^{(2)}, s_{ab}^{(1)}\}\,\Psi = 0,\qquad \Psi = \tilde \phi, {\cal B}, \bar {\cal B}, \eqno (A.8)\] 
which is just like our observation in (A.4). The non-vanishing and non-trivial  anticommutators in this regards are as follows
\[\{s_b^{(2)}, s_{ab}^{(1)}\}A_{\mu} = -\,i\,\partial_{\mu}(B-\bar{B}) = 2\,i\,m\,\partial_{\mu}\phi,\quad\{s_b^{(2)}, s_{ab}^{(1)}\}C = 2\,i\,\Box{C},\]
\[\{s_b^{(2)}, s_{ab}^{(1)}\}\bar{C} = -\,2\,i\,\Box{\bar{C}},\quad \{s_b^{(2)}, s_{ab}^{(1)}\}\phi = i\,m\,(B+\bar{B}) = -\,2\,i\,m\,(\partial.A),\]
\[\{s_b^{(2)}, s_{ab}^{(1)}\}B = 2\,i\,\Box{B},\quad\{s_b^{(2)}, s_{ab}^{(1)}\}\bar{B} = -\,2\,i\,\Box{\bar{B}},\eqno (A.9)\]
where we have used the restrictions (33) to demonstrate that the anticommutator $\{s_b^{(2)}, s_{ab}^{(1)}\}$ is {\it not } zero
(in spite of their imposition of (33) from {\it outside}).

At this stage,  we now take up the computation of the possible anticommutators amongst $s_{(a)d}^{(1, 2)}$ with our background knowledge  that the pairs 
$(s_d^{(1)}, s_{ad}^{(1)} )$ and $(s_d^{(2)}, s_{ad}^{(2)})$ absolutely anticommute {\it{without}} any use of the restrictions (33).
The non-trivial anticommutators from the $four$ fermionic operators $(s_{(a)d}^{(1, 2)} $) are as follows:
\[\{ s_d^{(1)}, s_d^{(2)}\},\qquad \{s_d^{(1)}, s_{ad}^{(2)}\}, \qquad \{ s_{ad}^{(1)}, s_d^{(2)}\}, \qquad \{s_{ad}^{(1)}, s_d^{(2)}\}. \eqno (A.10) \] 
It turns out that the following  general observation is correct:
$\{ s_{d}^{(1)}, s_{d}^{(2)}\}\,\Psi  = 0, \;\Psi = A_\mu, \bar C, \phi, \tilde\phi, B, \bar B, {\cal B}, \bar {\cal B},$
for the generic fields $\Psi$. However, we find that:
\[\{ s_{d}^{(1)}, s_{d}^{(2)}\}\, C = -\,4\,i\,\Box {\bar C}.\eqno (A.11)\]
The above anticommutator proves the fact that the fermionic operators   $s_{d}^{(1)}$ and  $s_{d}^{(2)}$ are {\it not} absolutely anticommuting in nature.
Next we focus on the evaluation  of $\{ s_{ad}^{(1)}, s_{ad}^{(2)}\}$ where we find that 
$\{ s_{ad}^{(1)}, s_{ad}^{(2)}\}\,\Psi  = 0, \;\; \Psi = A_\mu, \bar C, \phi, \tilde\phi, B, \bar B, {\cal B}, \bar {\cal B}$ for the generic field
$\Psi$ of our theory. However, we observe that the following is true, namely;
 \[\{ s_{ad}^{(1)}, s_{ad}^{(2)}\}\, \bar C = 4\,i\,\Box {C}.\eqno (A.12)\]
 Hence, $s_{ad}^{(1)}$ and  $s_{ad}^{(2)}$ do {\it not} absolutely anticommute with each-other (just like $s_{d}^{(1)}$ and  $s_{d}^{(2)}$).

 We take up now the anticommutator $\{s_{d}^{(1)}, s_{ad}^{(2)}\}$. In this context, we note that the following non-trivial anticommutators  are true, namely;
\[\{s_{d}^{(2)}, s_{ad}^{(1)}\}A_{\mu} = -\,i\,\varepsilon_{\mu\nu}\partial^{\nu}({\cal B} - \bar{\cal B}) = 2\,i\,m\,\varepsilon_{\mu\nu}\partial^{\nu} \tilde{\phi},\quad \{s_{d}^{(2)}, s_{ad}^{(1)}\}\,C = -\,2\,i\,\Box{C},\]
\[\{s_{d}^{(2)}, s_{ad}^{(1)}\}\bar{C} = 2\,i\,\,\Box{\bar{C}},\quad
\{s_{d}^{(2)}, s_{ad}^{(1)}\}\tilde{\phi} = i\,m\,({\cal B }+ \bar {\cal B}) = 2\,i\,m\,E,\]
\[\{s_{d}^{(2)}, s_{ad}^{(1)}\} {\cal B} = 2\,i\,\Box{\cal B},\quad
\{s_{d}^{(2)}, s_{ad}^{(1)}\}\bar{\cal {B}} = -\,2\,i\,\Box \bar{\cal{B}}.\eqno (A.13)\]
Thus, we find that {\it even } the impositions  of the restrictions (33) do {\it not} help in making $s_{d}^{(2)}$ and  $s_{ad}^{(1)}$
absolutely anticommuting in nature. However, we observe that: 
\[\{ s_d^{(2)}, s_{ad}^{(1)}\}\,\Psi = 0,\qquad \Psi =  \phi, {B}, \bar { B}. \eqno (A.14)\] 
In other words, we get the result that $s_{d}^{(2)}$ and  $s_{ad}^{(1)}$ absolutely anticommute {\it only} for the fields $\phi, {B}, \bar { B}$.
The last non-trivial  anticommutator  $\{s_{d}^{(1)}, s_{ad}^{(2)}\}$  is found to be absolutely anticommuting only for the fields: $\phi, {B}, \bar { B}$.
However, we find that the following non-trivial and non-zero anticommutators exist, namely;
\[\{s_{d}^{(1)}, s_{ad}^{(2)}\}A_{\mu} = i\,\epsilon_{\mu\nu}\partial^{\nu}({\cal B}- \bar{\cal B}) = -\,2\,i\,m\,\epsilon_{\mu\nu}\partial^{\nu}\tilde{\phi},\quad
\{s_{d}^{(1)}, s_{ad}^{(2)}\}C = -\,2\,i\,\Box{C},\]
\[\{s_{d}^{(1)}, s_{ad}^{(2)}\}\bar{C} = 2\,i\,\Box{\bar{C}},\quad
\{s_{d}^{(1)}, s_{ad}^{(2)}\}\tilde{\phi} = -\,i\,m\,({\cal B}+ \bar{\cal B}) = -\,2\,i\,m\,E,\]
\[\{s_{d}^{(1)}, s_{ad}^{(2)}\}{\cal B} = -\,2\,i\,\Box{\cal B},\quad
\{s_{d}^{(1)}, s_{ad}^{(2)}\}\bar{\cal {B}} = 2\,i\,\Box{\bar{ \cal B}}. \eqno (A.15)\]
We end this Appendix with the remarks that {\it all} the fermionic transformations $s_{(a)b}^{(1, 2)}$ and  $s_{(a)d}^{(1, 2)}$ do {\it not} 
absolutely anticommute amongst themselves. \\

 \vskip 1cm

\begin{center}
{\bf Appendix B: On the Absolute Anticommutativity of Nilpotent Charges }\\
\end{center}

\vskip 1cm

\noindent
We have already witnessed and verified that the pairs $(Q_b^{(1)}, Q_{ab}^{(1)}),\;(Q_d^{(1)}, Q_{ad}^{(1)}),\;(Q_b^{(2)},
 Q_{ab}^{(2)})$ and  $(Q_d^{(2)}, Q_{ad}^{(2)})$ absolutely anticommute with each-other (separately and independently).
 We have also noted that  {\it all} these charges are off-shell nilpotent $([Q_{(a)b}^{(1,\, 2)}]^2 = 0 = [Q_{(a)d}^{(1,\, 2)}]^2)$
 because of the following are true, namely;
\[ s_b^{(1)} Q_b^{(1)}  = 0,\qquad s_{ab}^{(1)}  Q_{ab}^{(1)}  = 0,\qquad s_d^{(1)}  Q_d^{(1)}  = 0,\qquad s_{ad}^{(1)}  Q_{ad}^{(1)}  = 0,\]
\[ s_b^{(2)} Q_b^{(2)}  = 0,\qquad s_{ab}^{(2)}  Q_{ab}^{(2)}  = 0,\qquad s_d^{(2)}  Q_d^{(2)}  = 0,\qquad s_{ad}^{(2)}  Q_{ad}^{(2)}  = 0,\eqno (B.1)\]
where we have used the relationship between the continuous symmetry transformations and their generators (cf. Eq. (6)). For example, we note 
that: $s_b^{(1)} Q_b^{(1)}  = -\,i\,\{Q_b^{(1)}, Q_b^{(1)}\}  = 0 \Rightarrow (Q_b^{(1)})^2 = 0.$ In fact, we have applied the fermionic transformations
 (26), (27), (35) and (37) {\it directly} on the charges (49) and (57) for the purpose of computations of $(B.1)$. 
 We use the expressions for the charges (cf. Eqs. (49), (57)) and nilpotent symmetry transformations (cf. Eqs. (26), (27), (35), (37)) to compute {\it all} the possible non-trivial anticommutators 
 amongst the conserved and off-shell nilpotent charges $(Q_{(a)b}^{(1,\, 2)})$. These {\it basic} non-trivial anticommutators for the (anti-)BRST charges are:
 \[\{Q_b^{(1)}, Q_b^{(2)}\}, \qquad \{Q_b^{(1)}, Q_{ab}^{(2)}\}, \qquad\{Q_b^{(2)}, Q_{ab}^{(1)}\}, \qquad\{Q_{ab}^{(1)}, Q_{ab}^{(2)}\}. \eqno (B.2)\]
 The above brackets  can be computed from the following {\it direct} applications of the nilpotent (anti-)BRST symmetries (26) and (27) on the charges (49) and (57), namely;
 \[ s_b^{(1)} Q_b^{(2)}  = -\,i\,\{Q_b^{(2)}, Q_b^{(1)}\} , \qquad  s_b^{(1)} Q_{ab}^{(2)}  = -\,i\,\{Q_{ab}^{(2)}, Q_b^{(1)}\}  ,\]
 \[ s_{ab}^{(1)} Q_{b}^{(2)}  = -\,i\,\{Q_{b}^{(2)}, Q_{ab}^{(1)}\}  ,\qquad s_{ab}^{(1)} Q_{ab}^{(2)}  = -\,i\,\{Q_{ab}^{(2)}, Q_{ab}^{(1)}\} 
 .\eqno (B.3)\]
The explicit computations of the l.h.s of the above equation are as follows 
\[s_b^{(1)}Q_b^{(2)} = 4\,m^2\,\int{dx\,C\,\dot{C}},\qquad
s_b^{(1)}Q_{ab}^{(2)} = \int{dx\,[2\,m^2\,(C\,\dot{\bar{C}}-\dot{C}\,\bar{C})+i(\bar{B}\,\dot{B}-\dot{\bar{B}}\,B)]},\]
\[s_{ab}^{(1)}Q_{b}^{(2)} = \int{dx\,[2\,m^2\,(C\,\dot{\bar{C}}-\dot{C}\,\bar{C}) + i\,(B\,\dot{\bar{B}}-\dot{B}\,\bar{B})]},\quad
s_{ab}^{(1)}Q_{ab}^{(2)} = 4\,m^2\,\int{dx\,\bar{C}\,\dot{\bar{C}}}.\eqno (B.4)\] 
Thus, it is crystal clear that the non-trivial anticommutators, listed in Eq. (B.2), are non-vanishing.
Hence, we conclude that, except the anticommutators $\{Q_b^{(1)}, Q_{ab}^{(1)}\}$ and $\{Q_b^{(2)}, Q_{ab}^{(2)}\}$, {\it rest} of the non-trivial 
anticommutators are non-zero (i.e. non-vanishing).

We perform similar exercise  with the nilpotent charges $Q_{(a)d}^{(1, 2)}$ and observe that the following non-trivial anticommutators amongst 
these (anti-)co-BRST charges, namely,
\[\{Q_d^{(1)}, Q_d^{(2)}\}, \qquad \{Q_d^{(1)}, Q_{ad}^{(2)}\}, \qquad\{Q_d^{(2)}, Q_{ad}^{(1)}\}, \qquad\{Q_{ad}^{(1)}, Q_{ad}^{(2)}\}, \eqno (B.5)\]
are to be evaluated using the basic principle of the continuous symmetry transformations and their  generators (cf. Eq. (6)).
In this connection, we find that: 
\[s_{d}^{(1)}Q_{d}^{(2)} = -\,i\,\{Q_d^{(2)}, Q_d^{(1)}\}  = 4\,m^2\,\int{dx\,\dot{\bar{C}}\,\bar{C}},\]
\[s_{d}^{(1)}Q_{ad}^{(2)} = -\,i\,\{Q_{ad}^{(2)}, Q_d^{(1)}\}  = \int{dx\,[2\,m^2\,(\dot{\bar{C}}\,C - \bar{C}\dot{C})+i\,(\dot{\bar{\cal B}}}\,{\cal B}
-\bar{\cal B}\,\dot{\cal B})],\]
\[s_{ad}^{(1)}Q_{d}^{(2)} = -\,i\,\{Q_d^{(2)}, Q_{ad}^{(1)}\}  = \int{dx\,[2\,m^2\,(\dot{\bar{C}}\,C - 
\bar{C}\dot{C})+i\,(\dot{\cal B}\,{\bar{\cal B}}-{\cal B}\,\dot{\bar{\cal B}}})],\]
\[s_{ad}^{(1)}Q_{ad}^{(2)} = -\,i\,\{Q_{ad}^{(2)}, Q_{ad}^{(1)}\}  = 4\,m^2\,\int {dx\,\dot C\,C}. \eqno (B.6)\] 
The above equation encapsulates the results that the non-trivial anticommutators amongst $Q_{(a)d}^{(1,2)}$ are {\it non-zero} establishing the fact that the absolute anticommutativity amongst the (anti-)co-BRST charges is  {\it{not}} true even if we impose  the  restrictions (33) from {\it outside}. Only the exceptions to these observations are:   
\[ s_d^{(1)} Q_{ad}^{(1)}  = -\,i\,\{Q_{ad}^{(1)}, Q_d^{(1)}\}  \equiv s_{ad}^{(1)}\,Q_d^{(1)},\]
 \[ s_{d}^{(2)} Q_{ad}^{(2)}  = -\,i\,\{Q_{ad}^{(2)}, Q_{d}^{(2)}\}  \equiv s_{ad}^{(2)}\,Q_d^{(2)},
 \eqno (B.7)\]
where there is {\it{no}} need of any kind of  restrictions from (33) because the pairs $(Q_d^{(1)}, Q_{ad}^{(1)})$ and $(Q_d^{(2)}, Q_{ad}^{(2)})$ absolutely anticommute ({\it separately} and {\it independently}).\\

\begin{center}
{\bf Appendix C: On the Derivation of $\kappa_1 = +\,1$}\\
\end{center}

\vskip 1cm

\noindent
We provide here an explicit computation of our result $\kappa_1 = +\,1$ in the case of determination of the secondary field $R_\mu (x)$
for the expansion (cf. Eq. (64)) 
\[B_\mu  (x, \bar\theta) = A_{\mu}(x) + \bar\theta\,R_\mu (x)\quad \equiv \quad A_{\mu}(x) + \bar\theta\, (\kappa_1\, \partial_\mu C(x)),\eqno (C.1)\]
where we have taken $R_\mu (x) = \kappa_1 \partial_\mu C$ (because of the restriction: $ B^\mu (x, \bar\theta)\,\partial_\mu C = A^\mu (x)\,\partial_\mu C (x)$
which leads to    $R_\mu (x) = \kappa_1 \partial_\mu C$). In fact, the constant $\kappa_1$ is just a numerical constant which has to be determined 
{\it precisely}. We further note that, a close look at Eqs. (63) and (65) shows that  $m\,\kappa_1 = \kappa_2 $ where $\kappa_2$ is a numerical constant in
 \[\Phi ^{(m)} (x, \bar\theta) = \phi (x) + \bar\theta\, (\kappa_2 \,C(x)). \eqno (C.2)\]
 Taking into account the top restriction in (66), we observe that we have: $m\,B_2 (x) = \kappa_2 B(x)$ which reduces to 
 $\,B_2 (x) = \kappa_1 B(x)$ due to our earlier derived relationship: $m\,\kappa_1 = \kappa_2$. At this  stage, the last restriction in (66) yields the following 
 \[ B_2 (x)\;\dot B (x)  =  \dot B_2 (x)\; B (x) \quad \Longrightarrow \quad \frac {1}{B_2}\; \frac {dB_2}{dt} = \frac {1}{B}\; \frac {dB}{dt}. \eqno (C.3)\]
Integrating the above equation w.r.t. the ``time" variable $t$, we obtain the following 
\[ ln \,B_2 (x) = ln \,(B(x)) + C,\eqno (C.4)\] 
where  $C$ is a numerical constant. Substituting $B_2 = \kappa_1\, B$, we obtain $ln (\kappa_1) = C$
which shows that $C$ can be made equal to zero by the choice $\kappa_1 = 1$. The latter choice immediately 
leads to: $\kappa_2 = m$. These lead to the explicit expressions for $(C.1)$ and $(C.2)$ as the  {\it ones} which lead to the derivation of $s_b ^{(1)}$, namely; 
\[B_\mu^{(b)}  (x, \bar\theta) = A_{\mu}(x) + \bar\theta\,(\partial_\mu C)(x)\quad \equiv \quad A_{\mu}(x) + \bar\theta\, (s_b ^{(1)} A_\mu (x)),\]
\[\Phi^{(b)} (x, \bar\theta) = \phi (x) + \bar\theta\, ( m \,C(x)) \quad\equiv \quad \phi (x) + \bar\theta\, ( s_b^{(1)}\,\phi(x)), \eqno (C.5)\]
which match with what we have already quoted in Eq. (68).\\

\vskip 1cm

\begin{center}
{\bf Appendix D: Bosonic and Ghost-Scale Symmetry Transformations }\\
\end{center}

\vskip 1cm

\noindent
To establish that the Lagrangian densities ${\cal L}_{(b_1)}$ and ${\cal L}_{(b_2)}$ represent the field-theoretic models for the Hodge theory, we define the 
bosonic symmetry transformations [59, 60]:
\[s^{(1)}_w = \{s_b^{(1)}, s_d^{(1)} \}  \equiv - \,\{s_{ab}^{(1)}, s_{ad}^{(1)} \}, \qquad
s^{(2)}_w = \{s_b^{(2)}, s_d^{(2)} \}  \equiv - \,\{s_{ab}^{(2)}, s_{ad}^{(2)} \}.\eqno (D.1)\] 
It is clear that the above symmetry transformations have  been defined with the help of {\it basic} fermionic symmetry transformations 
$s_{(a)b}^{(1,2)}$  and $s_{(a)d}^{(1,2)}$ for the Lagrangian densities  ${\cal L}_{(b_1)}$  and ${\cal L}_{(b_2)}$. The explicit forms of the 
bosonic symmetry transformations for all the fields (modulo a factor of $(-\,i)$) for {\it both} the Lagrangian densities  are:
\[s_w^{(1)} A_\mu = \partial_\mu {\cal B} + \varepsilon_{\mu\nu} \partial^\nu B, \;\,\;\qquad
s^{(1)}_w \phi = m\, {\cal B}, \,\;\;\qquad s^{(1)}_w \tilde \phi = m \, B,\]
\[s^{(1)}_w (\partial \cdot A) = \Box\, {\cal B}, \qquad s^{(1)}_w\, E = - \Box B,
\qquad s^{(1)}_w \bigl [B, {\cal B}, C, \bar C \bigr ] = 0, \] 
\[s_w^{(2)} A_\mu = \partial_\mu \bar {\cal B} + \varepsilon_{\mu\nu} \partial^\nu \bar B, \qquad
s^{(2)}_w \phi = - m\, \bar {\cal B}, \qquad s^{(2)}_w \tilde \phi = - m \, \bar B, \]
\[s^{(2)}_w (\partial \cdot A) = \Box\, \bar {\cal B}, \qquad s^{(2)}_w\, E = - \Box \bar B,
\qquad s^{(2)}_w \bigl [\bar B, \bar {\cal B}, C, \bar C \bigr ] = 0.\eqno (D.2)\] 
We note that the key feature of the above symmetry transformations is the observation that the (anti-)ghost fields do {\it not}
transform {\it at all} under $s_w^{(1,\, 2)}$. It is straightforward to check that the following are true, namely;
\[s_w^{(1)} \, {\cal L}_{(b_1)} = \partial_\mu \, \Bigl [ B \,\partial^\mu \,{\cal B} 
- {\cal B}\, \partial^\mu \,B
- m \,\varepsilon^{\mu\nu}\, \bigl (\phi \,\partial_\nu \,B + m\, A_\nu\, B \bigr ) 
- m \,\tilde \phi \,\partial^\mu \,B \Bigr ],\]
\[~~~~~~~s_w^{(2)} \, {\cal L}_{(b_2)} = \partial_\mu \, \Bigl [ \bar B\, \partial^\mu \,
\bar {\cal B} - \bar {\cal B} \,\partial^\mu \,\bar B
+ m \,\varepsilon^{\mu\nu} \,\bigl (\phi \,\partial_\nu \,\bar B - m \,A_\nu \,\bar B \bigr ) 
+ m \,\tilde \phi\, \partial^\mu\, \bar B \Bigr ],\eqno (D.3)\]
which demonstrate that the action integrals $S_1 = \int d^2 x \,{\cal L}_{(b_1)}$ and $S_2 = \int d^2 x \,{\cal L}_{(b_2)}$ remain {\it invariant} under
the bosonic symmetry transformations $s_w ^{(1,\, 2)}$.

In addition to the above continuous bosonic symmetry transformations $s_w^{(1, \,2)}$, the Lagrangian densities ${\cal L}_{(b_1)}$ and ${\cal L}_{(b_2)}$
respect the following  infinitesimal and continuous scale symmetry transformations $(s_g)$ [59, 60]
\[s_g C = +\,C,\quad s_g \bar C = -\,\bar C,\quad s_g \Psi = 0,\quad \Psi = A_\mu, \phi, \tilde\phi, B, \bar B, {\cal B}, \bar {\cal B},\eqno (D.4)\]
where the global scale parameter has been taken equal to {\it one } for the sake of brevity. In other words, we observe that the ghost and anti-ghost fields
(with ghost numbers + 1 and - 1, respectively) transform under the ghost-scale symmetry transformations {\it but} all the other fields (with ghost number zero)
do {\it not } transform {\it at all}. It can be checked that {\it all} the {\it six} continuous symmetries of the Lagrangian densities 
${\cal L}_{(b_1)}$  obey the algebra [59, 60]:
\[ (s_b^{(1)})^2 = (s_d^{(1)})^2 = (s_{ab}^{(1)})^2 = (s_{ad}^{(1)})^2 = 0, \;\; \{s_b^{(1)}, s_{ab}^{(1)} \} = 0, \;\;
\{s_d^{(1)}, s_{ad}^{(1)} \} = 0, \;\; \{s_b, s_{ad}^{(1)} \} = 0, \]
\[\{s_d^{(1)}, s_{ab}^{(1)} \} = 0, \quad \{s_b^{(1)}, s_{d}^{(1)} \} = s_w^{(1)} = -  \{s_{ab}^{(1)}, s_{ad}^{(1)} \}, \quad
[s_w^{(1)}, s_r^{(1)}] = 0, \quad r = (b, ab, d, ad, g), \]
\[ [s_g, s_b^{(1)} ] = + s_b^{(1)}, \qquad [s_g, s_d^{(1)} ] = - s_d^{(1)}, \qquad  [s_g, s_{ad}^{(1)} ] = + s_{ad}^{(1)},  
\qquad [s_g, s_{ab}^{(1)}] = - s_{ab}^{(1)}.  \eqno (D.5)\]
We perform the above exercise for ${\cal L}_{(b_2)}$ and obtain the following:
\[ (s_b^{(2)})^2 = (s_d^{(2)})^2 = (s_{ab}^{(2)})^2 = (s_{ad}^{(2)})^2 = 0, \;\; \{s_b^{(2)}, s_{ab}^{(2)} \} = 0, \;\;
\{s_d^{(2)}, s_{ad}^{(2)} \} = 0, \;\; \{s_b, s_{ad}^{(2)} \} = 0, \]
\[\{s_d^{(2)}, s_{ab}^{(2)} \} = 0, \quad \{s_b^{(2)}, s_{d}^{(2)} \} = s_w^{(2)} = -  \{s_{ab}^{(2)}, s_{ad}^{(2)} \}, \quad
[s_w^{(2)}, s_r^{(2)}] = 0, \quad r = (b, ab, d, ad, g), \]
\[[s_g, s_b^{(2)} ] = + s_b^{(2)}, \qquad [s_g, s_d^{(2)} ] = - s_d^{(2)}, \qquad  [s_g, s_{ad}^{(2)} ] = + s_{ad}^{(2)},  
\qquad [s_g, s_{ab}^{(2)}] = - s_{ab}^{(2)}.  \eqno (D.6)\]
The above algebra
is reminiscent of the algebra obeyed by the de Rham cohomologial operators of differential geometry  [7-11]
\[ d^2 = 0, \qquad \delta^2 = 0, \qquad \Delta = \{d, \delta \}, \qquad [\Delta, d] = 0, \qquad
[\Delta, \delta ] = 0, \eqno (D.7)\]
where the mapping between the symmetry operators and  cohomologial operators $(d, \delta, \Delta)$ is two-to-one as: 
\[(s_{b}^{(1)}, s_{ad}^{(1)}) \Rightarrow d,\quad (s_{d}^{(1)}, s_{ab}^{(1)}) \Rightarrow  \delta,\quad s_w ^{(1)}\Rightarrow \Delta,\]
\[(s_{b}^{(2)}, s_{ad}^{(2)}) \Rightarrow  d,\quad (s_{d}^{(2)}, s_{ab}^{(2)}) \Rightarrow  \delta,\quad s_w ^{(2)}\Rightarrow  \Delta. \eqno (D.8)\]
We conclude that the Lagrangian densities ${\cal L}_{(b_1)}$ and ${\cal L}_{(b_2)}$ represent a couple  of field-theoretic models for the Hodge theory
({\it separately} and {\it independently}) because the interplay between the discrete and continuous symmetries of these Lagrangian densities provide the 
physical realizations of the de Rham cohomological operators of differential geometry at the algebraic level as is evident from the mappings $(D.8)$.

\end{document}